\documentclass[fleqn,usenatbib]{mnras}

\usepackage{newtxtext,newtxmath}
\usepackage[T1]{fontenc}

\DeclareRobustCommand{\VAN}[3]{#2}
\let\VANthebibliography\thebibliography
\def\thebibliography{\DeclareRobustCommand{\VAN}[3]{##3}\VANthebibliography}

\usepackage{graphicx}	% Including figure files
\usepackage{bm,amsfonts,amsmath,tabularx,multirow,booktabs,array,mathtools}
\usepackage{comment}
\usepackage{xcolor,color,colortbl}
\usepackage{textgreek}
\usepackage[utf8]{inputenc}
\usepackage[british]{babel}
\usepackage{hyperref}
\hypersetup{
    colorlinks=true,
    linkcolor=blue,
    filecolor=blue,      
    citecolor=blue,
}
\usepackage[nameinlink,capitalise]{cleveref}
\makeatletter
\usepackage{etoolbox}
\patchcmd\H@refstepcounter{\protected@edef}{\protected@xdef}{}{}
\makeatother
\usepackage{color,colortbl}
\usepackage{tensind}
\tensordelimiter{?}
\usepackage{verbatim}
\usepackage[normalem]{ulem}
\usepackage{orcidlink}
\usepackage{soul,cancel}
\DeclareMathAlphabet{\pazocal}{OMS}{zplm}{m}{n}
\SetMathAlphabet\pazocal{bold}{OMS}{zplm}{bx}{n}

\allowdisplaybreaks
\newcommand{\ii}{\mathrm{i}}
\newcommand{\ee}{\mathrm{e}}

\newcommand{\cP}{\mathcal{P}}

\DeclareMathOperator{\sinc}{sinc}

\newcommand{\hi}{\textsc{Hi}}
\newcommand{\Ohi}{\Omega_\textsc{Hi}}
\newcommand{\bhi}{b_\textsc{Hi}}
\newcommand{\Thi}{\bar{T}_\textsc{Hi}}
\newcommand{\Ahi}{\mathcal{A}_\textsc{Hi}}
\newcommand{\fwhm}{\theta_\mathrm{FWHM}}
\newcommand{\Nfg}{N_\text{fg}}
\newcommand{\Nfgl}{N_\text{fg,L}}
\newcommand{\Nfgs}{N_\text{fg,S}}

\newcommand{\fum}{{\rm MHz}}
\newcommand{\dum}{h^{-1}\,{\rm Mpc}}
\newcommand{\kum}{h\,{\rm Mpc}^{-1}}
\newcommand{\pkum}{{\rm mK}^2\,(h^{-1}\,{\rm Mpc})^3}

\newcommand{\chess}{\textit{chess-board}}

\newcommand{\xv}{\bm{x}}

\newcommand{\kv}{\bm{k}}
\newcommand{\nv}{\hat{\bm{n}}}

\newcommand{\thv}{\bm{\theta}}

\let\Gamma\varGamma
\let\Delta\varDelta
\let\Theta\varTheta
\let\Lambda\varLambda
\let\Xi\varXi
\let\Pi\varPi
\let\Upsilon\varUpsilon
\let\Phi\varPhi
\let\Psi\varPsi
\let\Omega\varOmega

\defcitealias{MKxGAMA}{MK25}
\title[MeerKLASS detection of \hi\ auto-power spectrum]{\hi\ intensity mapping with MeerKAT: A measurement of the 21-cm auto-spectrum on cosmological scales}

\author[MeerKLASS Collaboration]{MeerKLASS Collaboration,\thanks{Corresponding authors:}
Matilde Barberi-Squarotti,$^{1,2,3,4}$\thanks{\href{mailto:matilde.barberi@unimi.it}{\texttt{Matilde.Barberi@unimi.it}} (MBS);}
Jos\'e L.\ Bernal,$^{5}$
Philip Bull,$^{6,7}$
\newauthor
Stefano Camera,$^{4,8,9,7}$
Isabella P.\ Carucci,$^{10,11}$\thanks{ \href{mailto:isabella.carucci@inaf.it}{\texttt{isabella.carucci@inaf.it}} (IPC);}
Zhaoting Chen,$^{12}$
Steven Cunnington,
$^{13,6}$
\newauthor
Brandon N.\ Engelbrecht,$^{7}$
Jos\'e Fonseca,$^{7,14,15}$
Keith Grainge,$^{6}$ 
Wenkai Hu,$^{16,7}$ 
Melis O.\ Irfan,$^{17,7}$
\newauthor
Yichao Li,$^{18,7}$
M\'ario G.\ Santos,$^{7,19}$
Marta Spinelli,$^{20,7}$
Jingying Wang,$^{21,7}$\thanks{\href{mailto:jywang@shao.ac.cn}{\texttt{jywang@shao.ac.cn}} (JW).}
\newauthor
Amadeus Witzemann,$^{7,6}$ and 
Laura Wolz$^{6}$
\\
$^{1}$Dipartimento di Fisica, Universit\`a degli Studi di Milano, via G.\ Celoria 16, 20133 Milano, Italy\\
$^{2}$INFN – Istituto Nazionale di Fisica Nucleare, Sezione di Milano, via G.\ Celoria 16, 20133 Milano, Italy\\
$^{3}$INAF -- Istituto Nazionale di Astrofisica, Osservatorio Astrofisico di Brera-Merate, via Brera 28, 20121 Milano, Italy\\
$^{4}$Dipartimento di Fisica, Universit\`a degli Studi di Torino, via P.\ Giuria 1, 10125 Torino, Italy\\
$^{5}$Instituto de F{\'i}sica de Cantabria (IFCA), CSIC-Univ. de Cantabria, Avda. de los Castros s/n, E-39005 Santander, Spain\\
$^{6}$Jodrell Bank Centre for Astrophysics, Department of Physics \& Astronomy, The University of Manchester, Manchester M13 9PL, UK\\
$^7$Department of Physics \& Astronomy, University of the Western
Cape, Robert Sobukwe Road, Cape Town 7535, South Africa\\
$^8$INFN – Istituto Nazionale di Fisica Nucleare, Sezione di Torino, via P.\ Giuria 1, 10125 Torino, Italy\\
$^9$INAF – Istituto Nazionale di Astrofisica, Osservatorio Astrofisico di Torino, 10025 Pino
Torinese, Italy\\
$^{10}$INAF -- Istituto Nazionale di Astrofisica, Osservatorio Astronomico di Trieste, Via G.B.\ Tiepolo 11, 34131 Trieste, Italy\\
$^{11}$IFPU -- Institute for Fundamental Physics of the Universe, Via Beirut 2, 34151 Trieste, Italy\\
$^{12}$Institute for Astronomy, The University of Edinburgh, Royal Observatory, Edinburgh EH9 3HJ, UK\\
$^{13}$Institute of Cosmology \& Gravitation, University of Portsmouth, Dennis Sciama Building, Portsmouth, PO1 3FX, UK\\
$^{14}$Instituto de Astrof\'isica e Ci\^encias do Espa\c{c}o, Universidade do Porto CAUP, 4150-762 Porto, Portugal\\
$^{15}$Departamento de F\'isica e Astronomia, Faculdade de Ci\^{e}ncias, Universidade do Porto, Rua do Campo Alegre 687, 4169-007 Porto, Portugal\\
$^{16}$State Key Laboratory of Radio Astronomy and Technology, National Astronomical Observatories, Chinese Academy of Sciences, Beijing 100101, China\\
$^{17}$Institute of Astrophysics, University of Cambridge, Madingley Road, CB3 0HA\\
$^{18}$Department of Physics, College of Sciences, Northeastern University, Wenhua Road, Shenyang 11089, China\\
$^{19}$South African Radio Astronomy Observatory (SARAO), Liesbeek House, Cape Town, 7700, South Africa\\
$^{20}$Observatoire de la Côte d’Azur, Laboratoire Lagrange, Bd de l’Observatoire, CS 34229, 06304 Nice cedex 4, France\\
$^{21}$State Key Laboratory of Radio Astronomy and Technology, Shanghai Astronomical Observatory, CAS, 80 Nandan Road, Shanghai 200030, China 
}

\date{Accepted XXX. Received YYY; in original form ZZZ}

\pubyear{\the\year{}}

\begin{document}
\label{firstpage}
\pagerange{\pageref{firstpage}--\pageref{lastpage}}
\maketitle

\begin{abstract}
We report a direct measurement of the 21-cm signal from neutral hydrogen (\hi) via the intensity mapping technique at redshift $z = 0.42$ through its power spectrum at cosmological scales $0.095\,\kum<k<0.245\,\kum$. We analyse intensity maps observed in the frequency range $971.2\,\fum<\nu<1023.6\,\fum$ during the 2021 observational campaign of the MeerKAT Large Area Synoptic Survey (MeerKLASS) spanning an area of approximately 236 $\deg^2$. Capitalising on the depth of this survey, we divide the data set into multiple observationally independent subsets. After performing blind contaminant separation independently on each subset, we exploit internal cross-correlations to suppress the instrumental noise plateau and isolate the cosmological signal from residual systematics. Although the subsets' auto-spectra show the presence of excess power that we deem linked to still unresolved systematics, the cross-spectra measurements satisfactorily pass a battery of null and robustness tests. We introduce a multi-subset formalism to infer the amplitude of the \hi\ auto-spectrum from our internal cross-spectra measurements while marginalising the potential signal loss that occurs during data processing, reaching a detection with signal-to-noise ratio of $6.2$. We constrain the clustering amplitude of the signal in terms of the \hi\ cosmic abundance and its linear bias, measuring $10^3\,\Ohi\bhi=1.271^{+1.080}_{-0.598}$. This direct measurement marks a milestone in demonstrating the effectiveness of the \hi\ intensity mapping technique for future surveys with the SKA Observatory, of which MeerKAT is a precursor.
\end{abstract}


%%%%%%%%%%%%%%%%%%%%%%%%%%%%%%%%%%%%%%%%%%%%%%%%%%

%%%%%%%%%%%%%%%%% BODY OF PAPER %%%%%%%%%%%%%%%%%%

\phantom{Ciao!}

%%%%%%%%%%%%%%%%%%%%%%%%%%%%%%%%%%%%%%%%%%%%%%%%%%
%%%%%%%%%%%%%%%%% INTRODUCTION %%%%%%%%%%%%%%%%%%
\section{Introduction}\label{intro}

Line intensity mapping, particularly using the 21-cm line from neutral hydrogen (\hi), offers a new way of making three-dimensional maps of the matter distribution over vast cosmic volumes \citep{Bharadwaj2001,Loeb2008,Chang2008,Wyithe2008,Seo2010,Ansari2012,Battye2013,Bull2015}. However, as with any novel observable, a variety of technical challenges must be overcome. These include instrumental effects, calibration-driven uncertainties, and low-level unflagged radio frequency interference (RFI), which, when combined with astrophysical foregrounds orders of magnitude brighter than the target signal, severely hinder the extraction of the cosmological signal \citep{Shaw2015,Carucci2020,2021MNRAS.506.5075M,Wang2022}. Nonetheless, several experiments worldwide are currently working hard to make cosmology with intensity mapping a reality. Although there are ongoing efforts dedicated to several spectral lines over a wide frequency range, the 21-cm line has a longer history of being studied. Hence, current experiments targeting \hi\ are comparatively larger in scale.

Part of such observational undertaking is the MeerKLASS project \citep{meerklass_wp}, which aims at observing cosmological 21-cm intensity maps using the 64 dishes of the MeerKAT telescope in single-dish mode, thus granting access to larger scales compared to standard interferometry. 
Previous efforts by the MeerKLASS collaboration have already led to detections of the 21-cm cosmological signal through cross-correlations with external spectroscopic galaxy surveys (see \citet{2025arXiv251027549C} for a recent review). The latter provide a well-characterised clustering signal that serves as a robust measurement for the Universe's large-scale structure, helping to mitigate systematics and boost signal-to-noise ratio. 
We reported a first detection in \citet{MKxWZ}, using the data from the 2019 MeerKLASS pilot survey \citep{MKcalibration} in cross-correlation with the overlapping WiggleZ Dark Energy Survey galaxy catalogue \citep{wigglez_def,wigglez_final}. We confirmed these results in a subsequent re-analysis in \citet{Carucci2024}, where an improved control of systematic errors and signal loss was achieved.

From September to December 2021, MeerKLASS conducted a second, deeper observing campaign in the L-Band, employing an upgraded self-calibration scheme. These data were analysed in cross-correlation with galaxies from the Galaxy and Mass Assembly (GAMA) spectroscopic catalogue \citep{gama_def,gama_dr4}, discussed in \citet[][`MK25' hereafter]{MKxGAMA} and in the follow-up stacking analysis presented in \citet{Zhaoting2025}. The present work is also based on those observations.

In parallel, \citet{Paul2023} reported a direct intensity mapping detection of the 21-cm signal at redshifts $z = 0.32$ and $0.44$ using different MeerKAT interferometric data \citep[see also an upper limit measurement by][]{Mazumder2025}. However, those measurements refer to small scales (approximately $1\,{\rm Mpc}$), far from the linear large scales required for cosmology.

Prior to MeerKLASS, other detections of the \hi\ cosmological signal were obtained from surveys performed by different telescopes, in cross-correlation with external data sets. Observations with the Green Bank Telescope (GBT) provided several cross-correlation detections at $0.6<z<1.0$ \citep{Chang2010,Masui2013,Switzer2013,Wolz2017,Wolz2022}. 
Similarly, \hi\ intensity maps with the Parkes radio telescope in cross-correlation with 2dF galaxies yielded a successful detection at a lower redshift ($z<0.1$) \citep{Anderson2018}. More recently, the Canadian Hydrogen Intensity Mapping Experiment (CHIME) presented the results of its first large-scale structure analysis, reporting detections of the \hi\ cosmological signal by stacking their \hi\ measurements on eBOSS galaxies and quasars at $0.8\lesssim z\lesssim1.4$ and cross correlating them with eBOSS Lyman-$\alpha$ forest measurements at $z\approx2.3$ \citep{2023ApJ...947...16A,2024ApJ...963...23A}. Encouragingly, CHIME also reported a detection of the 21-cm intensity mapping signal in auto-correlation at $z\approx1$ \citep{CHIME2025} at scales $k\gtrsim0.3\,{\rm Mpc}^{-1}$, significantly smaller than what we aim to with MeerKLASS.
Meanwhile, ongoing efforts with the Giant Metrewave Radio Telescope (GMRT) and its upgraded version, uGMRT, have delivered increasingly stringent upper limits on the \hi\ intensity mapping power spectrum at $z=0.37$ and $2.28$ \citep{2016MNRAS.460.2675R,2023MNRAS.520.2094E,2024MNRAS.529.3372E}. Drift-scan observations with the Five-hundred-meter Aperture Spherical radio Telescope (FAST) and their systematics characterisation are also progressing as reported in \citet{FAST2023} and \citet{2025ApJS..279...32Y}, further strengthening the viability of the technique of \hi\ line-intensity mapping after reionization across different instruments and frequency ranges.

There have also been detections with other line-intensity mapping tracers. They include observations of [\textsc{Cii}] \citep{2018MNRAS.478.1911P, 2019MNRAS.489L..53Y} and CO \citep{2024arXiv240607861R}, detected through the combination of cosmic infrared background maps with quasar and galaxy catalogues, or the stacking of unresolved optical observations with bright Lyman-$\alpha$ emitters to detect diffuse Lyman-$\alpha$ emission \citep{2022ApJ...929...90L, 2022ApJ...934L..26L, 2022ApJ...931...97K}. In particular, the auto-correlation of CO lines has been detected in the shot-noise regime---that is, small spatial scales---by COPSS \citep{2016ApJ...830...34K} and mmIME \citep{2020ApJ...901..141K}. 

In this paper, we report a direct detection of the \hi\ line-intensity mapping auto-correlation power spectrum on cosmological scales at redshift $z = 0.42$, characterised by a signal-to-noise ratio (SNR) of 6.2.

Taking advantage of the depth of the 2021 observational campaign (\citetalias{MKxGAMA})---the deepest \hi\ intensity mapping survey to date---we divide the data set into into subsets that are as observationally independent as possible. To remove contaminants more effectively, we employ a multiscale version of the Principal Component Analysis (PCA) technique, dubbed mPCA, which we recently introduced, optimised and tested with MeerKLASS data in \citet{Carucci2024}. We then perform an analysis based on \hi-only cross-subset power spectra. This strategy removes the noise bias from the power spectrum, and suppresses residual systematic contamination that is uncorrelated between subsets. Finally, we combine the internal cross-spectra from different subset combinations to constrain the amplitude of the \hi\ auto-correlation power spectrum.

We parameterise the \hi\ auto-spectrum amplitude as proportional to the product of the \hi\ cosmic abundance and linear bias, $\Ohi\,\bhi$, in analogy with previous works. We find $10^3\,\Ohi\bhi=1.271^{+1.080}_{-0.598}$, which is consistent with all the previous measurements reported by the MeerKLASS Collaboration through cross-correlations \citep[\citetalias{MKxGAMA};][]{MKxWZ,Carucci2024}.
 
The paper is structured as follows. In \cref{sec:data set} we present the data set used in the analysis, the strategy adopted to identify the subsets, and the pre-processing steps performed on the observed cubes. In \cref{sec:cleaning} we describe the contaminant separation method, and in \cref{sec:regrid} we detail the transformation from sky to comoving coordinates, needed to apply the power spectrum estimator, later discussed in \cref{sec:pestim} together with the covariance estimation details in \cref{sec:covariance}. The power spectrum model is outlined in \cref{sec:model}. \cref{sec:tablecloth} is dedicated to the assessment of the optimal level of foreground cleaning. The technique adopted for the fits and parameter inference is outlined in \cref{sec:mt_form}. \cref{sec:robustness} assesses the robustness of the analyses, whose final results are discussed in \cref{sec:discussion}. Finally, in \cref{sec:conclusion}, we summarise the results of this work and draw conclusions.

Throughout the analysis of this work, we fix the cosmological model to be that of \textit{Planck} 2018 \citep{2020A&A...641A...6P}. 

%%%%%%%%%%%%%%%%%%%%%%%%%%%%%%%%%%%%%%%%%%%%%%%%%%
%=================== DATA SET ===================%
%%%%%%%%%%%%%%%%%%%%%%%%%%%%%%%%%%%%%%%%%%%%%%%%%%

\section{Data} \label{sec:data set}

\begin{figure}
    \centering
    \includegraphics[width=1\columnwidth]{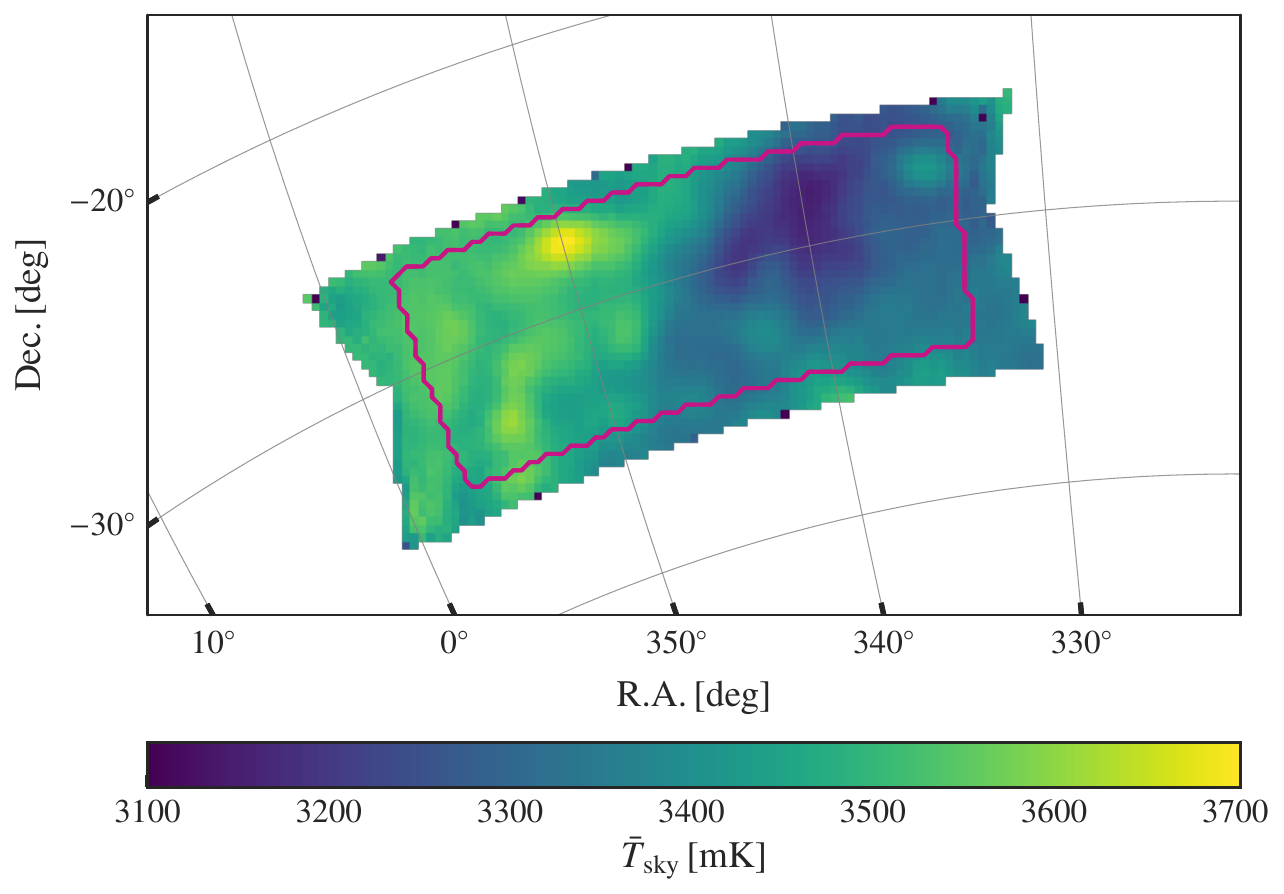}
    \caption{
    Full observed data set averaged along frequency. The final boundary of the region analysed in this work is highlighted in magenta. As a result of the trimming procedure (introduced later in \cref{sec:trim}), about $30\%$ of pixels are discarded; within the selected region 1 to 7 lines of sight are still to be flagged, as can be noticed in \cref{fig:cleanedmaps}.}
    \label{fig:trimming}
\end{figure}

We analyse the MeerKLASS L-band deep-field observations performed between $1^{\rm st}$ September and $29^{\rm th}$ December 2021, as described in \citetalias{MKxGAMA}. For our analysis, we employ data in the sub-band $971.2\,\fum<\nu<1023.6\,\fum$. The observations were made independently with each of the 64 dishes of the MeerKAT array operating with a fast, constant-elevation scanning strategy to help suppress pink noise, airmass, and ground spill fluctuations, as described in \citet{MKcalibration}. In practice, the dishes move back and forth in azimuth; two scans can be performed per night, corresponding to a rising and setting sky. This strategy results in a good sky coverage with close-to orthogonal scan crossings. Each scan---referred to as an observational `block'---lasts about $1.5$ hours. Before and after each block, we track a bright, compact celestial source to use as a bandpass and absolute flux calibrator. 

The complete set of observations of this field consists of 41 observational blocks, each observed by all 64 dishes, targeting the same sky patch (see footprint in \cref{fig:trimming}). We will refer to each observation by a single dish during a single block as a `single cube'. Every single cube is at first a set of raw data in arbitrary units, that is then calibrated using the {\tt KATcali} calibration pipeline \citep[\citetalias{MKxGAMA};][]{MKcalibration}. 

The calibration employs a Bayesian framework that models both instrumental effects and sky signals with the help of a noise diode which injects a signal in 20 s intervals into the observations. The procedure consists of two main steps. First, we track a celestial calibrator to calibrate the power of a periodically switched noise diode, which is assumed to have a stable output power thereafter; then, we calibrate the scan data using this calibrated diode power.
The pipeline consists of six levels of data processing. The overall flowchart is shown in Figure 1 of \citetalias{MKxGAMA}. We briefly summarise it below.
\begin{enumerate}
    \item At Level 1, a strong flagging is performed to discard time/channel sample points significantly contaminated by RFI due to telecommunication systems and satellites.
    \item At Level 2, we compare the raw signals---tracking the central pointing and four off-axis pointings of the celestial calibrator---with a model that incorporates the calibrator’s flux response at each beam position. From this fit, we derive the diode power, which we then use to calibrate the scan data.
    \item At Level 3, we use the diode power as the calibration reference to calibrate the scan data, yielding calibrated polarisation data (horizontal, HH, and vertical, VV).
    \item At Level 4, the calibrated data undergo additional RFI flagging and are then converted to intensity by averaging the HH and VV polarizations of $T_{\rm sky}$.
    
    \item At Level 5, the intensity maps are pixelised using the `Zenith equal area' map projection method: the coordinates of each observed point are converted into a pixel grid, with a pixel size of $0.3^\circ$ (chosen to be approximately one-third of the beam size). At this stage, the Level 5 outputs are single cubes, i.e., intensity maps of the target patch for each dish and block. To ensure pixel alignment, different cubes are gridded according to a unified standard.
    
    \item At Level 6, after a last low-RFI flagging, all the single cubes are combined into a final 3D cube by averaging all the data points falling within each pixel at each frequency. The map shown in \cref{fig:trimming} is an example of a Level-6 cube averaged along frequency.
\end{enumerate}

To enhance the data quality, Levels $3$---$6$ of the calibration process are iterated five more times as a self-calibration process, using the Level 6 output $T_{\rm sky}$ from the previous cycle as the input sky model for the next one.

The final output of the process are brightness temperature maps $T_{\rm sky}(\thv,\nu)$ with $\thv=({\rm R.A.},{\rm Dec.})$. The full patch covers the region with $330^\circ<{\rm R.A.}<360^\circ$ and $-36^\circ<{\rm Dec.}<-25.6^\circ$ for a total of $\approx236\,\deg^2$, with $n_{\rm R.A.}=133$ and $n_{\rm Dec.}=73$ pixels. After further flagging, only $n_\nu=252$ of the original 4096 equispaced $\delta\nu=0.209\,\fum$-width frequency channels of the L-band are kept for the cosmological analysis, and they cover the range $971.2\,\fum<\nu<1023.6\,\fum$, corresponding to redshifts $0.388<z<0.463$, hence leading to an effective redshift $z_{\rm eff} = 0.42$.

\begin{figure}
    \centering
\includegraphics[width=\columnwidth]{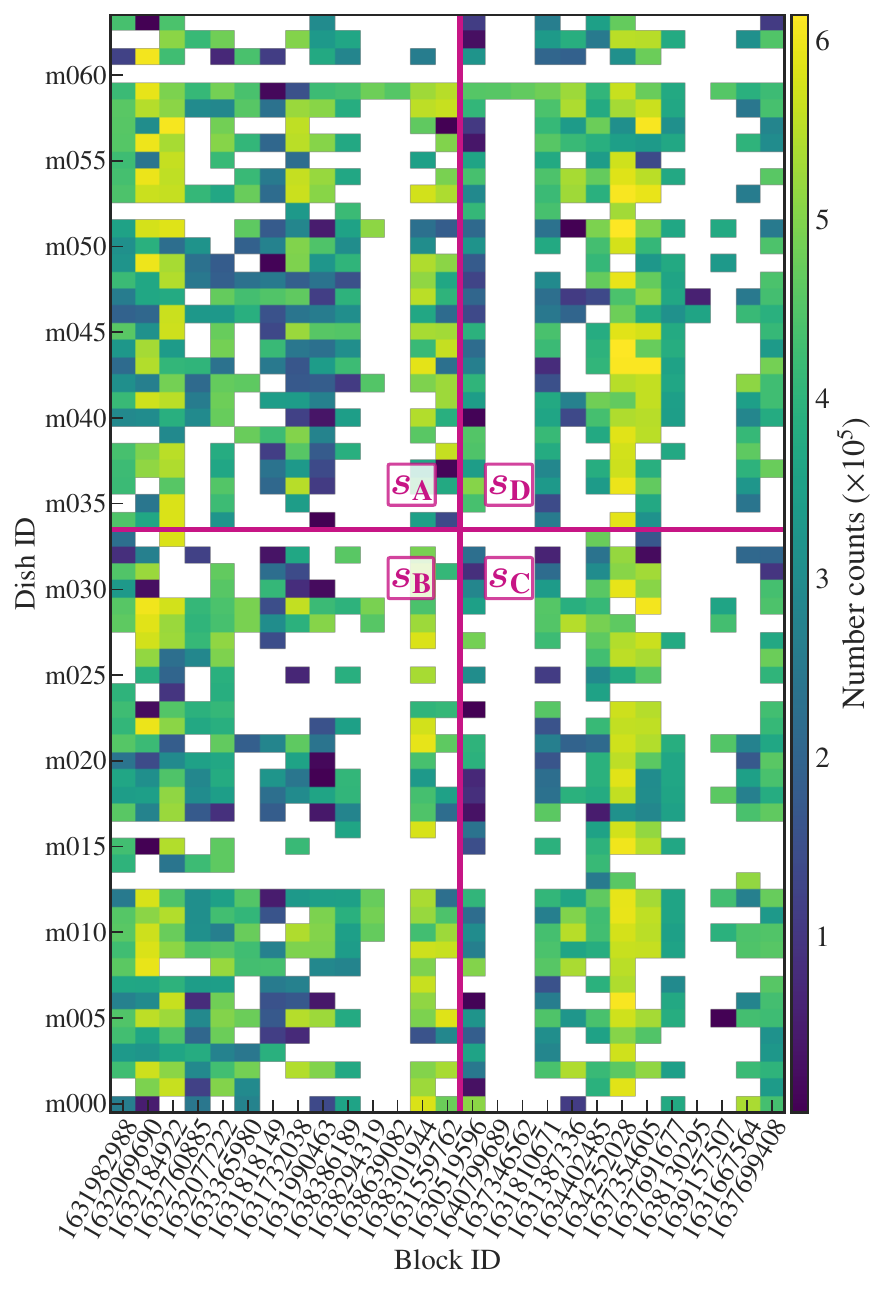}
    \caption{Hit map of the number counts of unflagged voxels of the Level 5 single cubes after the block reshuffling. The \chess\ division is depicted by the solid purple lines. Out of the 41 observed blocks, 14 are completely flagged and therefore not shown in this plot.}
    \label{fig:splitting}
\end{figure}

The amount of information encoded in each single cube after the five rounds of self-calibration can be quantified by the `hit' map (number count of the unflagged, good voxels, as we show in \cref{fig:splitting}), before they are merged in a Level-6 output. 

%%%%%%%%%%%%%%%%%%%%%%%%%%%%%%%%%%%%%%%%%%%%%%%%%%
%%%%%%%%%%%%%%%%%%%%%%%%%%%%%%%%%%%%%%%%%%%%%%%%%%

\subsection{Splitting the data set} \label{sec:split_sub}

In this work, we use the data sets produced by Level-5 of the last self-calibration loop, i.e., before the merging of the individual data blocks. We combine the data into subsets---identified by grouping different blocks and dishes---and run Level 6 independently for each subset. The latter procedure allows us to obtain independent subsets, all covering the same area on the sky, as done in \citet{Switzer2013} and \citet{Wolz2017, Wolz2022}. 
Before the splitting, we randomly shuffle the block order to maximise the diversity of observing conditions within each subset, rather than grouping together blocks observed under similar conditions (e.g., azimuth). The reason is that block-dependent systematics are not expected to behave as random noise, but may remain correlated between blocks observed under similar conditions, for example because of the telescope pointing and its associated observing environment. If such blocks are grouped together, the corresponding contamination may become more coherent within a subset and therefore less likely to average down when the blocks are combined.

The hit map in \cref{fig:splitting} shows the number counts of the Level-5 single cubes after shuffling, to be compared to the original, non-shuffled Figure 3 in \citetalias{MKxGAMA}. After shuffling, the subsets are then defined in such a way that their signal-to-noise ratios, which we quantify as the global number counts in the hits maps, are comparable.

Once subsets are defined and calibrated through the last Level 6 step, we treat each of them independently, in terms of contaminant separation and power spectrum estimation. We define a survey selection window function, $W\left(\thv,\nu\right)$, and a map of weights, $w(\thv,\nu)$ for each of them. For the latter, we use the normalised time stamp counts, i.e., the number of observational recordings for each 2 s pointing, in each pixel \citep{MKxWZ}, which corresponds to an inverse-noise-variance weighting, whereas the former is a simple binary window function---i.e., $W(\thv,\nu)$ is either 1 or 0 if a pixel at coordinates $\thv$ and frequency channel $\nu$ is filled or not, respectively. 

The strategy we adopt to split the data is to divide it both block-wise and dish-wise, yielding four subsets: $s_{\rm A}$, $s_{\rm B}$, $s_{\rm C}$, and $s_{\rm D}$, as shown in \cref{fig:splitting} with magenta lines. When carrying out this \chess\ division, the pairs $s_{\rm A}$-$s_{\rm D}$ and $s_{\rm B}$-$s_{\rm C}$ are made of single cubes that are observed by the same dishes, whereas the pairs $s_{\rm A}$-$s_{\rm B}$ and $s_{\rm C}$-$s_{\rm D}$ combine the same blocks observed by different dishes. The pairs $s_{\rm A}$-$s_{\rm C}$ and $s_{\rm B}$-$s_{\rm D}$ are the most independent: they share neither dishes nor blocks.

%%%%%%%%%%%%%%%%%%%%%%%%%%%%%%%%%%%%%%%%%%%%%%%%%%
%%%%%%%%%%%%%%%%%%%%%%%%%%%%%%%%%%%%%%%%%%%%%%%%%%

\subsection{Noise level} \label{sec:abba}

Below, we evaluate the instrumental thermal noise level in the maps. It helps characterise our data set and corroborate the equivalence of the subsets in terms of signal-to-noise ratio. In our analysis, we do not utilise the noise estimate for the cosmological analysis, as the subsets' cross-correlation will take care of its cancellation. However, the thermal noise will contribute to the covariance of our measurement, automatically taken into account in our data-driven estimation of the covariance.

We use the ABBA method \citep{MKcalibration}. We consider the difference between four neighbouring frequency channels in the calibrated maps, $T_{\rm sky}$. In each pixel $i$ of the $N_{\rm pix}$ pixels in the map, we have
\begin{multline}\label{eq:deltaT_ABBA}
    \Delta T_i(\nu^{\star}) = \frac{1}{2} \left[ T_i^{\rm sky}(\nu)+T_i^{\rm sky}(\nu+\delta\nu) \right] \\
    - \frac{1}{2} \left[ T_i^{\rm sky}(\nu-\delta\nu)+T_i^{\rm sky}(\nu+2\,\delta\nu) \right] \;,
\end{multline} 
where $\nu^{\star}$ stands for the frequency resulting from the combination of the four channels $\nu-\delta\nu$, $\nu$, $\nu+\delta\nu$, and $\nu+2\,\delta\nu$, with $\delta\nu=0.209\,\fum$ being the aforementioned frequency width of the MeerKAT L-band channels. Any signal that is constant or varying linearly across the channels should therefore cancel out in the computation of $\Delta T_i$, which consequently contains only noise.

Thence, we estimate the noise variance as the weighted root mean square over the $\Delta T_i(\nu^{\star})$ maps to account for non-uniform noise across the map. Namely,
\begin{equation} \label{eq:TRMS_ABBA}
    \Delta T_{\rm RMS}^2 = \frac{N_{\rm pix}}{N_{\rm pix}-1}\,\left[ \frac{\sum_i w_{{\rm N},i}\,\Delta T_i^2}{\sum_i w_{{\rm N},i}} - \left( \frac{\sum_i w_{{\rm N},i}\,\Delta T_i}{\sum_i w_{{\rm N},i}} \right)^2 \right] \;,
\end{equation}
with the $\nu^\star$-dependency left implicit in $w_{{\rm N},i}$ and $\Delta T_i$ to avoid clutter. The weights are standard inverse-variance weights,\footnote{Note that these weights are not the ones defined in \cref{sec:split_sub} and used for the estimation of the power spectra.} computed from the theoretical variance as $w_{{\rm N},i}(\nu^\star)=\sigma_{{\rm th},i}^{-2}(\nu^\star)$, with the ABBA theoretical variance in each pixel reading
\begin{multline}\label{eq:th_ABBA}
    \sigma_{{\rm th},i}^{2}(\nu^\star) = \frac{1}{4}\,\Big\{\left[ \sigma_{{\rm N},i}^{2}(\nu)+\sigma_{{\rm N},i}^{2}(\nu+\delta\nu) \right] \\
    + \left[ \sigma_{{\rm N},i}^{2}(\nu-\delta\nu)+\sigma_{{\rm N},i}^{2}(\nu+2\,\delta\nu) \right]\Big\} \;,
\end{multline} 
and the noise, at map level, being given by the radiometer equation \citep{2013tra..book.....W},
\begin{equation} \label{eq:radiometer}
    \sigma_{{\rm N},i}^2(\nu) = \frac{T_{\rm sys}^2(\nu)}{2\,\delta\nu\,t_{{\rm p},i}(\nu)}\;.
\end{equation}
Above, $T_{\rm sys}$ is the system temperature and $t_{\rm p}$ is the total observation time in each voxel, which can be obtained from the counts in the time stamps (each one lasting $2\,{\rm s}$). For the system temperature of the MeerKAT telescope, we use $T_{\rm sys}\simeq16\,{\rm K}$ across the whole frequency range covered, as estimated with the model provided in \citet{MKxWZ} and \citetalias{MKxGAMA}, which is tuned to match the assessment in \citet{MKcalibration}.

It is also possible to compute the root mean square of the theoretical variance,
\begin{equation} \label{eq:thRMS_ABBA}
     \sigma^2_{\rm th} (\nu^{\star}) = \left[\frac1{N_{\rm pix}}\,\sum_{i=1}^{N_{\rm pix}} \sigma^{-2}_{{\rm th},i}(\nu^{\star})  \right]^{-1}\;,
\end{equation}
which can be then compared with \cref{eq:TRMS_ABBA}.

\begin{table}
\centering
 \caption{Noise level estimates for the subsets corresponding to the \chess\ split. The values of the median of $\Delta T_{\rm RMS}(\nu^\star)$ and the theoretical variance are provided. In parentheses are the results of the estimates after the edges of the maps are trimmed as explained in \cref{sec:trim}.}
 \label{table:ABBAnoise}
 \begin{tabular}{ccc}
 \toprule
Subset & $\Delta T_{\rm RMS} \, {\rm [mK]}$ (trim) & $\sigma_{\rm th} \, {\rm [mK]}$ (trim) \\ 
  \midrule
$s_{\rm A}$ & $2.25 \, (2.08)$ & $2.19 \, (2.04)$ \\ 
$s_{\rm B}$ &$2.24 \, (2.06)$ & $2.11 \, (1.97)$ \\ 
$s_{\rm C}$ & $2.37 \, (2.19)$ & $2.25 \, (2.11)$ \\ 
$s_{\rm D}$ & $2.33 \, (2.16)$ & $2.30 \, (2.15)$ \\ 
  \bottomrule
 \end{tabular}
\end{table}

We report the noise level estimates in \cref{table:ABBAnoise}. The noise level is confirmed to be in agreement within about $10\%$ in all subsets. The ratio $\Delta T_{\rm RMS}(\nu^\star)/\sigma_{\rm th}(\nu^\star)$ is around 1.1, in agreement with the analysis by \citetalias{MKxGAMA}. Moreover, when dividing the data set in the four subsets, the noise increases by a factor $\smash{\sqrt{4}}$ as expected.

We perform other consistency tests regarding the reliability of the noise estimates in \cref{app:noise}.

%%%%%%%%%%%%%%%%%%%%%%%%%%%%%%%%%%%%%%%%%%%%%%%%%%
%%%%%%%%%%%%%%%%%%%%%%%%%%%%%%%%%%%%%%%%%%%%%%%%%%

\subsection{Map trimming} \label{sec:trim}

Given our scanning strategy, the edges of the observed patch are noisier than the inner area. Hence, pixels at the boundaries are more prone to be outliers in the temperature distribution of each map, and their presence degrades the performance of contaminant separation even when down-weighting them. Thus, we trim all the subsets' footprint, and the survey window functions and weight maps alike, practically setting to zero all pixels outside the chosen R.A.\ and Dec.\ limits in \cref{table:dim}. Additionally, if flagged (empty) pixels fall into the final target region, we flag the whole corresponding line of sight along frequency (i.e., that specific pixel in all maps), to preserve uniform sampling across the frequencies and ease the PCA decomposition. The final footprint is delimited in magenta in \cref{fig:trimming}.

After trimming the maps, we estimate again the noise level using the ABBA method and find a $10\%$ decrease compared to the `untrimmed' case (\cref{table:ABBAnoise}), corroborating our choice of discarding the pixels at the edges. In the remainder of the analysis, we present results for these final, trimmed maps.
\begin{table}
\centering
 \caption{R.A and Dec. limits of the final, trimmed footprint and physical, comoving dimension of the corresponding Cartesian cube (see \cref{sec:regrid} and \cref{fig:taper} for details on the Cartesian gridding).}
 \label{table:dim}
 \begin{tabular}{lc}
 \toprule
R.A.\ range & $333.2^{\circ},357.5^{\circ}$ \\ 
Dec.\ range & $-34.6^{\circ},-26.5^{\circ}$ \\ 
  \midrule
$L_x \; [\dum]$ & $660.34$ \\ 
$L_y \; [\dum]$ & $295.90$ \\ 
$L_z \; [\dum]$ & $233.66$ \\ 
  \bottomrule
 \end{tabular}
\end{table}

%%%%%%%%%%%%%%%%%%%%%%%%%%%%%%%%%%%%%%%%%%%%%%%%%%
%=================== CLEANING ===================%
%%%%%%%%%%%%%%%%%%%%%%%%%%%%%%%%%%%%%%%%%%%%%%%%%%

\section{Contaminant subtraction} \label{sec:cleaning}

In this work, we use multiscale Principal Component Analysis (mPCA) to separate contaminants from the cosmological signal in the cubes. We closely follow \citet{Carucci2024}, where the method was introduced and successfully applied to a previous MeerKLASS \hi\ intensity mapping data set at the same frequencies analysed here \citep{MKcalibration,MKxWZ}. 

This technique works with two main assumptions. \textit{(i)} Astrophysical foregrounds are spectrally smoother than the \hi\ signal, hence hold few degrees of freedom in the frequency direction, and contribute the most to the overall variance of the observed data set. However, their coupling with instrumental systematics corrupts their spectral smoothness. \textit{(ii)} Decomposing the maps in spatial scales helps breaking the `astrophysical, smooth regime' from the systematics-corrupted one, easing the separation of contaminants from the signal.

To optimise further the cleaning, we perform it independently for each subset cube $s_i$, as we expect different systematics to be at play in the different subsets, which would mix differently with the common astrophysical foregrounds and would need different mPCA solutions.

In practice, we first redundantly decompose $s_i$ into its large and small angular scale using the isotropic undecimated wavelet dictionary, also called the `starlet' transform \citep{4060954}, 
\begin{equation} \label{eq:redundantly}
    s_i = s_{{\rm L},i} + s_{{\rm S},i}\;.
\end{equation}
In other words, $s_{{\rm L},i}$ and $s_{{\rm S},i}$ have the same $N_{\rm pix} \times N_\nu$ dimensions as their parent full-scale $s_i$.
Secondly, we perform PCA to the above terms separately and find the optimal `clean' solutions, $s_{{\rm L, clean},i}$ and $s_{{\rm S, clean},i}$, as we will describe later. Lastly, we sum the two cleaned wavelet-decomposed cubes to define the final, cleaned subset $s_{{\rm clean},i}$,
\begin{equation}
    s_{{\rm clean},i} = s_{{\rm L, clean},i} + s_{{\rm S, clean},i}\;.
\end{equation}
In general, PCA can be thought of as a change of basis, a decomposition of a data cube $s(\bm\theta,\nu)$ in new $N_\nu$ `principal components', ordered in decreasing variance; that is, by subtracting PCA components from the original cube, we increasingly reduce its variance asymptotically. The foreground subtraction is done by discarding the first $N_{\rm fg}$ PCA components---that should characterise and isolate the contaminants---and leaving the remaining $N_\nu-N_{\rm fg}$ PCA components in the cleaned data set.

Hence, when performing mPCA, we need to set the number of components to be removed during the cleaning. For each subset $s_i$, this translates into setting the pair $N_{\rm fg,L}$ and $N_{\rm fg,S}$, for its large and small scale cleanings. In our work, we compare the cross-correlation measurements among different subsets with the expected theoretical model to inform the optimal $N_{\rm fg,L}$ and $N_{\rm fg,S}$ parameter space. The latter approach has been used successfully with the cross-correlation with an external data set \citep[\citetalias{MKxGAMA};][]{Carucci2024}.

\begin{figure}
    \centering
    \includegraphics[width=1\columnwidth]{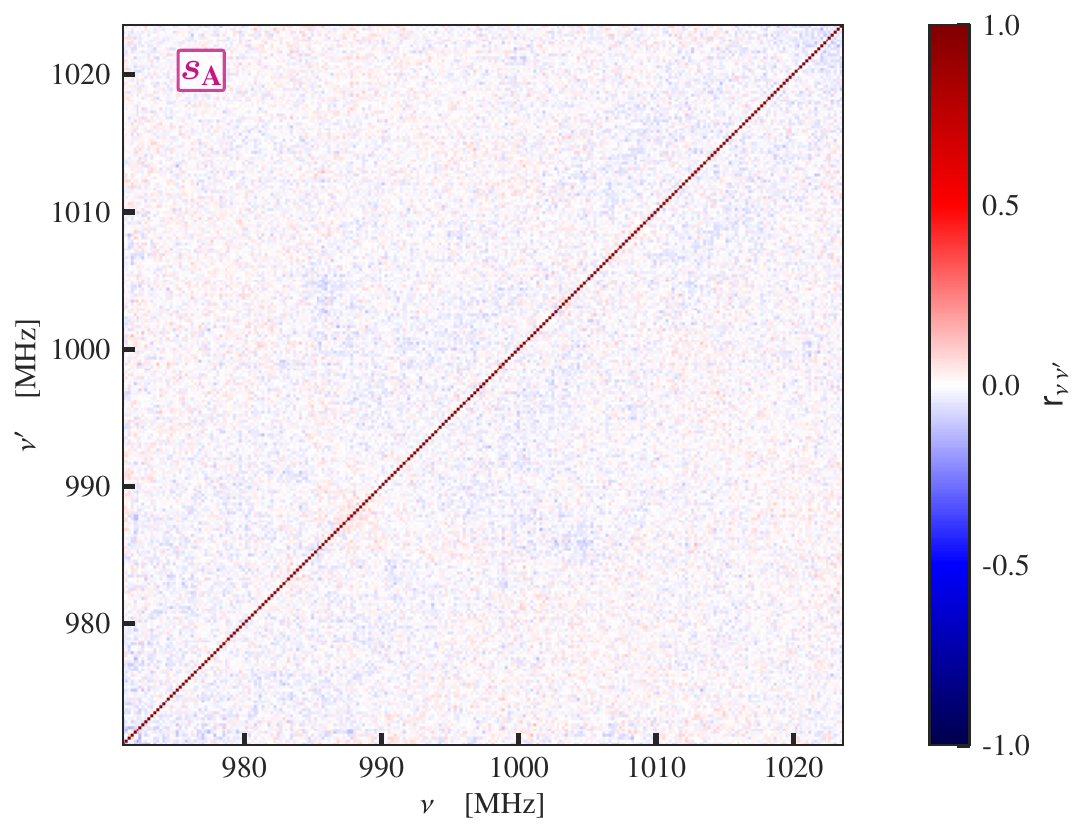}
    \caption{Frequency-frequency correlation matrix for the cleaned $s_{\rm A}$ subset; the adopted values of $\Nfgl$ and $\Nfgs$ are in \cref{table:cleaning}. The colour bar highlights the absence of correlation in white, while red and blue correspond respectively to correlation and anti-correlation. We do not find significant differences between $s_{\rm A}$ and the other subsets. 
    }
    \label{fig:freq_corr}
\end{figure}

As a first assessment of the foreground subtraction performance, we compute the frequency-frequency correlation matrix of cleaned cubes, $\mathsf{r}_{\nu\nu'}$. Indeed, we expect cosmological signal (and noise) to be uncorrelated in frequency (definitely above the $> 2$---$3$ MHz scale, i.e., $\approx10$ channels), unlike astrophysical foregrounds which are highly correlated (see also the discussion in \cref{app:radial_valid}). An example of such matrices is shown in \cref{fig:freq_corr}; as expected, we find a significant correlation along the diagonal, while off-diagonal terms are subdominant (their correlation is weaker than $15\%$, with an average of $2.5\%$), supporting the effectiveness of the cleaning performed. 

To conclude this section, we must note that in the contaminant separation process, we expect leakages between the two parts of the separation: residual contaminants in the cleaned cube and cosmological signal leaked in the subtracted components. An optimised cleaning pipeline---the algorithm of choice and its pre-processing steps to ease it---does its best to minimise these leakages. Nevertheless, we need to characterise and address leakages after the contaminant subtraction, for as unbiased an analysis as possible of the \hi\ content in our data cubes, as we will later do.

%%%%%%%%%%%%%%%%%%%%%%%%%%%%%%%%%%%%%%%%%%%%%%%%%%
%================== REGRIDDING ==================%
%%%%%%%%%%%%%%%%%%%%%%%%%%%%%%%%%%%%%%%%%%%%%%%%%%

\section{Cartesian gridding} \label{sec:regrid}

After data has been cleaned with mPCA, we measure the power spectra through fast Fourier transforms on a Cartesian grid, which we can define given the assumed cosmology \citep{2020A&A...641A...6P}. We work at the effective redshift $z_{\rm eff}=0.42$, given by the median frequency of the band $\nu_{\rm med} = 997.38\,\fum$. Given the small size of our footprint and our signal-to-noise regime, we work in the plane-parallel approximation and neglect projection or dilation effects \citep[e.g.,][]{AP1979}.

We transform the cleaned data cubes, $s_{{\rm clean},i}(\thv, \nu)$, from celestial coordinates (R.A., Dec., $\nu$) into comoving Cartesian coordinates, $s_{{\rm clean},i}(\xv)$ with $\xv=(x, y, z)$ following the method provided in \citet{Cunnington2024} and also used in \citetalias{MKxGAMA}.\footnote{Note that cubes of shape $n_{\rm R.A.}\times n_{\rm Dec.}\times n_{\nu}$ are gridded into cubes of shape $n_x\times n_y\times n_z$, in exactly this order. There is no ambiguity between the Cartesian $z$ coordinate and the redshift since we purposefully align the $z$-axis along the line-of-sight direction.} We employ a Monte Carlo integration technique that distributes sampling particles across the map and assigns them a Cartesian coordinate. In particular, $N_{\rm sp}$ sampling particles are randomly distributed in each cube voxel, whose brightness temperature is assigned to them, before being converted in comoving coordinates and being allocated in the Cartesian voxel that corresponds to their coordinates. Finally, the voxels of the comoving box are normalised according to the number of sampling particles falling inside.

We define the Cartesian boxes to have the same resolution as the original cube and therefore with $(n_x, n_y, n_z) =(133, 733, 252)$ cells in each dimension. We find that $N_{\rm sp}=20$ is a sufficient number of sampling particles to cover the whole Cartesian box adequately; we also include an empty buffer of $10$ additional sampling particles in each direction not to lose sampling particles falling just around the original footprint.

We use the nearest-grid-point (NGP) mass assignment as interpolation scheme and we perform a compensation for field smoothing induced by the particle assignment \citep{2005ApJ...620..559J}. This correction is performed on the Fourier transform of the Cartesian-gridded field $\tilde{s}^{\rm NGP}_{{\rm clean},i}(\kv)$,
\begin{equation}
    \tilde{s}_{{\rm clean},i}(\kv) = \frac{\tilde{s}^{\rm NGP}_{{\rm clean},i}(\kv)}{\mathcal{W}_{\rm NGP}(\kv)} \;,
\end{equation}
where the nearest grid point window function \(\mathcal{W}_{\rm NGP}(\kv)\) is the Fourier transform of the top-hat function used to interpolate particles into cells of size $\smash{L_i/n_i}$, where $L_i$ and $n_i$ are, respectively, the size in $\dum$ (see \cref{table:dim}) and the number of cells in each dimension $x^i=(x,y,z)$. In other words,
\begin{equation}
    \mathcal{W}_{\rm NGP}(\kv) = \prod_{i=1}^3 \sinc\left(\frac1{2}\frac{k_i\,L_i}{n_i}\right) \;.
\end{equation}

Compared to \citetalias{MKxGAMA}, we additionally include interlacing---also in Fourier space---to remove the effects of aliasing due to the discretisation, i.e., excess of signal at small scales when approaching the Nyquist frequency of the grid \citep{2016MNRAS.460.3624S}. The combination of the NGP mass assignment scheme with compensation and interlacing allows us to reach sub-percent accuracy on the scales of interest \citep{Cunnington2024}, which are anyway well below the Nyquist frequency, making the use of higher-order mass assignment schemes unnecessary.

Analogously, the binary window and the set of weights are gridded on the same three-dimensional grid, to obtain the window $W_i(\xv)$ and the weights $w_i(\xv)$ in comoving coordinates consistently.

\begin{figure}
\centering 
\includegraphics[width=\columnwidth]{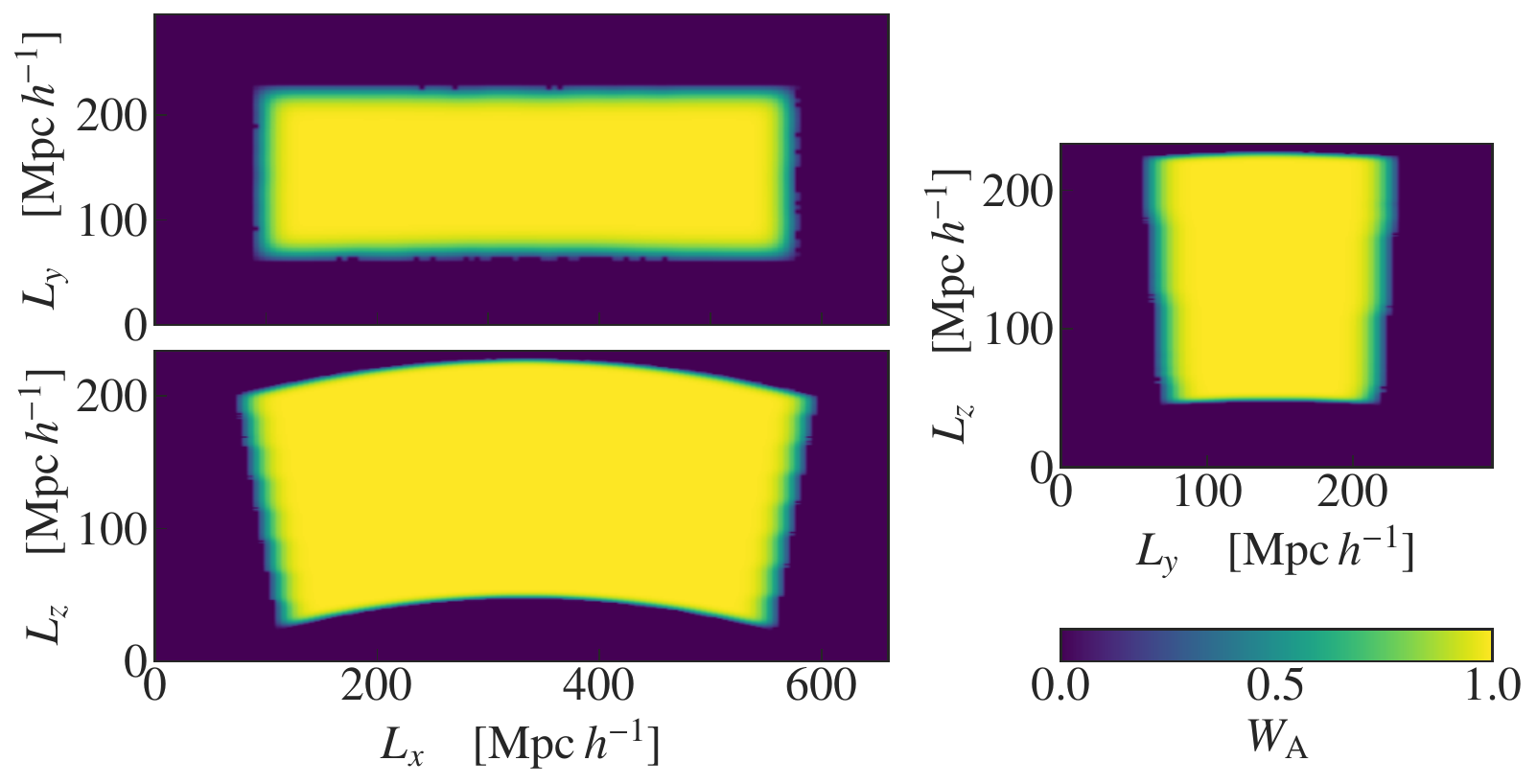}
\caption{Slices of the Cartesian-gridded and tapered window in the $(x,y)$-, $(x,z)$- and $(y,z)$-planes, each one taken at the central coordinate along the orthogonal axis. We show the subset $s_{\rm A}$ case as the others do not show any differences. The tapering smooths the edges in all three directions, closely following the boundaries of the observed region. These windows do not exhibit any holes; in the case of the subsets with a complete coverage in celestial coordinates, this confirms that $N_{\rm sp}$ is sufficient to sample the data cube fully; for the other cases, since the flagged pixels are few and non-adjacent, we find no gaps either in the Cartesian-gridded windows.} 
\label{fig:taper} 
\end{figure}

After the coordinate transformation, the fringes of the Cartesian-gridded fields are smoothed to avoid sharp discontinuities when performing the fast Fourier transform. The Cartesian volume containing our data has non-orthogonal boundaries preventing us from directly tapering the box (see \cref{fig:taper}). Instead, we tailor the tapering by starting from the binary survey window function $W_i(\xv)$ of each subset and iteratively applying a Gaussian smoothing to it. We show the smoothing effect of the final tapering in \cref{fig:taper}. Unlike in \citetalias{MKxGAMA}, the tapering is applied along all three directions rather than only in the radial one because we have a more extensive coverage in the angular direction and do not suffer from an excessive down weighting of pixels.\footnote{\citetalias{MKxGAMA} analysed the region overlapping with the GAMA galaxy survey, resulting in a smaller footprint.}

%%%%%%%%%%%%%%%%%%%%%%%%%%%%%%%%%%%%%%%%%%%%%%%%%%
%================ POWER SPECTRUM ================%
%%%%%%%%%%%%%%%%%%%%%%%%%%%%%%%%%%%%%%%%%%%%%%%%%%

\section{Power spectrum estimation} \label{sec:pestim}

We follow \citetalias{MKxGAMA} for computing the power spectrum of the \hi\ intensity maps of each cleaned subset $s_i$ in auto- or in cross-correlation with another $s_j$.

After the gridding in Cartesian dimensions, the cube undergoes the fast Fourier transform
\begin{equation} \label{eq:fourierfield}
    \widetilde F_i(\kv)=\sum_{\xv} s_{{\rm clean},i}(\xv)\,W_i(\xv)\,w_i(\xv)\,\ee^{-\ii\,\kv{\cdot}\xv} \;.
\end{equation}
The power spectrum estimator is
\begin{equation}\label{eq:PkHI}
    \widehat P_{ij}(\kv) = \frac{V_\text{cell}}{\sum_{\xv} w_i(\xv)\,w_j(\xv)}\,\operatorname{Re}\left\{\widetilde F_i(\kv)\,\widetilde F^{*}_j(\kv)\right\} \;,
\end{equation}
where $V_{\rm cell}=L_x\,L_y\,L_z/(n_x\,n_y\,n_z)$ is the comoving volume of each Cartesian voxel and the weights $w_i(\xv)$ were discussed in \cref{sec:data set}.
The subscripts refer to the subsets. The power spectra are finally spherically averaged into band-powers of radius $k=|{\kv}|$ to obtain their Legendre monopole used in the remainder of the analysis.

We set the $k$-range as follows. In the angular (i.e., perpendicular to the line-of-sight) direction, the tightest resolution limit is given by the telescope beam. Hence, we set the smallest scale to $k_{\rm max}=0.245\,\kum $ to exclude $\gtrsim90\%$ damped signal (see later \cref{sec:beam}). Specifically, we set $k_{\rm max}=k_{\|{\rm max}}=k_{\perp{\rm max}}$, corresponding to an isotropic cut in $k$-space. This choice avoids fully non-linear scales, as we assume linear clustering throughout our analysis. Moreover, adding smaller-scale bins, either by expanding the $k$ limits isotropically or restricting the higher-$k$ bins to purely radial modes, would yield negligible additional information, because the SNR of the measurement drops significantly at these scales. Finally, this threshold excludes the small-scale regime where unmodeled couplings between $k_\perp$ and $k_\|$, such as those induced by the plane-parallel approximation and beam chromaticity, may have a greater impact. This chosen $k_{\rm max}$ is well below the Nyquist limit in all directions.

At large scales, before spherical averaging, we apply anisotropic cuts to optimise signal-to-noise \citepalias{MKxGAMA}. These cuts are defined in the $(k_{\|},k_\perp)$ space, with $k_{\|}$ and $k_\perp$ the parallel to the line-of-sight and transverse components of the wave vector $\kv$. We set $k_{\|{\rm min}}=0.07\,\kum$ to exclude the $k$-region where we expect most foreground residuals and the severest signal loss (see later the radial-direction analysis in \cref{sec:pknu_residual}). The mixing between the cosmological signal and contaminants at large angular scales also induces a loss of signal in the perpendicular direction \citep{Switzer2015,2015MNRAS.447..400A,Carucci2020}. Hence, we apply a cut in $k_{\perp}$, too. Evading numerical instabilities in the power spectrum estimate---since only perpendicular modes would live in the first $k$-bins---leads us to set $k_{\perp{\rm min}}= 0.02\,\kum$. The above considerations result in the one-dimensional $k_{\rm min}=0.095\,\kum$. Then, we define linearly-spaced $k$-bins and set the number of bins to $n_k=9$, yielding a bin-width $\Delta k \approx 0.017\,\kum$, such that each bin is sufficiently populated to allow for a stable power spectrum estimation.

Cross-correlation power spectra between distinct subsets provide an estimate of the \hi\ auto-spectrum free from thermal noise and with mitigated residual contaminants (stemming from dish- or time-dependent systematics), which instead affect the auto-spectra. For this reason, we will focus on possible ways to construct such observables.

After splitting the data set into four subsets using the \chess\ division (subsets $s_{\rm A}$ to $s_{\rm D}$ in \cref{fig:splitting}), we identify six cross-spectra that can be classified into three types.

\paragraph*{Cross-spectra across blocks and dishes.}
The cross-spectra $P_{\rm AC}$ and $P_{\rm BD}$ involve subsets that are observed at different times and by different dishes; therefore, we expect that contributions from contaminations due to the receivers or coming from spurious sky signals arising at specific observing times will be minimised. Since calibration solutions are time- and dish-dependent, calibration errors that are time- or dish- dependent will be mitigated as well by these cross-correlations. In short, $P_{\rm AC}$ and $P_{\rm BD}$ are the most robust combinations we can analyse; we shall thus refer to them as the `super-cross' combinations. 

\paragraph*{Cross-spectra across blocks or dishes.}
The other two types of cross-correlation are the cross-spectra across dishes---$P_{\rm AD}$ and $P_{\rm BC}$---or across blocks---$P_{\rm AB}$ and $P_{\rm CD}$. In these cases, subsets are still independent, but the former spectra mitigate potential systematics from the receivers, whereas the latter downplay block-dependent systematics. Differences among all cross-spectra will help us identify the origin of spurious contributions in the data cubes. We expect the cross-correlations over blocks to be more reliable than the ones across dishes because we are more concerned about block-dependent systematics, like low-intensity and short-lived RFI that could have eluded flagging.

%%%%%%%%%%%%%%%%%%%%%%%%%%%%%%%%%%%%%%%%%%%%%%%%%%
%================== COVARIANCE ==================%
%%%%%%%%%%%%%%%%%%%%%%%%%%%%%%%%%%%%%%%%%%%%%%%%%%

\section{Covariance estimation} \label{sec:covariance}

We adopt a data-driven covariance matrix estimation using the jackknife method \citep{10.1214/aos/1176350154}, which we also validate by performing a comparison with the analytical Gaussian variance---an assumption based on the high level of white noise in the subsets (see \cref{table:ABBAnoise}).

\paragraph*{Jackknife covariance.}

To estimate jackknife errors, we follow \citet{{2009MNRAS.396...19N}}, and point to \citet{2021MNRAS.505.5833F} and \citet{2022MNRAS.514.1289M} for more recent discussions. We divide each cleaned data cube in celestial coordinates into a three-dimensional grid of $\smash{N_{\rm jack} = N_{\rm R.A.} \times N_{\rm Dec.} \times N_{\nu}}$ contiguous, non-overlapping sub-cubes along the right ascension, declination, and frequency axes, yielding a total of $N_{\rm jack} = 224$ sub-cubes for each $s_{{\rm clean},i}$. We then construct $N_{\rm jack}$ jackknife replicas of each cube using the delete-one procedure: the $J$-th replica is obtained by setting the $J$-th sub-cube to zero while retaining all remaining sub-cubes, so that each replica uses a fraction $(N_{\rm jack}-1)/N_{\rm jack}$ of the full survey volume. Once Cartesian-gridded, we measure the power spectrum $\smash{\widehat P^{\rm jack}}$ for each realisation, applying the jackknife mask also to the window and weights, both in the case of an auto-correlation or a cross-correlation between subsets.

The expression of the jackknife covariance matrix is given by 
\begin{multline} \label{eq:jackknifeCov}
    \varSigma[\widehat P_{ij}(k),\widehat P_{ij}(k')] = \frac{N_{\rm jack}-1}{N_{\rm jack}-n_k-2}\frac{N_{\rm jack}-1}{N_{\rm jack}} \\
    \times\sum_{J=1}^{N_{\rm jack}} \left[\widehat P_{ij}^{{\rm jack},J}(k) - \langle\widehat P_{ij}^{{\rm jack},J}(k)\rangle  \right]\left[ \widehat P_{ij}^{{\rm jack},J}(k') - \langle\widehat P_{ij}^{{\rm jack},J}(k')\rangle\right]    \;,
\end{multline}
where the mean $\smash{\langle\widehat P_{ij}^{{\rm jack},J}(k)\rangle}$ is computed over the $N_{\rm jack}$ jackknife realisations. In \cref{eq:jackknifeCov}, the first prefactor is the \citet{2007A&A...464..399H} factor, introduced to avoid a biased estimation of the inverse of the covariance matrix, while the second accounts for the the fact that the $N_{\rm jack}$ realisations are not fully independent.

\begin{figure}
    \centering
    \includegraphics[width=1\columnwidth]{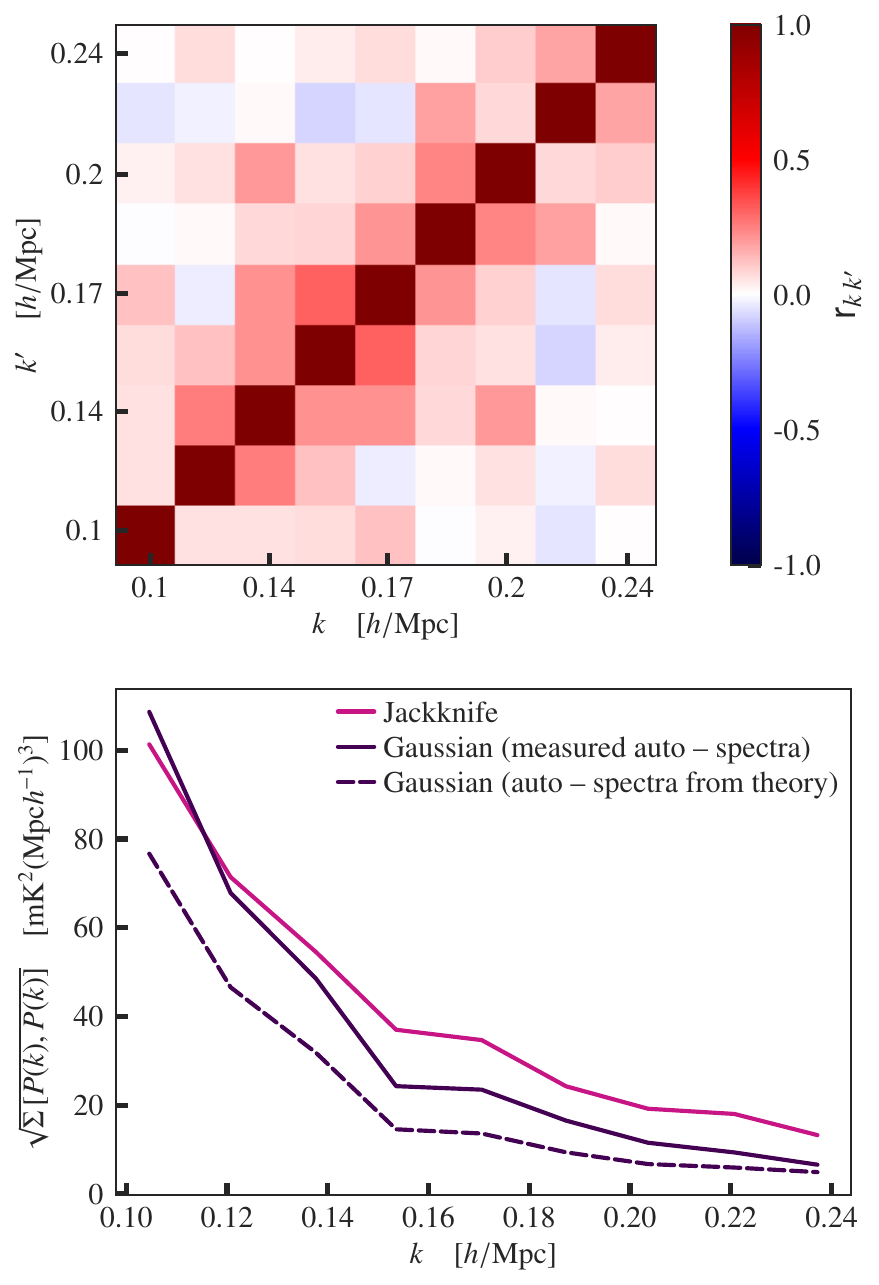}
    \caption{\textit{Top panel:} Correlation matrix for the power spectrum estimated from the cross-correlation of the subsets $s_{\rm B}$ and $s_{\rm D}$ cleaned with the values of $\Nfgl$ and $\Nfgs$ provided in \cref{table:cleaning}. The colour bar highlights in white the absence of correlation, while positive (negative) correlation are represented in red (blue). This matrix is very diagonal: besides a $\approx30\%$ cross-correlation between the second and third $k$-bins, the other adjacent bins show correlations on the order of $\approx10\%$ that decreases to few percents for more distant bins. \textit{Bottom panel:} Comparison between the square root of diagonal elements of the jackknife covariance and the Gaussian errors for $P_{\rm AC}$; the latter ones are calculated either using the measured $\widehat P_{\rm AA}$ and $\widehat P_{\rm CC}$ in \cref{eq:PvarGauss} (solid line) or a theoretical prediction as explained in the text (dashed line).}

    \label{fig:covJK}
\end{figure}

As shown in the top panel of \cref{fig:covJK}, where the correlation matrix $\mathsf{r}_{kk'}$ for $P_{\rm BD}$ is displayed, the cross-bin correlations are very weak at all scales---no more than $\approx30\%$ even for adjacent $k$-bins---implying that the jackknife covariance matrix is approximating a diagonal one.

\paragraph*{Gaussian variance.} 
Following the Gaussian approximation, we can express the variance of power spectra measurements as
\begin{equation} \label{eq:PvarGauss}
    \varSigma_{\rm G}[\widehat P_{ij}(k),\widehat P_{ij}(k)] = { \frac{\widehat P_{ii}(k)\,\widehat P_{jj}(k) + \widehat P_{ij}(k)\,\widehat P_{ji}(k)}{N_k} }\;,
\end{equation}
where $N_k$ is the number of independent modes within the sampled cosmic volume at a given spherically averaged $k$-bin. 
 
\medskip 

When comparing the Gaussian errors with the diagonal elements of the jackknife covariance, we find that the latter are larger at all scales (see the bottom panel of \cref{fig:covJK}). This difference arises because the jackknife covariance captures the non-Gaussianities in the maps due to, e.g., non-linear clustering and foreground cleaning systematics. In fact, even the Gaussian errors capture not only the thermal noise component but also any additional fluctuation or residual systematics that may contribute to the total uncertainty, since they are computed using the measured auto-power spectra of the subsets; these power spectra, especially at low-$k$, exhibit an excess of signal (as we will explicitly show later in \cref{sec:MCMCfit}), probably due to residual contaminants, which in turn inflates the estimated error bars. To test this interpretation, we recomputed the Gaussian variance using as input for the auto-spectra the theoretical predictions for the cosmological signal (consistent with the best-fit results that will be shown later) combined with the ABBA-estimated thermal noise contribution; in this case, the resulting Gaussian errors are significantly tighter, at large scales in particular, confirming that the broader error bars obtained when including the measured auto-power spectra are indeed driven by additional noise in the data. 

Hence, the jackknife covariance reflects both the off-diagonal correlations (top panel of \cref{fig:covJK}) and an enhanced variance along the diagonal (bottom panel of \cref{fig:covJK}). 
This comparison supports our conclusion that the jackknife covariance effectively characterises the full noise budget of the measurements, including eventual cross-$k$-bin correlations and better capturing systematics and observational effects. We will therefore use the jackknife errors in the remainder of the analysis---unless otherwise stated, for comparison purposes.

To conclude this section, we acknowledge that our estimated covariances are suitable for quantifying the detection significance of our measurements and for constraining physical quantities within the observed volume. They do not include uncertainties related to modes of the signal larger than the size of the survey, i.e., through supersample covariance \citep{2013PhRvD..87l3504T}, for correctly constraining global quantities \citep{2026arXiv260904488P}. We will address this in future analyses, since here we are more concerned about demonstrating the goodness of our detection rather than its constraining power.

%%%%%%%%%%%%%%%%%%%%%%%%%%%%%%%%%%%%%%%%%%%%%%%%%%
%===================== MODEL ====================%
%%%%%%%%%%%%%%%%%%%%%%%%%%%%%%%%%%%%%%%%%%%%%%%%%%

\section{Power spectrum model}
\label{sec:model}

In this section, we outline our theoretical model used for parameter inference. We use \texttt{classylss}\footnote{\texttt{\url{https://classylss.readthedocs.io/en/stable/index.html}}.} to calculate the linear matter power spectrum and all cosmological quantities (e.g., the growth rate $f$ and the Hubble parameter $H$). Below, we will omit the dependence on $z_{\rm eff} = 0.42$, unless necessary to avoid ambiguity.

In the plane-parallel wave approximation, we can model the \hi\ power spectrum as
\begin{equation}\label{eq:pmodel0}
    P_{\hi\hi}^{\rm th}(\kv) = \Thi^2\,\left( \bhi+f\,\mu^2 \right)^2\,P_{\rm m}(k) \;,
\end{equation}
where $P_{\rm m}$ is the linear matter power spectrum, $\mu=\kv \cdot \nv$ the cosine of the angle between the wave-vector $\kv$ and the line of sight $\nv$, and $f\mu^2$ accounts for linear redshift-space distortions \citep[RSD,][]{1987MNRAS.227....1K}. Then, $\bhi$ is the linear bias of the \hi\ field, and $\Thi$ the mean brightness temperature at which the \hi\ radiates. The latter depends on the overall \hi\ cosmic abundance, $\Ohi$, and thence on redshift. We adopt the common ansatz \citep{Furlanetto2006}
 \begin{equation}
    \Thi(z)=180\,h\,\Ohi (z)\,\frac{(1+z)^2}{H(z)/H_0} \, {\rm mK} \equiv \Ohi (z)\,A_\hi(z)\;.\label{eq:A_HI-def}
\end{equation}
Note that, in the last equality, we have factorised out the dependence upon the cosmological \hi\ abundance by introducing the quantity $A_\hi(z)$. 

For the \hi\ bias we can assume the following parametric relation \citep{2018ApJ...866..135V}
\begin{equation}\label{eq:bhiFid}
    b_{\hi}(z)=0.84+0.69\,z-0.05\,z^2 \; .
\end{equation}
Although this expression is not explicitly employed in our fitting procedure, it will be used later in the analysis to characterise the \hi\ cosmic abundance and decouple it from the clustering properties (see \cref{sec:discussion_modelling,sec:omegahi_constraints}).

Given the signal-to-noise ratio of our measurements (later quantified), we neither model non-linear bias, nor non-linear matter clustering, nor non-linear RSD caused by the pairwise velocity dispersion of the \hi\ field. 

Finally, to simplify our model further and in analogy with the literature \citep{Wolz2022,MKxWZ}, we include matter-only RSD. We will discuss the limits of our modelling choices in \cref{sec:discussion_modelling}.
Thus, at $z_{\rm eff}$, \cref{eq:pmodel0} becomes
\begin{equation} \label{eq:pmodel}
     P_{\hi\hi}^{\rm th}(\kv) = A_\hi^2\,(\Ohi\bhi)^2\,\left( 1+f\,\mu^2 \right)^2 \,P_{\rm m}(k)\;,
\end{equation}
which is our adopted theoretical model. Summarising, having fixed cosmology, the amplitude of the power spectrum of temperature fluctuations due to \hi\ depends solely on the combination of its abundance and clustering properties, $\Ohi\bhi$. In the following, we will incorporate observational effects that modify the observed power of \cref{eq:pmodel}.

%%%%%%%%%%%%%%%%%%%%%%%%%%%%%%%%%%%%%%%%%%%%%%%%%%
%%%%%%%%%%%%%%%%%%%%%%%%%%%%%%%%%%%%%%%%%%%%%%%%%%

\subsection{Beam damping}\label{sec:beam}

We approximate the overall beam effect as a Gaussian damping associated to the expected resolution at the median frequency of the cube,
\begin{equation} \label{eq:Bbeam}
    \mathcal{B}_{\rm beam}(\kv) = {\rm exp} \left[-\frac12\,\left(1-\mu^2\right)\,k^2\,R_{\rm b}^2 \right]\;.
\end{equation}
Above, $R_{\rm b}=d_{\rm c}(z_{\rm eff})\,\fwhm/\sqrt{8\,\ln 2}$ is the standard deviation of the telescope beam in comoving units, $d_{\rm c}(z_{\rm eff})$ being the comoving distance at the effective redshift of the survey and $\fwhm$ the full width at half maximum of the beam, given by
\begin{equation}
    \fwhm = \frac{c}{\nu_{\rm med}\,D_{\rm dish}} \;,
\end{equation}
with $c$ the speed of light and $D_{\rm dish}=13.5\,\mathrm{m}$ the diameter of the MeerKAT dishes.
Further discussion on the handling of the beam effects is in \cref{sec:beam_discus}.

%%%%%%%%%%%%%%%%%%%%%%%%%%%%%%%%%%%%%%%%%%%%%%%%%%
%%%%%%%%%%%%%%%%%%%%%%%%%%%%%%%%%%%%%%%%%%%%%%%%%%

\subsection{Pixelisation and channelisation}\label{sec:pix_and_chan}

Other sources of damping come from having collected the intensity measurements in finite voxels, which leads to the distortion on the perpendicular modes due to the size of the pixels and the one on the parallel modes due to the width of the frequency channels (in order of relevance). We model these two effects with the following damping kernels \citep{Blake2019}
\begin{align}
    \mathcal{B}_{\rm pix}(\kv,z_{\rm eff}) &=
    \sinc\left(k\sqrt{1-\mu^2}\, L_{\rm pix}(z_{\rm eff})/2\right)\;,\label{eq:PixelDamping}\\
    \mathcal{B}_{\rm chan}(\kv) &=
    \sinc\left(k\,\mu\, L_{\rm chan}/2\right)\;,\label{eq:ChannelDamping}
\end{align}
where $L_{\rm pix}$ and $L_{\rm chan}$ are the pixels size and the channel depth in comoving units. The angular pixel size of the MeerKLASS L-band survey is $\Delta\theta=0.3^\circ$, hence
\begin{equation}
    L_{\rm pix}(z_{\rm eff})=d_{\rm c}(z_{\rm eff})\,\Delta\theta \;.
\end{equation}
Furthermore, $L_{\rm chan}$ is approximated as the the mean of the comoving distance computed at the boundaries of the frequency channels, namely
\begin{equation}
    L_{\rm chan}= \frac1{N_\nu}\,\sum_{i=1}^{N_\nu} \left[ d_{\rm c}(z_i) - d_{\rm c}(z_{i+1})\right]\;.
\end{equation}

%%%%%%%%%%%%%%%%%%%%%%%%%%%%%%%%%%%%%%%%%%%%%%%%%%
%%%%%%%%%%%%%%%%%%%%%%%%%%%%%%%%%%%%%%%%%%%%%%%%%%

\subsection{Signal-loss damping}\label{sec:signal_loss}

As mentioned in \cref{sec:cleaning}, the contaminant cleaning procedure can result in a loss of cosmological signal.\footnote{Potentially, minor signal loss might come from other pipeline segments too, such as bandpass calibration. Investigations are ongoing.} A significant part of the observational \hi\ intensity mapping works had to deal with this and estimated it differently \citep[see, e.g., \citetalias{MKxGAMA},][]{Masui2013,Switzer2015,Anderson2018,Cunnington2023,MKxWZ}. 
Theoretical works have also quantified this effect, for instance, employing simulations \citep[e.g.,][]{Cunnington2023}. Common to all analyses is the suppression effect in the signal power spectrum, especially at the largest scale and often behaving as a power-law for the scales probed and precision available (see also discussion in \cref{app:tf_valid}).

Here, we address signal loss phenomenologically at the power spectrum level. Indeed, we expect the power loss to be concentrated on the largest scales, where foregrounds and signal overlap the most, and to become more negligible as we move to smaller scales. Leveraging past findings in the \hi\ intensity mapping literature, we model the suppression of the signal power spectrum due to the signal loss as
\begin{equation} \label{eq:Dsl}
     \mathcal{D}(k)
     =\left(\frac{k}{\kum}\right)^{\beta}
     \;,
\end{equation}
and fit the $\beta$ parameter as a nuisance in the final measurements. We further validate the expression above with simulations in \cref{app:tf_valid}. We will demonstrate later that the extra signal-loss freedom given by letting $\beta$ free accommodates our uncertainty on signal loss.

%%%%%%%%%%%%%%%%%%%%%%%%%%%%%%%%%%%%%%%%%%%%%%%%%%
%%%%%%%%%%%%%%%%%%%%%%%%%%%%%%%%%%%%%%%%%%%%%%%%%%

\subsection{Noise contributions}\label{sec:thermshot_noise}

The system temperature of the receiver produces thermal noise that we model as a scale-independent additive term in the \hi\ power spectrum,
\begin{equation} \label{eq:p_thermn}
    P_{\rm TN}=V_{\rm cell}\,\sigma^2_{\rm TN} \;,
\end{equation}
where $\sigma^2_{\rm TN}$ is the variance of the temperature noise, which we estimate via the ABBA method described in \cref{sec:abba} (see \cref{eq:deltaT_ABBA}) and further validate in \cref{app:noise}. We will include the term in \cref{eq:p_thermn} when estimating the auto-spectra of the subsets, and neglect it for the cross-spectra where the two different noise contributions cancel out.

We ignore other sources of noise in the modelling, such as Poissonian shot noise or stochastic noise, given our measurement uncertainties and the scale-range probed \citep{2017MNRAS.471.1788C, 2018ApJ...866..135V, 2019MNRAS.484.1007W,2023PhRvD.108h3528O,2024PhRvD.110f3555F}. Further discussion is in \cref{sec:discussion_modelling}.

%%%%%%%%%%%%%%%%%%%%%%%%%%%%%%%%%%%%%%%%%%%%%%%%%%
%%%%%%%%%%%%%%%%%%%%%%%%%%%%%%%%%%%%%%%%%%%%%%%%%%

\subsection{Modelling the observed power spectrum}

Once we include the observational correction terms as in \cref{sec:beam,sec:signal_loss,sec:pix_and_chan,sec:thermshot_noise}, the model of the signal power spectrum becomes
\begin{equation} \label{eq:Pmodel_DB}
    P_{\hi\hi}(\kv) = \mathcal{D}(k)\,\left[ \mathcal{B}^2(\kv)\,P_{\hi\hi}^{\rm th}(\kv) +  P_{\rm TN}\right]\;,
\end{equation}
where $\mathcal{B}(\kv)=\mathcal{B}_{\rm beam}(\kv)\,\mathcal{B}_{\rm pix}(\kv)\,\mathcal{B}_{\rm chan}(\kv)$ is a global, small-scale damping term and $P_{\hi\hi}^{\rm th}$ is given by \cref{eq:pmodel}. The signal-loss kernel $\mathcal{D}$ is applied to the full spectrum, also including the noise, since the latter is similarly affected by the loss induced by the coupling with foreground modes during the foreground cleaning process. Also, it does not appear squared because it does not act on the temperature overdensity field (as $\mathcal{B}$ does), but rather on the power spectrum itself \citep{Cunnington2023, Zhaoting2025}. Finally, since we deal with an auto \hi\ power spectrum either when cross-correlating subsets or computing their auto-spectra, from now on we will omit the `\hi' subscript to avoid clutter.

We apply four further steps to the model in \cref{eq:Pmodel_DB} to make it a more accurate description of our power spectrum measurements, following what has been done in previous MeerKLASS analyses \citep[\citetalias{MKxGAMA};][]{MKxWZ}. 
In short, we follow the estimator procedure in \cref{sec:pestim} and \textit{(i)} we convolve the model in \cref{eq:Pmodel_DB} with the survey window function to account for the survey footprint. \textit{(ii)} We sample it in the same three-dimensional $k$-space grid of data cubes. \textit{(iii)} We normalise the sum of the weights and the window to ensure the consistency between the amplitude of the \hi\ power spectrum model and the estimator described in \cref{sec:pestim}. Lastly, since we focus on the monopole of the power spectrum, \textit{(iv)} we spherically average the model to $P(k)$, following what is done for the measurements.

Including also the window convolution and weights as outlined above, the model now reads
\begin{multline} \label{eq:pmod_obs}
    P_{ij}^{\rm obs}(\kv)=
    Q_{ij}(\kv)\circledast
    \left\{\mathcal{D}_{ij}(k)\,\left[ \mathcal{B}^2(\kv)\,P_{\hi\hi}^{\rm th}(\kv) +  P_{{\rm TN},i}\,\delta_{ij}\right]\right\}\;.
\end{multline}
The subscripts indicate the auto-correlation of a subset $s_i$ with itself ($i=j$) or the cross-correlation between two different subsets $s_i$ and $s_j$ ($i\neq j$), and $\delta_{ij}$ is the Kronecker-delta symbol. In \cref{eq:pmod_obs} we have defined
\begin{equation}
    Q_{ij}(\kv)=\frac{\widetilde W_i(\kv)\,\widetilde W^\ast_j(\kv)}{\sum_{\xv} w_i(\xv)\,w_j(\xv)\,W_i(\xv)\,W_j(\xv)}\;,
\end{equation}
with $\smash{\widetilde W_i(\kv)}$ the Fourier transform of the re-gridded window, and $\circledast$ denoting convolution. We also notice that the exponential of the signal-loss damping term (\cref{eq:Dsl}) will depend on the involved subsets, since signal loss depends on the contaminant separation process; i.e., we deal with subset-pair dependent $\beta_{ij}$ and, consequently, $\mathcal{D}_{ij}$.

%%%%%%%%%%%%%%%%%%%%%%%%%%%%%%%%%%%%%%%%%%%%%%%%%%
%================= GUIDANCE FITS ================%
%%%%%%%%%%%%%%%%%%%%%%%%%%%%%%%%%%%%%%%%%%%%%%%%%%

\section{Optimising the contaminant subtraction parameters} \label{sec:tablecloth}
When using mPCA, we must choose the numbers of components $\Nfgl$ and $\Nfgs$ associated with contaminants at large and small scales, respectively, to retrieve the cosmological signal. We perform different {$\Nfgl$, $\Nfgs$} cleanings, resulting in distinct cleaned cubes that undergo the process previously described to produce cross-spectrum measurements. Finally, we compare each measurement with the theoretical model described in \cref{sec:model}, and setting $\beta_{ij}=0$, i.e., ignoring signal loss in the modelling, to avoid being prone to over-cleaning and large reconstruction from the $\beta$ correction. We choose the conclusive {$\Nfgl$, $\Nfgs$} set as the one that maximises the signal-to-noise ratio (keeping signal loss as low as possible yet removing enough contaminants to `reveal' the subsets' cross-correlation) while optimising the reduced $\chi^2$, $\chi_{\rm red}^2=\chi^2/n_{\rm dof}$ (where $n_{\rm dof}$ denotes the number of degrees of freedom), with the theoretical template (since contaminant separation alters the spectrum in a $k$-dependent way). 

In further detail, we expect to get closer to a good match when removing more PCA modes, that is removing more contaminants from data to reveal a cleaner signal. However, after we hit a `cleaning sweet spot' \citep[\citetalias{MKxGAMA};][]{MKxWZ,Carucci2024}, signal loss starts dominating, and the correlation signal starts being washed out, reducing the spectrum amplitude and approaching its agreement with zero.
We will later test whether the final inferred amplitude is dependent on the $N_{{\rm fg,L}},\,N_{{\rm fg,S}}$ choices.

\begin{figure}
    \centering
    \includegraphics[width=0.8\columnwidth]{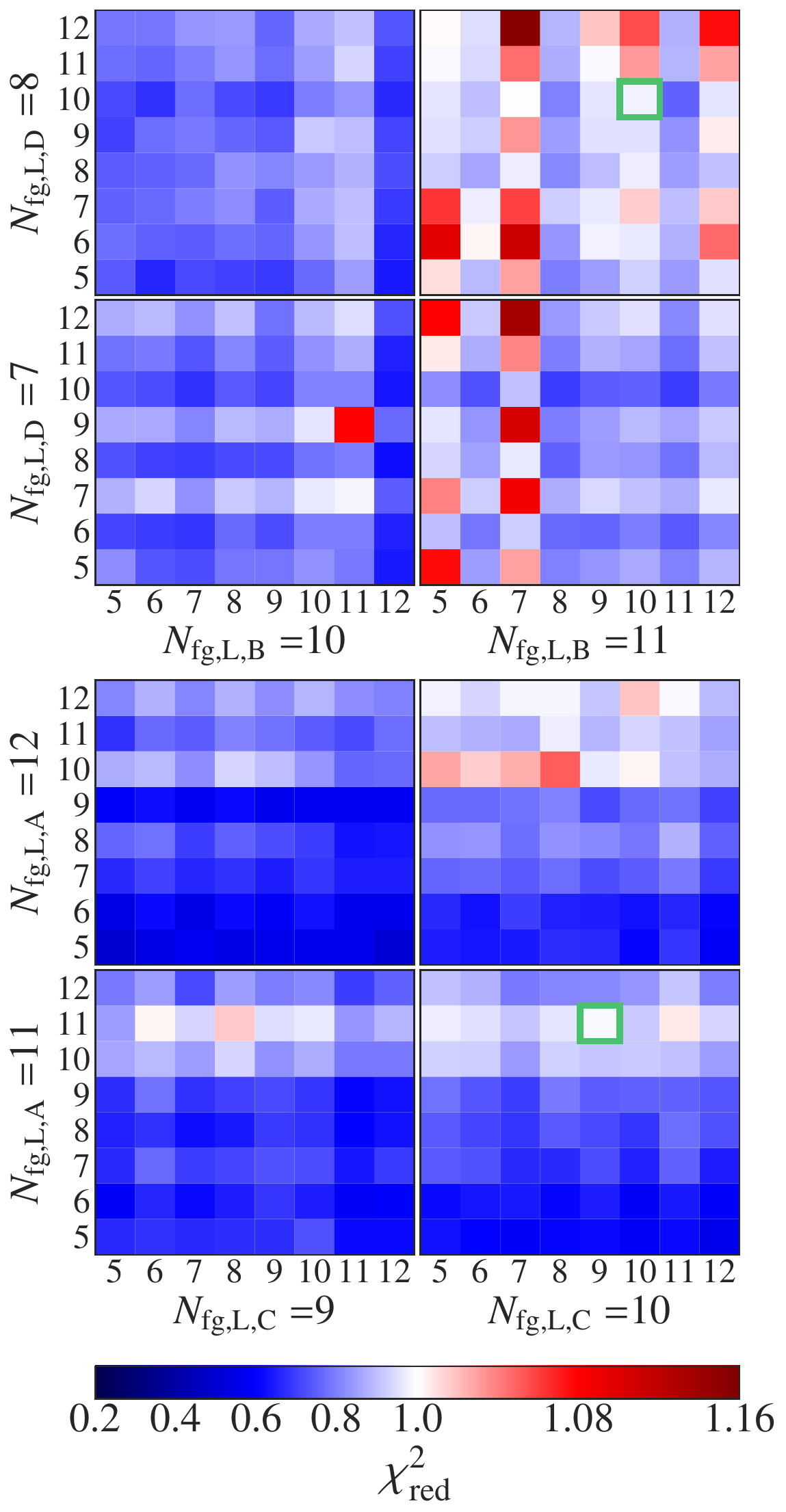}
    \caption{The $\chi^2_{\rm red}$ analyses varying the mPCA parameters for the super-cross power spectra of the \chess\ division---in a neighbourhood of the optimal cleaning region---, $s_{\rm B}\times s_{\rm D}$ on the top, $s_{\rm C}\times s_{\rm A}$ on the bottom. For each cross-analysis, each pixel corresponds to a $\{N_{{\rm fg,L},i}\,N_{{\rm fg,S},i}\,N_{{\rm fg,L},j}\,N_{{\rm fg,S},j}\}$ combination, colour-coded according to its $\chi^2_{\rm red}$ with the white colour highlighting the $\chi^2_{\rm red}=1$ value. In particular, each of the four sub-panels correspond to fixing the two $\Nfgl$, within which we vary the corresponding $\Nfgs$ in the smaller pixels, as per the axes' labels. With green, squared contours, we highlight the final choices of the $\Nfg$ parameters.
    }
    \label{fig:tablecloth}
\end{figure}
\begin{table}
\centering
 \caption{Adopted mPCA parameters used for the contaminant separation process of each subset.}
 \label{table:cleaning}
 \begin{tabular}{lcccc}
 \toprule
 Subset & $s_{\rm A}$ & $s_{\rm B}$ & $s_{\rm C}$ & $s_{\rm D}$ \\
  \midrule
$\Nfgl$ & 11 & 11 & 10 & 8 \\ 
$\Nfgs$ & 11 & 10 & 9 & 10\\ 
  \bottomrule
 \end{tabular}
\end{table}
For each of the $6$ cross-spectra $\smash{\widehat P_{ij}(k)}$ for the \chess\ divisions we deal with $4$ foreground removal parameters: $\{N_{{\rm fg,L},i}\,N_{{\rm fg,S},i}\,N_{{\rm fg,L},j}\,N_{{\rm fg,S},j}\}$. To efficiently explore the 4-dimensional $\Nfg$ parameter space for each cross-spectrum, we proceed as follows.

We fix the signal-loss parameter in the model $\beta_{ij}=0$; otherwise, the damping kernel $\mathcal{D}$ would absorb the amplitude variations among the different spectra, and the evident minima in the $\chi^2_{\rm red}$ behaviour would diminish or disappear. We will let the $\beta$ parameters vary in the final inference analysis, once we have fixed the cleaning parameters. Thus, we are left with only one free parameter to fit, the amplitude $\Ohi\bhi$. We least-square-estimate (LSE) the $\beta_{ij}=0$ model with the measurements, varying the 4 $N_{\rm fg}$ parameters. The match between measurements and model is deemed good when the reduced $\chi^2_{\rm red}$ is close to unity. Based on the number of degrees of freedom and the cumulative $\chi^2$ distribution, we set the $\chi^2_{\rm red}$ threshold corresponding to the $68\%$ confidence level to assess when the model is a good description of the measured cross-spectrum.
 \begin{figure}
    \centering
    \includegraphics[width=1\columnwidth]{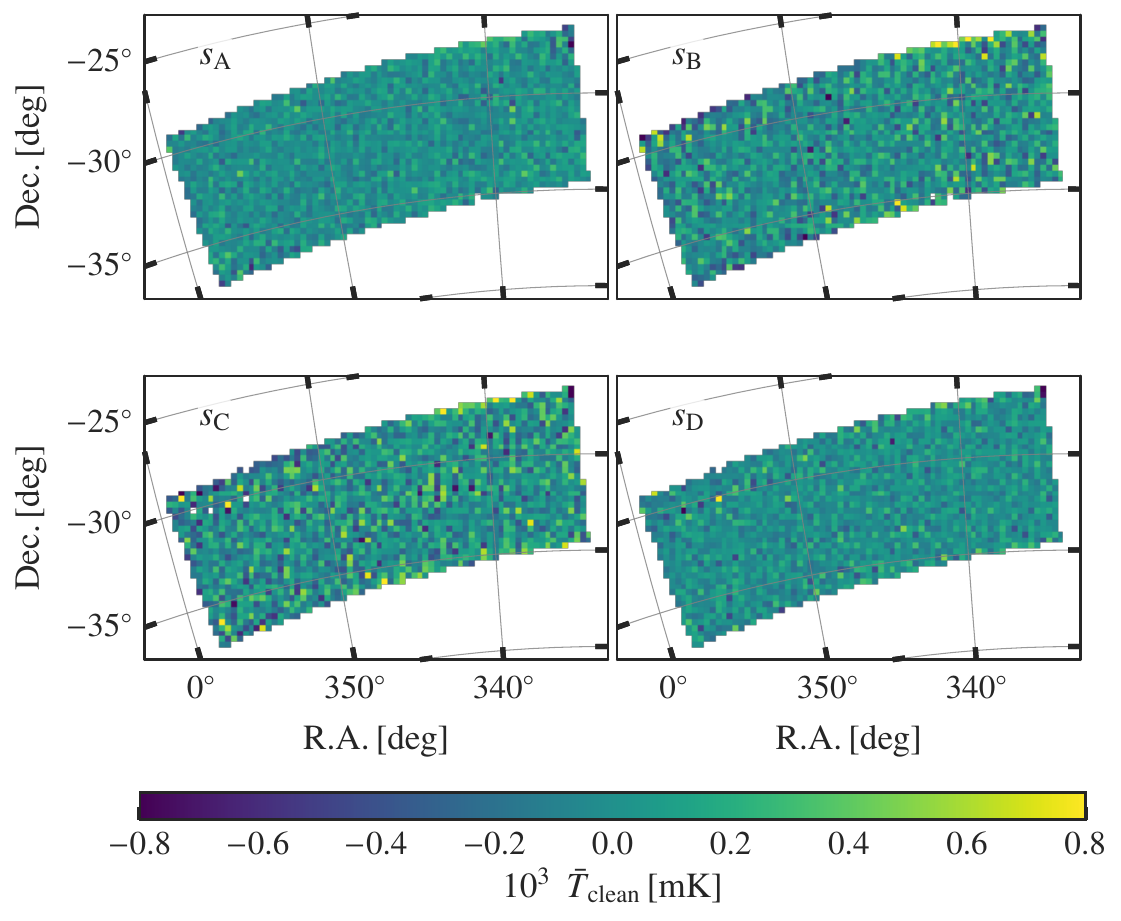}
    \caption{The four $s_{\rm A}$ to $s_{\rm D}$ subsets' intensity maps of the \chess~division, cleaned by mPCA using the parameters in \cref{table:cleaning}, and averaged along the frequency dimension.}
    \label{fig:cleanedmaps}
\end{figure}

As an example, \cref{fig:tablecloth} shows the $\chi^2_{\rm red}$ analysis for the two super-cross spectra. In each panel, the four sub-panels correspond to the two large-scale $\Nfgl$ cleaning parameters of the two involved subset cubes. Namely, each block corresponds to a given $N_{{\rm fg,L},i}$--$N_{{\rm fg,L},j}$ combination. Within each block, the smaller pixels refer to the small-scale cleaning $\Nfgs$, with the same $x$ and $y$ axis convention of the $\Nfgl$ referring to the $i$ and $j$ subset. To each $\{N_{{\rm fg,L},i}\,N_{{\rm fg,S},i}\,N_{{\rm fg,L},j}\,N_{{\rm fg,S},j}\}$ combination, we associate and colour-code a $\chi^2_{\rm red}$: white for $\chi^2_{\rm red}=1$, diverging to blue for low values and to red for high ones. %We mark in grey the combinations that result into a fit outside the $68\%$ confidence region. 
We want to adopt the $\Nfg$ combinations of each pair of involved subsets that maximise the quality of the fits (i.e., for which $\chi^2_{\rm red}\approx 1$). 

For some $\{N_{{\rm fg,L},i}\,N_{{\rm fg,S},i}\,N_{{\rm fg,L},j}\,N_{{\rm fg,S},j}\}$ combinations, we work in a very low signal-to-noise regime and the $\chi^2_{\rm red}\lesssim1$ because error bars are so large that they allow the model to be a good description of the data nevertheless. For this reason, looking at the fitted $\Ohi\bhi$ amplitude is indicative as those `over-fitted' cases typically correspond to fitting null spectra, i.e., with amplitudes extremely close to zero. Additionally, since we want to eventually combine the information from all cross-spectra (that is, we want to work with unique cleaned subsets in the rest of the analysis), we fix the `good' cleaning of a subset $s_i$ (i.e., its $N_{{\rm fg,L},i}$ and $N_{{\rm fg,S},i}$ parameters) to be the same in all spectra involving $s_i$. When multiple possibilities are equivalent, we opt for the lowest values of $\Nfg$ to minimise signal loss \citep{cunnington2021,Carucci2024}. 
Finally, when we have converged to a plausible region of the 4-dimensional $\Nfg$ space, we check whether relaxing the $\beta_{ij}=0$ assumption (i.e., fitting also for $\beta_{ij}$) can cause significant variations in the fitted spectra amplitudes or in the $\chi^2_{\rm red}$ behaviour, which, typically, just shows a `diluted' sweet spot evidence as the $\beta_{ij}$ compensate for larger signal loss or larger residual contaminants in the less-optimal cleaning solutions.

With the above considerations, we can set the final $\Nfgl$ and $\Nfgs$ parameters unambiguously, resulting in the values listed in \cref{table:cleaning} and highlighted in \cref{fig:tablecloth}, which we will use for the remainder of the analysis. The final, cleaned versions of the subsets we work with are shown in the frequency-averaged maps of \cref{fig:cleanedmaps}.

As a final consideration on how we set the $\Nfgl$ and $\Nfgs$ parameters, we acknowledge that selecting them based on a theoretical model might risk circularity, as the cleaning is designed to reproduce the chosen model. However, since we work within a fixed cosmological model, only the power spectrum amplitude can change. The $\Nfg$ parameters cannot accommodate any spectrum. To ensure the guidance fits we perform are appropriate, and that our final results do not depend on the cleaning parameter choice, we will conduct specific tests (see especially \cref{sec:Nfg_test} and \cref{sec:xGAMA}, and corresponding discussions in \cref{sec:discussion}).

%%%%%%%%%%%%%%%%%%%%%%%%%%%%%%%%%%%%%%%%%%%%%%%%%%
%================= MULTI-SUBSET =================%
%%%%%%%%%%%%%%%%%%%%%%%%%%%%%%%%%%%%%%%%%%%%%%%%%%

\section{Multi-subset formalism} \label{sec:mt_form}

Once subsets have been cleaned of contaminants, we can proceed with the inference analysis.

When working within the \chess\ division, subsets are observationally-independent data sets that trace the same underlying \hi\ field. Hence, we can combine their cross-spectra in a single data vector and fit it against the common-to-all power spectrum model previously discussed. Such a `multi-subset' analysis thus coherently adds information from the different spectra. In practice, we adapt to our case the well-known multi-tracer formalism \citep{2004MNRAS.347..645P, 2009PhRvL.102b1302S, 2009JCAP...10..007M, 2009MNRAS.397.1348W, 2013MNRAS.432..318A,2015ApJ...812L..22F}.

A full multi-subset data vector should include both the auto- and cross-spectra. However, we decide to omit auto-spectra for two reasons. 
\textit{i)} We would need an accurate model to account for the extra noise and spurious power that we measure in the auto (more on this later), which we do not have; and \textit{ii)} even if systematic effects were absent, the division into multiple subsets would lead to noise-dominated data sets, since the cosmological signal is subdominant compared to the thermal noise level in most auto-spectrum sampled bins.

We validate the last statement by comparing information matrix forecasts \citep{dafebb01-f6c3-3bb0-8577-2b844df40cce,1997ApJ...480...22T} for the complete multi-subset analysis and the one that includes only the cross-spectra. In the forecast, we adopt highly optimistic assumptions for the auto-spectra, assuming contributions only from white thermal noise through a constant parameter $P_{\rm TN}^0$ common to all the subsets and simply scaling it with the number of dishes and the global integration time for each subset, that is\ $P_{\rm TN}=P_{\rm TN}^0/(N_{\rm dishes}\,t_{\rm obs})$. Even under these idealised conditions, the forecasted uncertainty on the amplitude of the \hi\ power spectrum only differs by $\approx8\%$ between the two configurations. 

In short, we find no advantage in including the auto-spectra in the analysis; hence, we will rely solely on the cross-spectra to ensure greater robustness without loss of constraining power.

We also decide not to perform the multi-subset analysis combining all the six possible cross-spectra, rather we focus on the super-cross correlations, since we expect their combination to be the most robust against systematics (we test and discuss this later in \cref{sec:constraints}). This choice is supported by an information matrix forecast indicating that the loss of constraining power due to the reduced data vector size remains within $25\%$. The benefit from the full data vector is limited because each $P_{ij}(k)$ requires a nuisance parameter $\beta_{ij}$, which expands the parameter space. Most importantly, the potential gain from combining all cross-spectra is an optimistic upper bound since Fisher forecasts cannot capture the data's true complexity. As we will show in \cref{sec:splitting_test}, the cross-spectra other than the super-cross are noisier and harder to model. 

Hence, our \chess\ data vector reads
\begin{equation}\label{eq:Pcb}
    {{\bm P}_{\rm cb} = \begin{pmatrix} P_{\rm BD} & P_{\rm AC}\end{pmatrix}}^{\sf T}\;,
\end{equation}
with corresponding covariance matrix
\begin{equation}
\label{eq:multi_xchess}
    \mathsf{\Sigma}_{\rm cb}
    = \begin{bmatrix} 
        \varSigma(P_{\rm BD},P_{\rm BD}) & \varSigma(P_{\rm BD},P_{\rm AC}) \\
        & \varSigma(P_{\rm AC},P_{\rm AC})
    \end{bmatrix} \;.
\end{equation}

The entries composing the matrix in \cref{eq:multi_xchess} represent the covariance matrices between different power spectra, estimated using the jackknife method.

%%%%%%%%%%%%%%%%%%%%%%%%%%%%%%%%%%%%%%%%%%%%%%%%%%
%%%%%%%%%%%%%%%%%%%%%%%%%%%%%%%%%%%%%%%%%%%%%%%%%%

\subsection{Posterior inference}\label{sec:MCMCfit}

\begin{figure}
    \centering
    \includegraphics[width=\columnwidth]{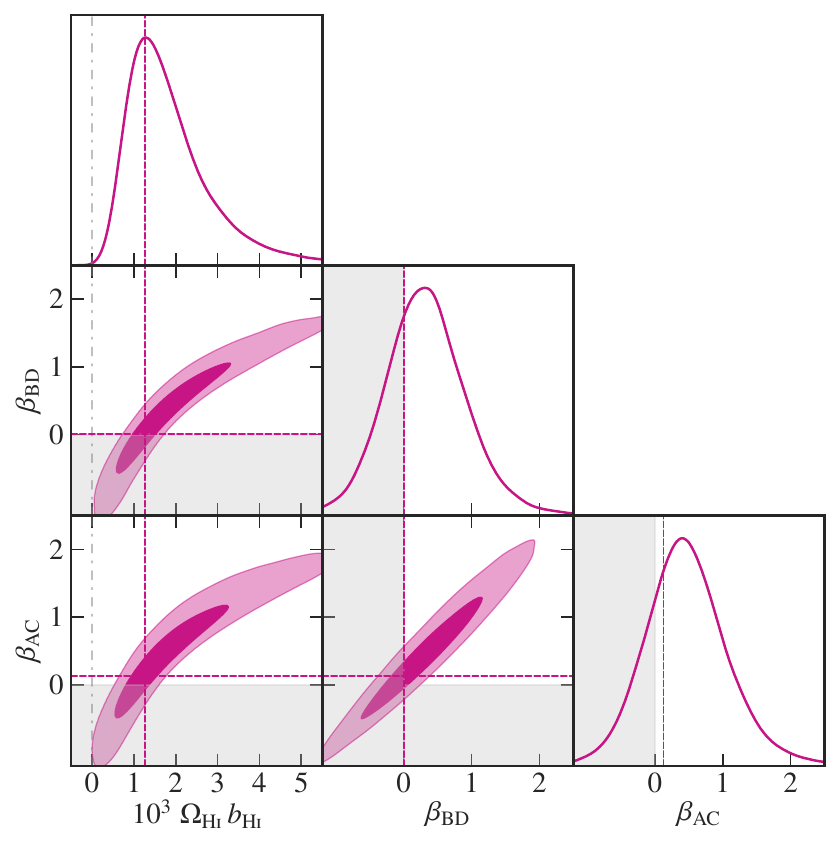}
    \caption{Corner plot showing the multivariate posterior distribution of the multi-subset fits performed on the ${\bm P}_{\rm cb}$ data vector. The coloured regions represent the $68\%$ and $95\%$ confidence level contours and the dotted lines indicate the MAP values for the free parameters of the fit. The grey shaded areas highlight the negative regions of the $\beta_{ij}$ parameter space. The vertical dash-dotted line in the $10^3\,\Ohi\bhi$ panels highlights the level of null amplitude.} 
    \label{fig:corner}
\end{figure}

\begin{table}
\centering
 \caption{Best-fit values, $\chi_{\rm red}^2$ and $p$-values resulting from the multi-subset fits. %performed using the jackknife covariance. 
For the least-squares fits we provide the value of the overall amplitude of the power spectrum $\Ohi\bhi$, while for the MCMC fits that include the signal-loss kernel we report the MAP values of $\Ohi\bhi$ and also the $\beta_{ij}$ parameters. 
}
 \label{table:fits}
 \begin{tabular}{ccccc}
 \toprule
 $10^3\,\Ohi\bhi$ & $\beta_{\rm BD}$ & $\beta_{\rm AC}$ & $\chi_{\rm red}^2$ & $p$-value \\
  \midrule
$1.178\pm0.092$ & -- & -- & $0.93$ & $0.53$ \\ 
$1.271^{+1.080}_{-0.598}$ & $0.01^{+0.85}_{-0.30}$ & $0.13^{+0.89}_{-0.28}$ & $0.98$ & $0.47$\\ 
  \bottomrule
 \end{tabular}
\end{table}

We fit data vectors ${\bm P}_{\rm cb}$ against the \hi\ power spectrum model (see \cref{sec:model}), and we vary three free parameters: the amplitude $\Ohi\bhi$, plus the nuisance parameters $\{\beta_{\rm BD},\,\beta_{\rm AC}\}$. 

We perform the multi-variate analysis with a Markov chain Monte Carlo (MCMC) method, employing the publicly available library \texttt{emcee},\footnote{\texttt{\url{https://emcee.readthedocs.io/en/stable/}}.} and sampling the parameter space using wide uniform priors on the free parameters. Specifically, we set $10^3\,\Ohi\bhi\in\left[0,\,10\right]$, as well as $\beta_{ij}\in\left[-5,\,5\right]$ for both $ij$ cross-spectra. % included in the multi-subset data vector.
We expect the $\beta_{ij}$ to be positive, indicating that signal loss is more severe than the contribution of potential residual contaminants in the data. However, we decide to leave the $\beta_{ij}<0$ region accessible to confirm our expectations. 

We ensure the chains converge by assessing the autocorrelation time changes less than $1\%$ every $100$ iterations. We also verify that the chains achieve $\smash{\widehat R-1<10^{-2}}$, with $\smash{\widehat R}$ the Gelman-Rubin coefficient \citep{1992StaSc...7..457G,2021BayAn..16..667V}. 

Results are in \cref{fig:corner}, where we report the full corner plot and we quote the best-fit values of $\Ohi\bhi$ and the signal-loss parameters $\beta_{ij}$ in \cref{table:fits}. They are defined as the maxima of the posterior distribution (maximum-a-posteriori, MAP), with uncertainty being the highest posterior density region within the $68\%$ confidence level. The latter choice was driven by the skewness of the posterior distributions resulting from the MCMC runs (see \cref{fig:corner}). To verify the effect of including the signal-loss kernel, we compare those results with simpler least-square fits with a $\beta_{ij}=0$ model, again in \cref{table:fits}. 
The robustness of the fit is demonstrated by the values of $\smash{\chi_{\rm red}^2}=\chi^2/n_{\rm dof}$ and the corresponding $p$-values, defined as the probability of obtaining a $\chi^2$ value greater than or equal to the one observed under the null hypothesis that the adopted model correctly describes the data. The number of degrees of freedom $n_{\rm dof}$ given the number of free parameters in the fit $n_{\rm free}$ is $n_{\rm dof}=n_k-n_{\rm free}$. In both cases the $\smash{\chi_{\rm red}^2}$ remains close to the ideal unity value, while the corresponding $p$-values ($\approx0.5$) confirm the good statistical consistency between the model and the data, with no evidence of significant systematic discrepancies.

\begin{figure*}
    \centering
    \includegraphics[width=0.8\textwidth]{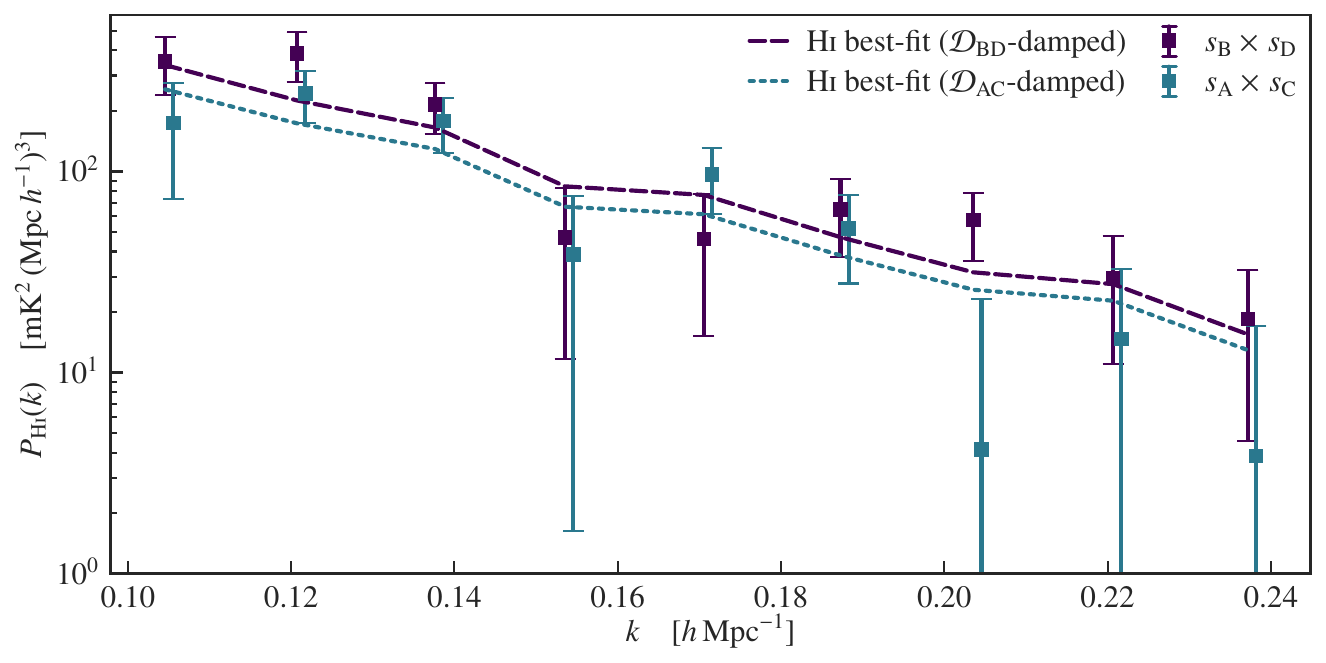}
    \caption{Measured cross-spectra of the subsets from the \chess\ division (we only show the ones from the super-cross combinations). The error bars are the square root of the diagonal elements of the jackknife covariance matrices. For comparison, we plot with dashed and dotted lines and corresponding colours the best-fit \hi\ power spectra we computed using the $\Ohi\,\bhi$ and $\beta_{\rm BD}$ or $\beta_{\rm AC}$ resulting from the multi-subset multivariate MCMC fits (see \cref{table:fits}).}
    \label{fig:measured supercross}
\end{figure*}

\begin{figure*}
    \centering
    \includegraphics[width=2\columnwidth]{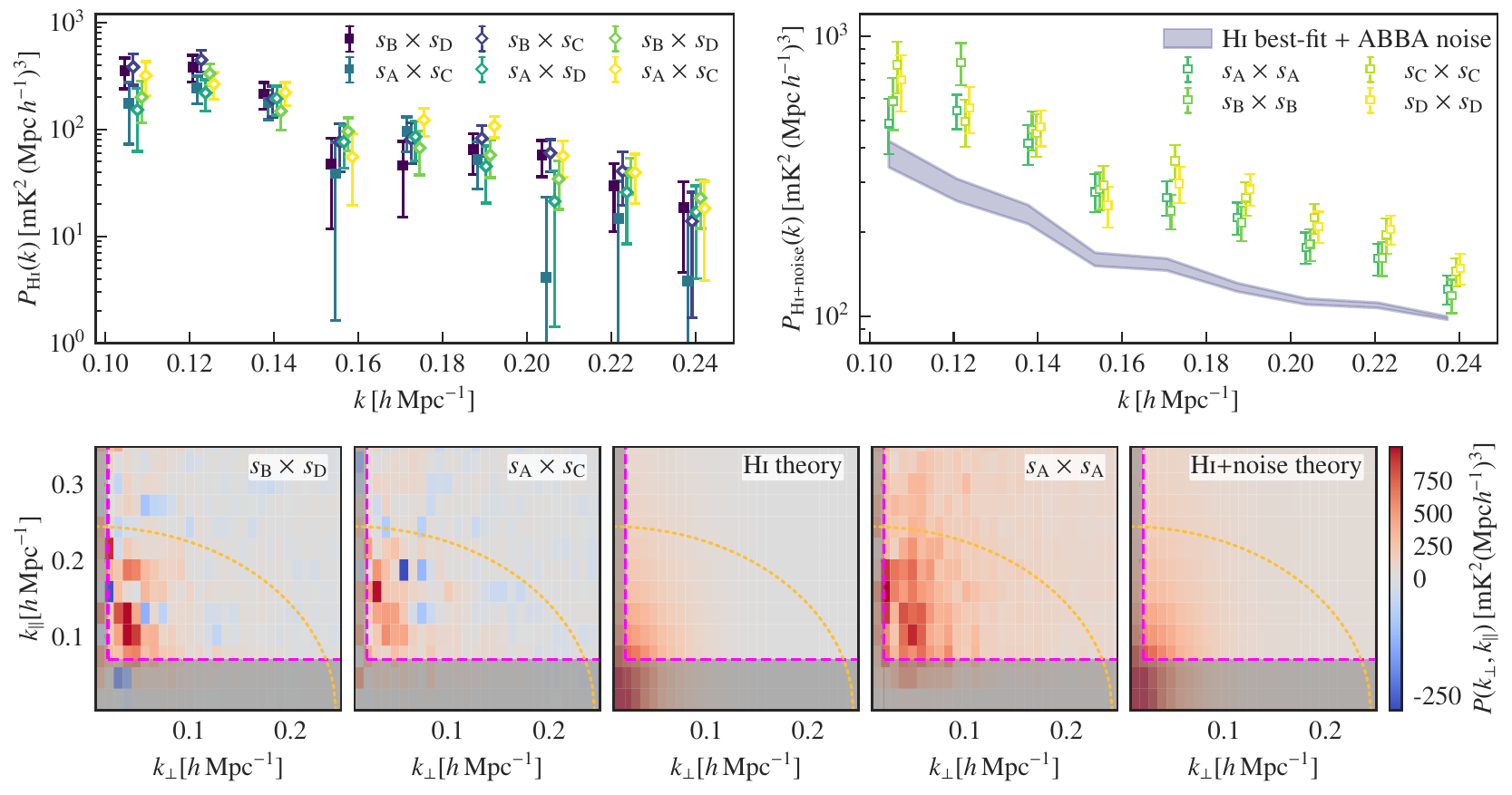}
    \caption{
    \textit{Top panels:} Measured cross- and auto-spectra (left and right panel, respectively) of the subsets from the \chess\ division. The error bars are the square root of the diagonal elements of the jackknife covariance matrices. On the left, we distinguish the super-cross power spectra (filled squares) from the cross-dish and cross-block power spectra (empty diamonds); we highlight the agreement between the different cross-correlation strategies. On the right, we produce a sole theoretical prediction as a power range, fixing $\Ohi\,\bhi$ to the inferred value and spanning between the minimum and maximum combination resulting from using one of $\beta_{ij}$ values plus a scale-independent contribution from thermal noise estimated as \cref{eq:p_thermn} and using the ABBA noise variance, averaged across the subsets. There is an evident and scale-dependent excess of power in the auto-spectra measurements on the right, which drives our decision to exclude those measurements from our multi-subset analysis. 
    \textit{Bottom panels:} Two-dimensional power spectra of the two super-cross combination (first two panels) and a theoretical expectation for the \hi\ cosmological signal (third panel). An example of two-dimensional auto-power spectrum is represented in the fourth panel, to be compared with a prediction of \hi\ cosmological signal plus noise in the rightmost panel. The theoretical predictions are obtained using the best-fit value of $\Ohi\bhi$ from the main analysis and fixing $\beta_{ij}=0$ for simplicity; the noise level is estimated analogously to the one-dimensional case. The clustering power is colour-coded with the same range for all panels. The dashed magenta lines and darker regions highlight the region excluded in the analysis through the cuts in $k$-space. The dotted yellow line depicts the one-dimensional $k_{\rm max}$ used in the analysis. 
    % }
    }
    \label{fig:allspectra}
\end{figure*}

Using the MCMC-constrained values, we can evaluate the best-fit \hi\ power spectra that we show in \cref{fig:measured supercross} (dashed lines) alongside the measured super-cross power spectra (filled symbols and error bars). 
The super-cross $P(k)$ are also compared with the cross-block and cross-dish power spectra in the top left panel of \cref{fig:allspectra}, where the super-cross and the other measurements are distinguished by the use of filled square markers and open diamonds, respectively; the fact that all cross-correlations yield consistent results confirms that the pipeline is effectively immune to dominant contaminants, thereby validating the super-cross $P(k)$ as a reliable probe for our analysis.

With the best-fit curves, adding to them the estimated thermal noise averaged among the four subsets (see \cref{sec:abba} and \cref{eq:pmod_obs}), we can predict the auto-spectra of the \chess\ subsets, as we do in the top right panel of \cref{fig:allspectra} (shaded area spanning the minimum and maximum predictions) together with the measurements (empty symbols and error bars). We will discuss these results and their significance in \cref{sec:discussion}. 

In the auto-spectra case shown in the bottom left panel of \cref{fig:allspectra}, the discrepancy between theory and measurements is prominent, confirming that we lack an understanding and modelling of the extra power exhibited by those spectra and supporting our choice not to include them in the analysed data vector ${\bm P}_{\rm cb}$. Most importantly, this unexplained extra power is most likely caused by residual contaminants which is a main focus in the next section where we assess whether it is uncorrelated between subsets, regardless of its origin.

Before moving on to testing our results, we look more closely at those clustering measurements by also showing their cylindrical power spectra $P(k_\perp, k_\parallel)$, colour-coded in the top panels and bottom right panels of \cref{fig:allspectra}. % (however, the final inference analysis used the spherically-symmetric spectra already presented). 
The first row shows the two super-cross, followed by a theoretical prediction for the \hi\ signal. Next, we show one of the auto-spectra, along with its theoretical prediction (\hi\ plus uncorrelated noise).

Again, the discrepancy between the auto-subset $P(k)$ and the expected signal is prominent. Instead, for both super-cross spectra, the amplitude decreases sharply, beyond the bare effect of thermal noise cancellation. Compared to $s_{\rm A}\times s_{\rm C}$, the $s_{\rm B}\times s_{\rm D}$ case shows higher overall amplitude and oscillates more, with high and negative peaks that might compensate each other when fitted together, as it was also the case with the one-dimensional spectrum shown in the panel above. The grey-shaded areas in those $P(k_\perp, k_\parallel) $ plots correspond to the region removed from the inference analysis, based on considerations drawn from some of the tests we will show in the next section.

To conclude the inference section, we acknowledge the simplifying assumptions of the theoretical model and will discuss their implications in \cref{sec:discussion_modelling}. In this respect, we emphasise that the main objective of this work is to detect a cosmological signal in the intensity maps, measured by the inferred amplitude of its spectrum. Consistency of this amplitude across different fitting strategies and analysis choices provides a reliable benchmark for validation. We will further stress-test our measurement in \cref{sec:robustness}.

%%%%%%%%%%%%%%%%%%%%%%%%%%%%%%%%%%%%%%%%%%%%%%%%%%
%================== VALIDATION ==================%
%%%%%%%%%%%%%%%%%%%%%%%%%%%%%%%%%%%%%%%%%%%%%%%%%%

\section{Validation tests}
\label{sec:robustness}

To validate our measurements, we conduct a battery of diverse checks. We group them into 12 tests, summarised at the end of this section in \cref{table:tests}, each designed to verify a specific hypothesis, from $\mathcal{H}_{1}$ to $\mathcal{H}_{12}$.

%%%%%%%%%%%%%%%%%%%%%%%%%%%%%%%%%%%%%%%%%%%%%%%%%%
%%%%%%%%%%%%%%%%%%%%%%%%%%%%%%%%%%%%%%%%%%%%%%%%%%

\subsection{Splitting strategy} \label{sec:splitting_test}

\begin{figure}
    \centering
    \includegraphics[width=\columnwidth]{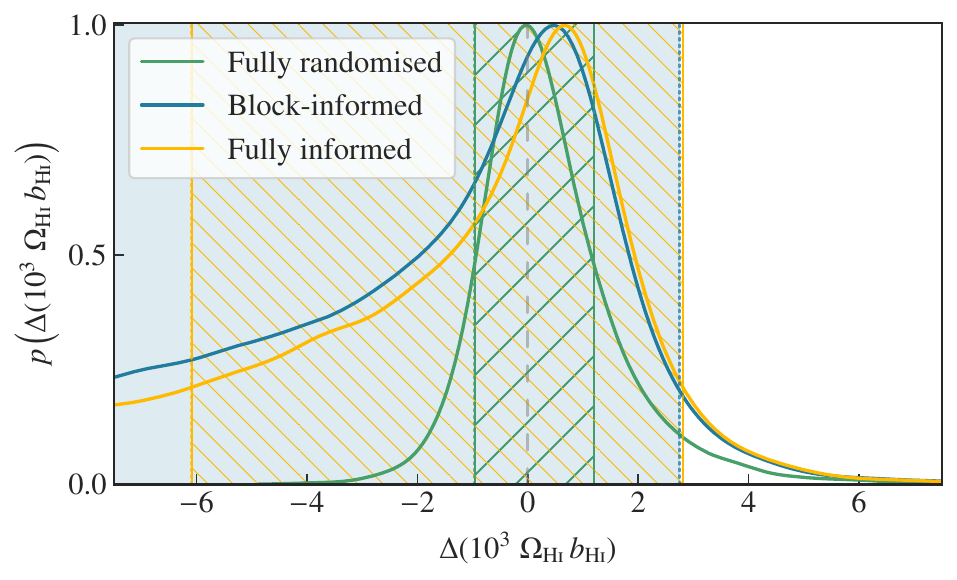}
    \caption{Difference posterior distributions of the $\Ohi\bhi$ parameter between the re-analyses with different splitting strategies and the reference results. Perfect consistency is indicated by the vertical grey dashed line corresponding to zero tension. The fully-randomised, block-informed, and fully-informed cases are represented with different colours. 
    }
    \label{fig:deltaOb_post}
\end{figure}

We start by asking whether our results depend on the way we partitioned the data set, $\mathcal{H}_{1}$. 
We perform consistency tests by varying the data-splitting strategy used to define the subsets. Besides our reference configuration (shuffled blocks and ordered dishes), we analysed three alternative splitting criteria: \textit{i)} a fully randomised \chess\ (both block and dish orders randomised before identifying subsets); \textit{ii)} a block-informed version (ordered blocks and shuffled dishes); and \textit{iii)} a fully-informed version (ordered blocks and dishes). The ordering of the dishes reflects a distinction between core and external receivers.\footnote{This is not a perfect division in our reference analysis, because we prioritised equivalent noise levels (the number of hits) across subsets over strict physical separation.} The ordering of the blocks is an observation-driven criterion that groups observations by antenna pointing in local coordinates. This ensures each group is consistently exposed to RFI and spurious signals from a similar direction. For every scenario involving randomisation, the reshuffling process is performed anew to ensure the random groupings are unique to each test. This prevents the same data sets from being consistently paired across different subset definitions and provides a more rigorous check of the statistical independence of our results. The subsets are identified, as in the main analysis, to ensure comparable noise levels across them. For each alternative \chess\ configuration, the entire analysis pipeline is repeated, including optimisation of the foreground cleaning via mPCA by guidance fits.\footnote{Scale cuts are not varied and are kept the same as the ones optimised for the main analysis.}

We observe the greatest difference relative to the reference analysis with the block-informed \chess\ and the fully-informed one. Most of their guidance fits results in a high $\smash{\chi^2_{\rm red}}$, suggesting that grouping blocks of the same type fails to minimise block-related systematics, which appear to be the dominant source of contamination. Indeed, the cleaning solutions in these \chess\ scenarios are associated with inflated covariance matrices and lower SNR, likely due to residual contamination in the subsets.

However, when performing Bayesian inference on the full parameter set ($\Ohi\bhi$ plus the signal-loss nuisance parameters), we find that the multi-subset formalism using the two super-cross power spectra for each \chess\ is a robust mitigation strategy regardless of the splitting method and its related contamination variations. By leveraging the most independent pairs, we effectively decorrelate most systematic residuals at the estimator level. Even in the block-informed and fully-informed cases, where non-randomised blocks inflate the covariance, the super-cross $P(k)$ yields unbiased results, showing no significant tension in the final parameters. In particular, the MAP values of $\Ohi\bhi$ agree with the reference case, although the reduced detection significance (halved SNR of the new super-cross-combined data vectors relative to the reference ${\bm P}_{\rm cb}$) makes them consistent with zero.

Using the \texttt{tensiometer}\footnote{\texttt{\url{https://tensiometer.readthedocs.io/en/latest/}}.} package, we compared the reference marginalised distribution of the $\Ohi\bhi$ parameter to the distribution inferred from the new re-analyses. The level of agreement is quantified by the probability of a shift in the posterior distribution, $P_{\rm shift}$, and its effective Gaussian-equivalent significance $\sigma_{\rm eff}$ \citep{2019PhRvD..99d3506R}. The fully-randomised case is consistent within $0.05\,\sigma_{\rm eff}$ with a shift probability of $P_{\rm shift}=0.02$. The other two cases show similar features, with a long tail towards negative shifts. This reflects that $\Ohi\bhi$ in these cases is consistent with zero at $68\%$ of confidence level (C.L.). Nevertheless, their posteriors remain consistent within $0.14-0.24\,\sigma_{\rm eff}$ with the reference case. 
We show the parameter difference distributions in \cref{fig:deltaOb_post}, with the $68\%$ C.L. regions highlighted by vertical bands. As discussed, no significant differences are found in the posterior distributions in all three alternative splits, which remain in excellent agreement. However, the reduction in constraining power when adopting splitting strategies that preserve block ordering is reflected in the broadening of the left tail of the distributions.

A key finding of this test is the identification of the origin of the dominant systematic contamination. By comparing `informed' splitting strategies with randomised configurations, we find that block-dependent systematics have a significantly greater impact on the data than dish-dependent effects.

Finally, these consistency tests demonstrate that our results are robust to the choice of splitting strategy, as the super-cross technique successfully decorrelates residual spurious signals. 

%%%%%%%%%%%%%%%%%%%%%%%%%%%%%%%%%%%%%%%%%%%%%%%%%%
%%%%%%%%%%%%%%%%%%%%%%%%%%%%%%%%%%%%%%%%%%%%%%%%%%

\begin{figure}
    \centering
\includegraphics[width=\columnwidth]{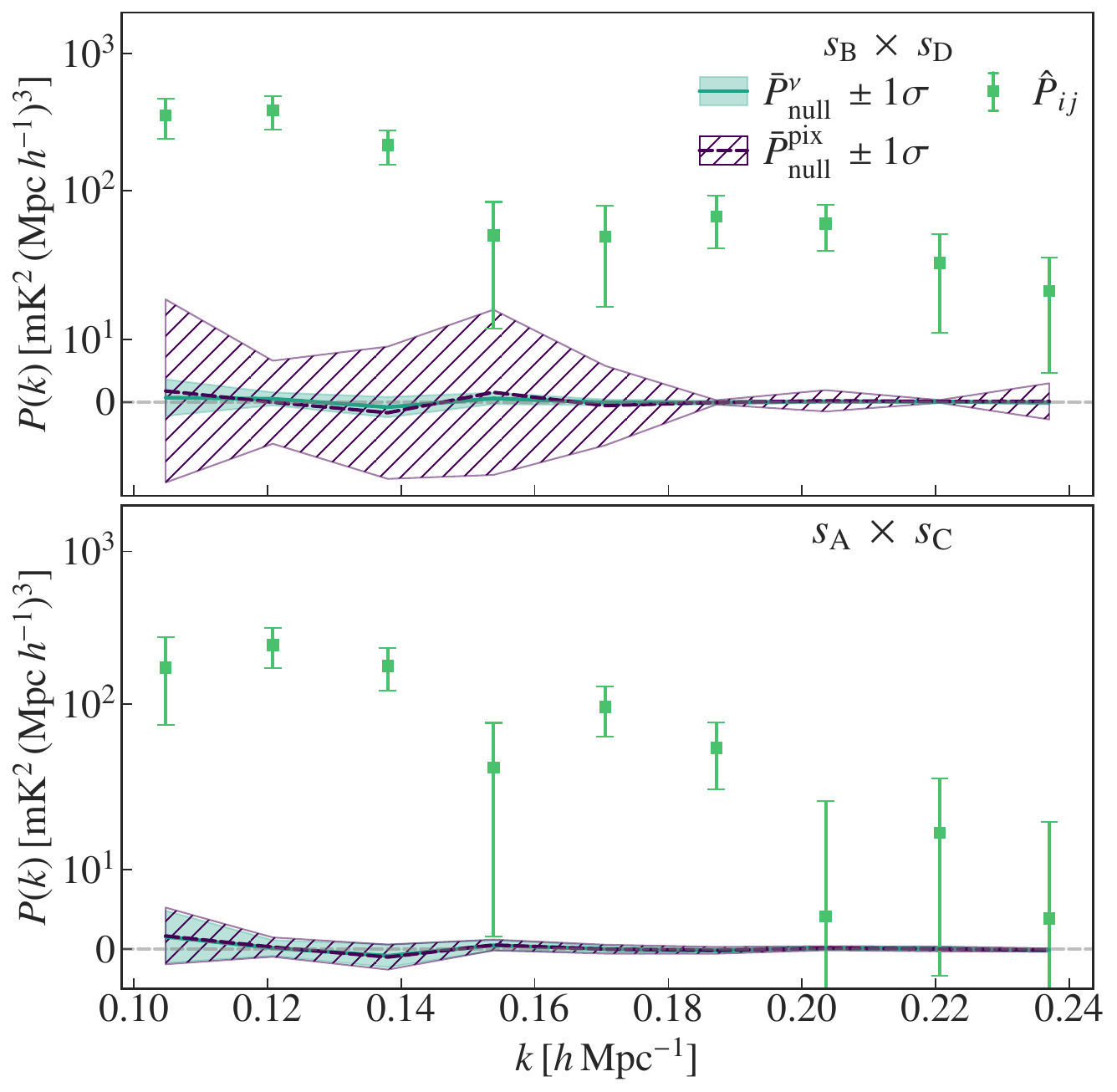}
    \caption{
    Results of the null tests for the two super-cross power spectra (the other cross-spectra show similar behaviours). 
    The measured $P(k)$ (points with error bars) are compared with the mean and scatter of the corresponding power spectra for both the null test performed in frequency (solid line and shaded region for the $1\,\sigma$ scatter) and in pixel space (dashed line and hatched region). %, with the standard deviation represented with shaded and hatched regions respectively. 
    $\smash{\bar{P}_{\rm null}^\nu}$ and $\smash{\bar{P}_{\rm null}^{\rm pix}}$ are very consistent with zero at all scales.
    For a clearer representation, we highlight the zero-level with a dashed grey line and we make use of a symmetric logarithmic scale on the vertical axis, that deviates to a linear mapping for $\left|P(k)\right|<10\,\pkum$. 
    }
    \label{fig:nulltest}
\end{figure}

\subsection{Null tests of the spectra measurements} \label{sec:null}

In the next test, we ask whether the detections of the internal cross-correlations among subsets are genuine, $\mathcal{H}_{2}$.

We randomly shuffle the \hi\ cleaned maps, either in frequency or in pixel space, and we then estimate the power spectrum combining a shuffled subset with a pristine one. The shuffling in either direction should wash out any cosmological correlation and, therefore, lead to a null power spectrum. In \cref{fig:nulltest}, we show examples of how the cross-correlation of subsets $i$--$j$ after one of them has been shuffled is in agreement with zero. Here, $\smash{\bar{P}_{\rm null}^\nu}$ and $\smash{\bar{P}_{\rm null}^{\rm pix}}$ are the averages over $200$ power spectra involving $200$ shuffled cubes, in the radial or angular direction respectively. They are compared to the non-shuffled $\widehat P_{ij}$ (filled points and error bars in \cref{fig:nulltest}). The extra scatter in the $\smash{\bar{P}_{\rm null}^{\rm pix}}$ of the $s_{\rm B}\times s_{\rm D}$ case, although significant only for a couple of central $k$-bins, might point to the presence of a persistent, coherent signal in the frequency direction at that intermediate scale. We will look more closely at this in another tests that focus on the parallel-direction information (\cref{sec:pknu_residual}).

To quantify the outcome of these null tests, for each cross-correlation and each shuffling direction, we compute the following generalized $\chi^2$ statistics,
\begin{align}
    \chi^2_{\rm data} & = \left(\widehat{\bm P} - \bar{\bm P}_{\rm null}\right)^{\sf T}\,{\sf\Sigma}_{\rm tot}^{-1}\,\left(\widehat{\bm P} - \bar{\bm P}_{\rm null}\right)\;, \\
    \chi^2_{{\rm null},i} & = \left(\bm P_{{\rm null},i} - \bar{\bm P}_{\rm null}\right)^{\sf T}\,{\sf\Sigma}_{\rm tot} ^{-1}\,\left(\bm P_{{\rm null},i} - \bar{\bm P}_{\rm null}\right)\;,
\end{align}
where ${\sf\Sigma}_{\rm tot}={\sf\Sigma}+{\sf\Sigma}_{\rm null}$ is the total covariance accounting for both the covariance of the measurements and of the null power spectra, ${\sf\Sigma}_{\rm null}$, estimated as sample covariance from the $200$ ${P}_{\rm null}$ realisations\footnote{The $ij$ subscripts are dropped for clarity.}. The significance is quantified as the fraction of null realisations whose $\smash{\chi^2_{\rm null}}$ exceeds $\smash{\chi^2_\mathrm{data}}$. The results demonstrate that the measured cross-power spectra are inconsistent with the null hypothesis in all cases. For both $s_{\rm B} \times s_{\rm D}$ and $s_{\rm A} \times s_{\rm C}$, and independently of the shuffling direction, none of the $200$ null realisations reaches $\smash{\chi^2_{{\rm data}}}$ and the $\smash{\chi^2_{{\rm null},i}}$ values cluster near zero ($\smash{\bar{\chi}^2_{\rm null}}\lesssim0.3$). This establishes a lower bound on the significance of $99.5\%$ C.L.\ ($p < 5 \times 10^{-3}$), consistent across both shuffling directions.

Overall, these results support the presence of a robust spatial correlation among the information carried by the subsets. Moreover, a spurious contribution driven by radial or angular systematics would be suppressed by one shuffling direction but not by the other, which is not observed. 

%%%%%%%%%%%%%%%%%%%%%%%%%%%%%%%%%%%%%%%%%%%%%%%%%%
%%%%%%%%%%%%%%%%%%%%%%%%%%%%%%%%%%%%%%%%%%%%%%%%%%

\subsection{`Patchwork' tests of the spectra measurements} \label{sec:frankenstein}

\begin{figure}
    \centering
\includegraphics[width=\columnwidth]{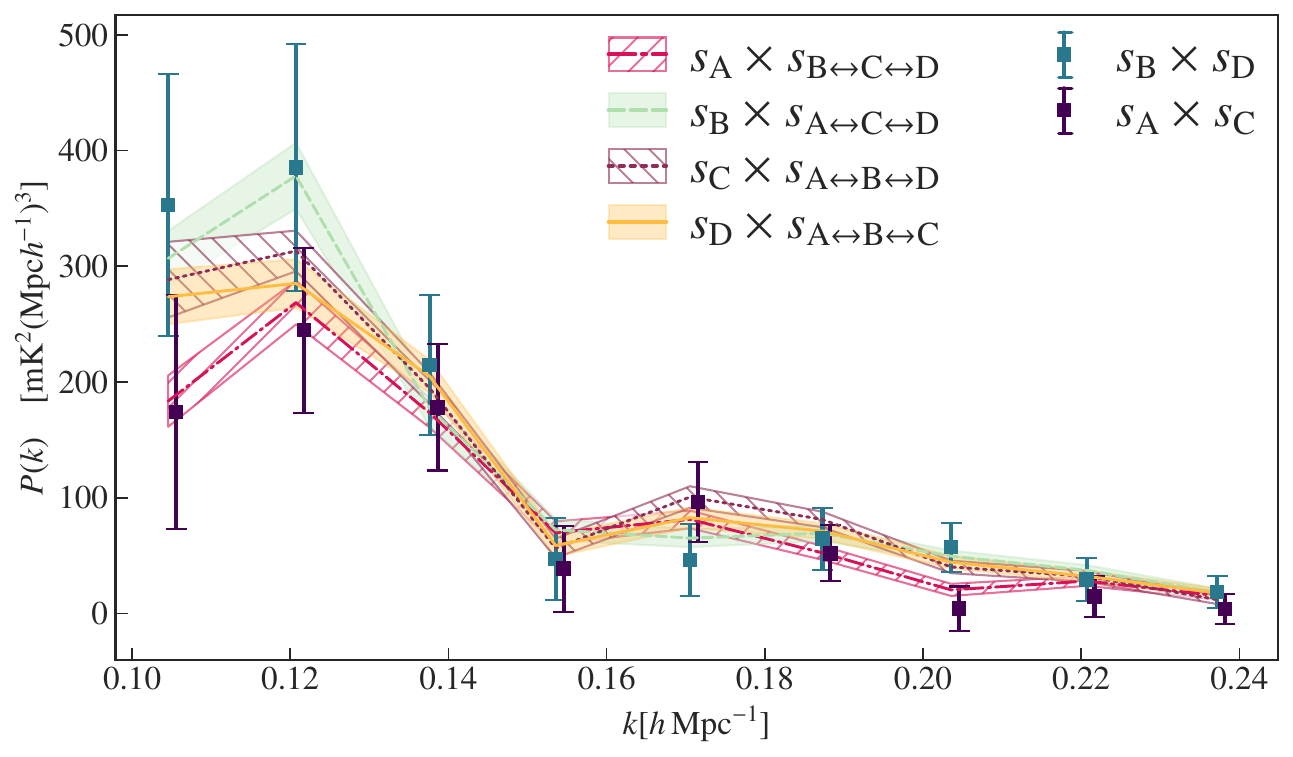}
    \caption{Results of the patchwork tests for all the subsets. The measured $\hat{P}_{ij}(k)$ of the super-cross (points with error bars) have to be compared with the mean of 100 shuffled spectra and their $1\,\sigma$ scatter.}
    \label{fig:franktest}
\end{figure}

Next, we address $\mathcal{H}_{3}$: The clustering signal measured by the internal cross-correlations is sky-driven (that is, not resulting from spurious spatial structures in the maps), hence subset-independent. To assess $\mathcal{H}_{3}$, we conceived the `patchwork' test. To reveal the eventual residual systematic contributions in the subsets, we create new patchwork subsets by randomly replacing voxels from 3 of the 4 other subsets. In the patchwork subset, we preserve the angular position (pixel) and frequency position of the sky information, while subset-specific extra contributions (i.e., potential residual systematics) are shuffled. We cross-correlate each patchwork subset with the fourth, uninvolved subset and check whether the cross-correlation signal remains consistent with the internal cross-correlations. If strong systematic contributions are present, we would see a drop in power by replacing these voxels. The test is indeed passed, as we show in \cref{fig:franktest}: error bars correspond to the two super-cross spectra and the four shaded regions are the $1\,\sigma$ scatter of $4\times100$ realisations of the patchwork shuffling.

To be more quantitative, for each patchwork type we follow an approach similar to the one presented in \cref{sec:null} and compute
\begin{equation}
    \chi^2_{\rm pw} = \left(\widehat{\bm P} - \bar{\bm P}_{\rm pw}\right)^{\sf T}\,{\sf\Sigma}_{\rm tot}^{-1}\,\left(\widehat{\bm P} - \bar{\bm P}_{\rm pw}\right)\;,
\end{equation}
with $\smash{\bar{\bm P}_{\rm pw}}$ the mean over the 100 patchwork power spectra and ${\sf\Sigma}_{\rm tot}={\sf\Sigma}+{\sf\Sigma}_{\rm pw}$ accounts for the covariance of the data, ${\sf\Sigma}$, and the sample covariance from the different realisations, ${\sf\Sigma}_{\rm pw}$. The patchwork power spectra are found to be mutually consistent: ${\sf\Sigma}_{\rm pw}$ gives a negligible contribution to ${\sf\Sigma}_{\rm tot}$ and $\smash{\bar{\chi}^2_{\rm pw}} \approx 0.5$. This would not be the case in the presence of strong systematics, which would cause different voxel reassignments to produce incoherent power spectra and a broad distribution of $\smash{\chi^2_{\rm pw}}$. The observed coherence among realisations already constitutes a positive outcome of the test. 

We further compare $\smash{\chi^2_{\rm pw}}$ against the analytical $\smash{\chi^2(n_k)}$ distribution, which describes the expected fluctuations of data given $\Sigma_{\rm tot}$\footnote{As ${\sf\Sigma}_{\rm pw}$ is subdominant with respect to ${\sf\Sigma}$, the empirical $\smash{\chi^2}$ distribution of the individual realisations is strongly compressed near zero and does not represent a valid reference for $\smash{\chi^2_{\rm pw}}$. The latter is consequence of the mutual consistency of the realisations, that do not manifest the expected scatter of a single noisy measurement. Hence, we use the theoretical $\smash{\chi^2(n_k)}$ distribution to compare against $\smash{\chi^2_{\rm pw}}$.}. Half of the cases fall in the range $0.4 \lesssim p \lesssim 0.8$, indicating good compatibility between the data and the patchwork mean. Only one case gives $p \approx 0.14$, which remains well above the conventional $p < 0.05$ threshold for rejection; the other cases yield $p \gtrsim 0.95$, reflecting the large measurement uncertainties. Overall, these results support the validity of $\mathcal{H}_3$ for all four patchwork types and both cross-correlations; the patchwork realisations are mutually coherent, and in agreement with the measured spectra, demonstrating that the clustering signal is not driven by subset-specific systematic residuals.

%%%%%%%%%%%%%%%%%%%%%%%%%%%%%%%%%%%%%%%%%%%%%%%%%%
%%%%%%%%%%%%%%%%%%%%%%%%%%%%%%%%%%%%%%%%%%%%%%%%%%

\subsection{Accuracy of the component separation} \label{sec:rleakage}

\subsubsection{Quantifying leakages between components} \label{sec:rleakage_lk}

The underlying assumption of mPCA is that it is possible to linearly decompose the maps into their individual components, which we assign to the contaminant part or the cosmological signal plus uncorrelated noise. However, as previously noted, the coupling between different components in the observed data hinders the possibility of a perfect separation. How much contaminant has leaked into the signal? And how much signal is lost to the contaminant? To give a statistical answer to these questions we now test $\mathcal{H}_{4}$: the cleaning has succeeded in decomposing each subset in two uncorrelated parts.

\begin{figure}
        \centering
        \includegraphics[width=\columnwidth]{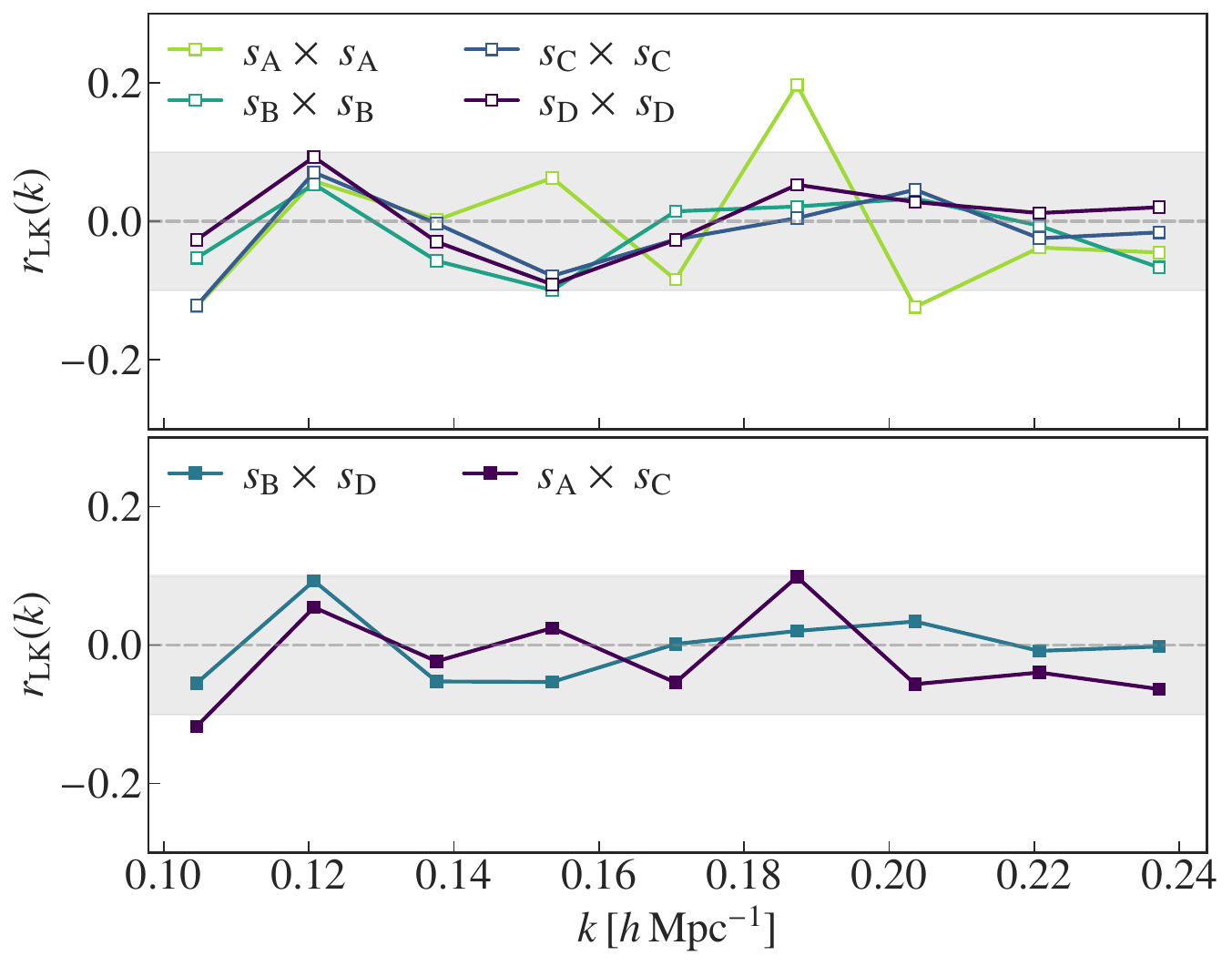}
        \caption{Cross-correlation coefficient among the signal and contaminants part of each component separation process (as in \cref{eq:rLK}). Empty points in the top panel refer to all the subsets of the \chess\ division; filled points in the bottom panel refer to the two super-cross. A dashed line highlights the no-correlation limit ($r_{\rm LK}=0$), and the shaded region the $\pm10\%$ level of correlation. 
        }
        \label{fig:rleak}
\end{figure}

We compute the spatial correlation of the two contaminant-separation terms. For each cleaned subset $s_{{\rm clean},i}$ and its foreground counterpart $s_{{\rm fg},i}\coloneqq s_i-s_{{\rm clean},i}$, we compute the leakage (`LK') cross-correlation coefficient
\begin{multline} \label{eq:rLK}
    r_{{\rm LK,}ij}=\frac1{2}\left[\frac{\cP(s_{{\rm fg},i},s_{{\rm clean},j})}{\sqrt{\cP(s_{{\rm fg},i},s_{{\rm fg},i})\,\cP(s_{{\rm clean},j},s_{{\rm clean},j})}} \right. \\+ \left.\frac{\cP(s_{{\rm fg},j},s_{{\rm clean},i})}{\sqrt{\cP(s_{{\rm fg},j},s_{{\rm fg},j})\,\cP(s_{{\rm clean},i},s_{{\rm clean},i})}}\right] \; ,
\end{multline}
where the symbol $\cP$ stands for the power spectrum estimator defined in \cref{sec:pestim}. Besides expecting $s_{{\rm clean},i}$ and $s_{{\rm fg},i}$ to be uncorrelated ($r_{{\rm LK,}ii}\approx0$ at all scales), we check for a null $r_{{\rm LK,}ij}$ too, as the cleaned maps of one subset should not correlate with the foreground maps of other subsets. To ensure that in the $r_{{\rm LK,}ij}$ case the arithmetic mean in \cref{eq:rLK} does not mask a significant correlation in one of the individual cross-terms, we also monitor each term separately and compute their geometric average, which aligns very well with the arithmetic one; this rules out the possibility of a large systematic in one term being diluted by a near-zero value in the other. \Cref{fig:rleak} shows that the cross-correlation coefficients remain consistently modest with $|r_{\rm LK}|$ lower than about $10\%$ (see grey-shaded band) throughout the whole scale range analysed, with $|r_{\rm LK}|<5\%$ in most cases. Furthermore, the fact that $r_{\rm LK}$ oscillates around zero, rather than maintaining a consistent sign, indicates that these correlations are stochastic in nature and do not represent a systematic leakage. 

This confirms that mPCA effectively separates two almost spatially uncorrelated cubes with subdominant leakage between them. 

Beyond providing a diagnostic for the orthogonality of the mPCA decomposition, these correlation coefficients allow us to place a quantitative upper bound on the foreground contamination in our final results. By modelling the residual leakage in each cleaned subset as a fraction of the removed components, $s_{{\rm leak},i}\approx \epsilon_i\,s_{{\rm fg},i}$, we can relate the systematic bias in the cross-power spectrum, $\smash{P_{{\rm leak},ij}\coloneqq\cP(s_{{\rm leak},i},s_{{\rm leak},j}}$), to the measured $r_{{\rm LK}}$. This approximation is justified by the operating principle of mPCA, which identifies and isolates the high-variance, spatially coherent structures of the foregrounds. Since the $s_{{\rm fg},i}$ cubes represent the only physical contaminants present in the data, any residual leakage resulting from an incomplete separation is expected to inherit the same structural and morphological properties, effectively acting as a rescaled trace of the removed component. Supporting this assumption is the observed Gaussianity of the voxel distribution the cleaned maps (discussed in \cref{sec:lowpass}): the fact that the structured (non-Gaussian) part of the foregrounds has been entirely transferred to $s_{{\rm fg},i}$ implies that any remaining contaminant can be approximated as a first-order linear fraction of these identified components, as there are no other significant independent structures left to contribute to the leakage.

The auto-correlation coefficients $r_{{\rm LK},ii}$ can directly constrain the leakage power to be a small fraction of the total variance, as we can derive that
\begin{equation} \label{eq:contam_auto}
    \frac{P_{{\rm leak},ii}}{P_{ii}}\approx\epsilon_i^2\frac{P_{{\rm fg},ii}}{P_{ii}}\approx r_{{\rm LK,}ii}^2 \; .
\end{equation} 
To extend this bound to the cleaned cross-power spectra $P_{ij}(k)$, we leverage the Cauchy-Schwarz inequality,
\begin{equation}\label{eq:contam_cross}
    P_{{\rm leak},ij}\leq\sqrt{P_{{\rm leak},ii}\,P_{{\rm leak},jj}}=|r_{{\rm LK,}ii}\,r_{{\rm LK,}jj}|\sqrt{P_{ii}\,P_{jj}} \; ,
\end{equation}
yielding a conservative upper bound for the relative contamination in the cross-correlation.

These tests demonstrate that the remaining foreground contamination has been reduced to negligible levels compared to the cosmological signal. Specifically, the contamination level remains within the sub- and a few-per-cent ranges across all scales of the measured $P_{ij}(k)$, since the relative leakage is proportional to $r_{{\rm LK},ii}^2$. Moreover, compared with the statistical uncertainty of the measurement, the systematic bias is consistently lower than $10\%$ of the error bars. 

\subsubsection{Cross-correlation with non-cosmological frequencies} \label{sec:rleakage_fg}

To further assess the robustness of our foreground removal and quantify lingering contamination in the cleaned data cubes, we perform a null-consistency test by cross-correlating the cleaned subsets with unclean data from a non-cosmological frequency range centred at 1460.9 MHz. At these frequencies, any detected 21-cm signal is artificial. This higher-frequency window also contains RFI and observational systematics in addition to astrophysical emissions. This test relies on the high spectral coherence of foreground signals. As demonstrated in \cref{app:radial_valid}, pure foreground maps maintain a correlation coefficient near unity even across large frequency separations. Consequently, any significant residual foregrounds in the cosmological range should appear as a detectable signal when correlated with these unclean frequencies outside the cosmological band. Therefore, analysing potential correlations between maps at frequencies within or outside the cosmological band allows us to test $\mathcal{H}_5$. The data at the non-cosmological frequencies are divided into four subsets $s_{{\rm astro},i}$ that correspond exactly to the subsets used in the cosmological range in terms of dish- and block-grouping. This ensures that, beyond foregrounds, time-dependent or scan-related systematics manifesting in both frequency ranges are captured consistently across subsets. Each $s_{{\rm astro},i}$ subset is then regridded into a Cartesian box using the same parameters and pipeline as the cleaned cosmological data, treating the astrophysical data operationally as cosmological. 

We use the cross-correlation coefficient defined as
\begin{multline} \label{eq:rnoncosmo}
    r_{{\rm FG,}ij}=\frac1{2}\left[\frac{\cP(s_{{\rm astro},i},s_{{\rm clean},j})}{\sqrt{\cP(s_{{\rm astro},i},s_{{\rm astro},i})\,\cP(s_{{\rm clean},j},s_{{\rm clean},j})}} \right. \\+ \left.\frac{\cP(s_{{\rm astro},j},s_{{\rm clean},i})}{\sqrt{\cP(s_{{\rm astro},j},s_{{\rm astro},j})\,\cP(s_{{\rm clean},i},s_{{\rm clean},i})}}\right] \; ,
\end{multline}
as metrics for this test. We evaluate the cross-correlation coefficient not only between different subsets at different frequencies but also between the same subset across the two frequency chunks. This latter case is the most conservative combination because any subset-specific contaminant is expected to correlate most strongly with it. Indeed, foreground cleaning is performed independently on each subset, so the specific modes removed differ. If an individual subset remains more contaminated after cleaning, those residuals might not find a correlating counterpart when performing internal cross-correlations within the cosmological frequency band. However, when cross-correlating with an unclean, non-cosmological subset, the dominant and coherent foreground signal ensures any residual contamination is effectively captured.

\begin{figure}
        \centering
        \includegraphics[width=\columnwidth]{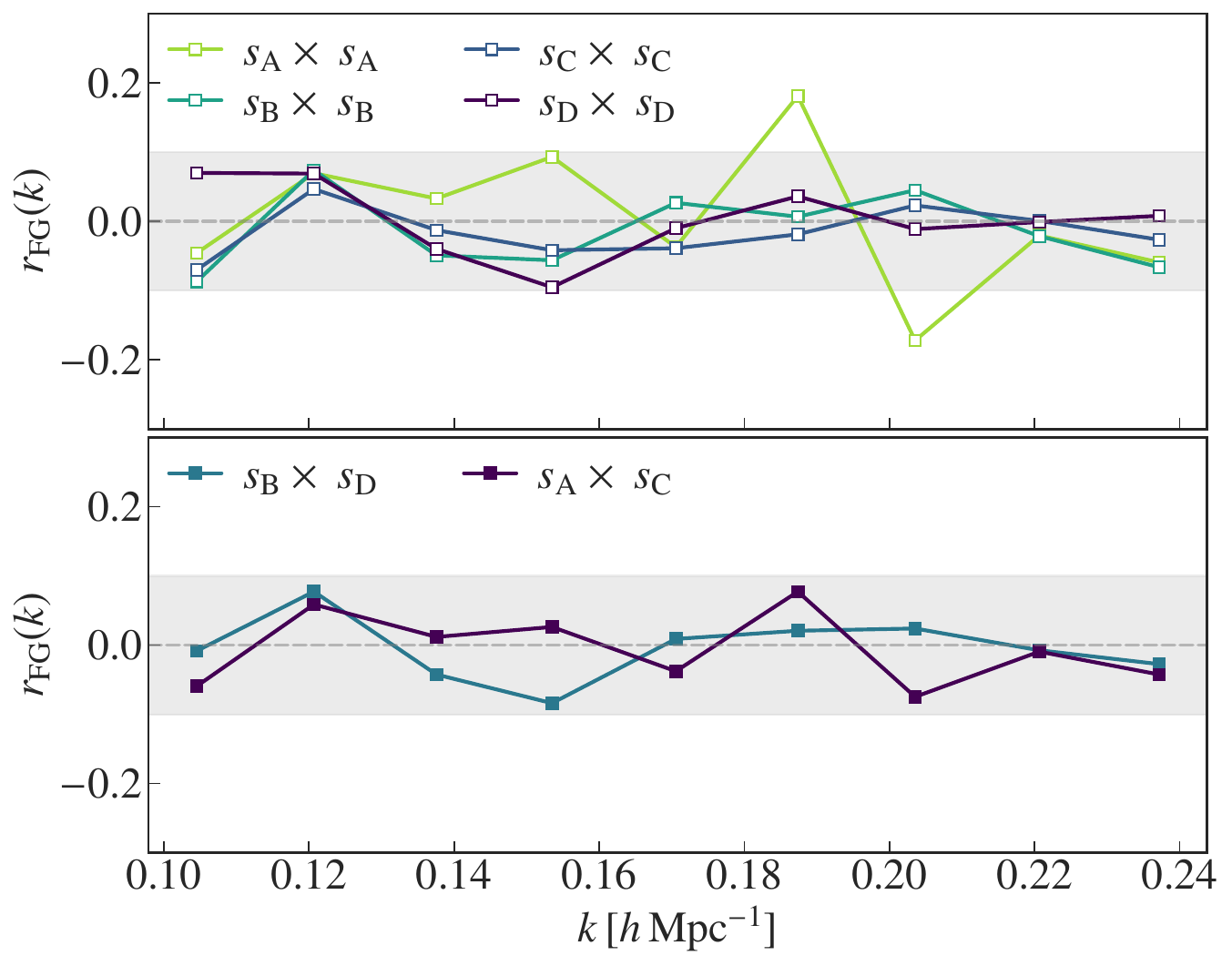}
        \caption{Cross-correlation coefficient (as in \cref{eq:rnoncosmo}) of all the subsets of the \chess\ division (empty points, top panel) and for the two super-cross (filled points, bottom panel), where the second subset of the cross-correlation is from the non-cosmological frequencies data set. A dashed line highlights the no-correlation limit ($r_{\rm FG}=0$), and the shaded region the $\pm10\%$ level of correlation.}
        \label{fig:rnoncosmo}
\end{figure}

The results are shown in \cref{fig:rnoncosmo}. For both the cross-subset and auto-subset cases the $r_{{\rm FG,}ij}$ are remarkably consistent, with the measured coefficients oscillating around zero across the majority of the $k$-range. Notably, $s_{\rm A}$ exhibits a slightly higher level of spurious correlation compared to the others, aligning with the outcome of other tests. The larger fluctuations observed in these cases are at $k\approx 0.2\,\kum$, corresponding to scales where the signal is anyway heavily suppressed.

This test strengthens the hypotheses that measured power spectra are not significantly biased by foreground residuals.

%%%%%%%%%%%%%%%%%%%%%%%%%%%%%%%%%%%%%%%%%%%%%%%%%%
%%%%%%%%%%%%%%%%%%%%%%%%%%%%%%%%%%%%%%%%%%%%%%%%%%

\subsection{Radial-direction analysis} \label{sec:pknu_residual}

The large scales along the line-of-sight direction are those affected the most by the mode-mixing between astrophysical foregrounds and the \hi\ signal \citep{2015MNRAS.447..400A,Carucci2020}. As such, they are also the most informative to highlight residual foregrounds in the cleaned maps. Indeed, \textit{i)} we expect foregrounds to be highly correlated along frequency, with the highest correlation amplitude at the largest radial separations and slowly decreasing monotonically at smaller scales (due to their smooth frequency spectra), compared to the \hi\ signal that is characterised by an almost flat radial power spectrum ($\mathcal{H}_{6}$).\footnote{In our Cartesian setup, also along the radial direction the \hi\ field experiences the damping effect of the telescope beam \citep{Villaescusa2017,Carucci2020}.} From the same argument, it follows that, \textit{ii)} if we divide the data cube in two, the high and the low-frequency sub-cubes (dubbed `front' and `back') and correlate them, we would measure zero correlation if the splits contain cosmological signal only; otherwise, we would measure a positive correlation if residual foregrounds are present ($\mathcal{H}_{7}$). For a more exhaustive discussion, we refer the reader to \cref{app:radial_valid}, where we demonstrate the points above with simulations. In particular, \textit{(i)} and \textit{(ii)} can be appreciated in the top and bottom panel of \cref{fig:Pknu_sims}, respectively.

We do not perform any cuts in $k$-space for the tests in this section, aiming instead to scan the full radial scale range for systematics. Indeed, we use these tests to define our final cut in the parallel direction. This approach allows us to strengthen the choice for filtering data at specific $k_\|$ in the primary analysis, ensuring the reliability of the retained parallel modes.

\begin{figure}
    \centering
    \includegraphics[width=\columnwidth]{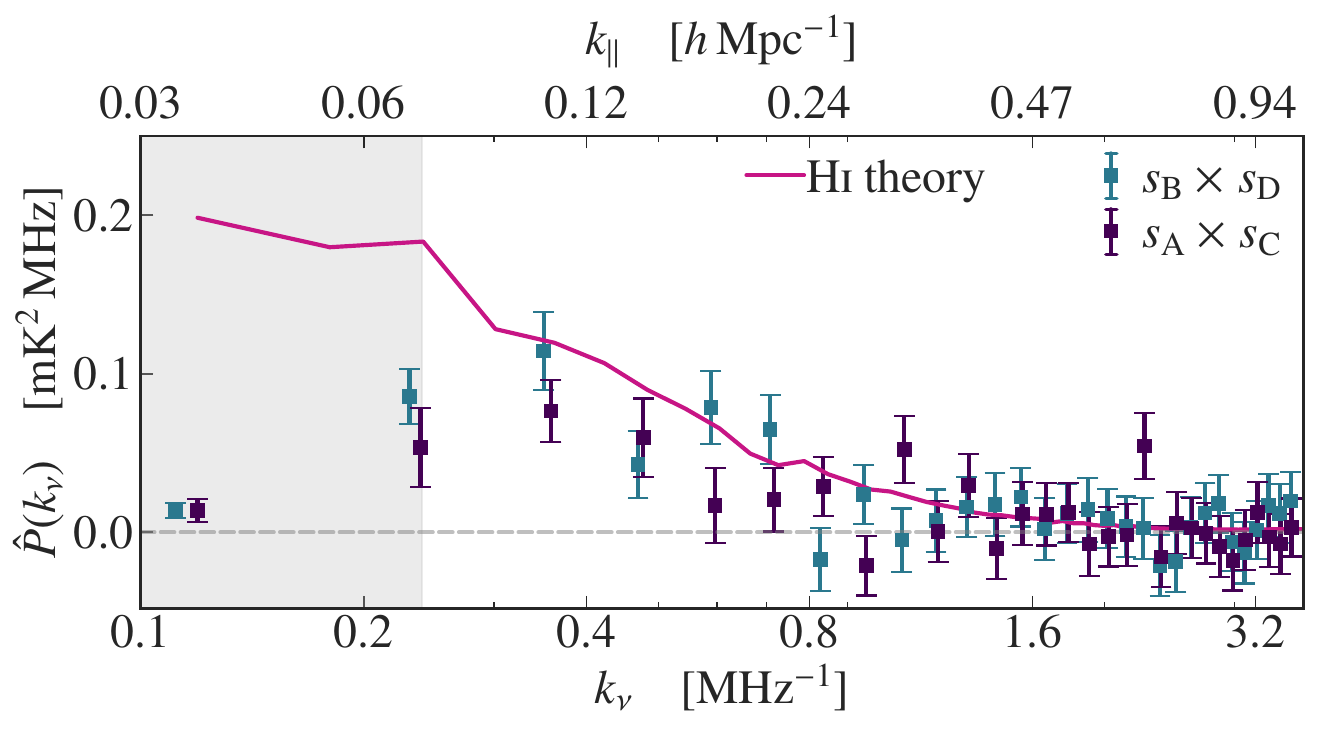}
    \caption{
    Cross-$P(k_\nu)$ for the super-cross combinations, which bracket all \chess\ division combinations. %As in \cref{fig:radial_Pk_frbk}, 
    The grey shaded regions corresponds to $k_\|<0.07\,\kum$, excluded from the primary analysis. The solid magenta line shows a theoretical prediction derived from the simulations presented in \cref{app:radial_valid}, hence following the same gridding as data, and with the amplitude rescaled to match the $\Ohi\bhi$ best-fit value obtained in the main analysis.}
    \label{fig:radial_Pk}
\end{figure}

As a radial power spectrum estimator, we adopt the one-dimensional power spectrum in frequency space $\smash{P(k_\nu)}$, where $k_\nu=2\,\pi/\nu$.\footnote{Note that in this context we refer to $\nu$ as the frequency real axis (starting from 0) and not to measured physical frequencies.}
This formalism does not need any gridding of the observed maps. In practice, we apply the Fourier transform directly to the cleaned maps $s_{\rm clean}(\thv, \nu)$: for each pixel (at coordinates $\thv$), the Fourier transform is 
\begin{equation}
    \tilde{s}_{{\rm clean},i}(\thv,k_\nu)=\sum_\nu\,s_{{\rm clean},i}(\thv,\nu)\,\ee^{-\ii\,k_\nu\,\nu}
\end{equation}
and the power spectrum is obtained by averaging over all the lines of sight,
\begin{equation}
    \widehat P_{ij}(k_\nu)\coloneqq\delta\nu\,\left\langle\tilde{s}_{{\rm clean},i}(\thv,k_\nu)\,\tilde{s}_{{\rm clean},j}^\ast(\thv,k_\nu)\right\rangle_{\bm\theta} \;.
\end{equation}
Assuming no redshift evolution in the analysed cube, we can relate $k_\nu$ to $k_\|$ through
\begin{equation}\label{eq:knu2kpar}
    k_\| = \frac{\nu_\hi\,H(z_{\rm eff})}{c\,(1+z_{\rm eff})^2} \,h^{-1}\, k_\nu \;,
\end{equation}
with $\nu_\hi = 1420\,\mathrm{MHz}\equiv c/(21\,\mathrm{cm})$ being the rest-frame frequency of the \hi\ 21-cm emission.

\subsubsection{Cross-$\widehat P(k_\nu)$} \label{sec:Pknu}

\begin{figure}
        \centering
        \includegraphics[width=\columnwidth]{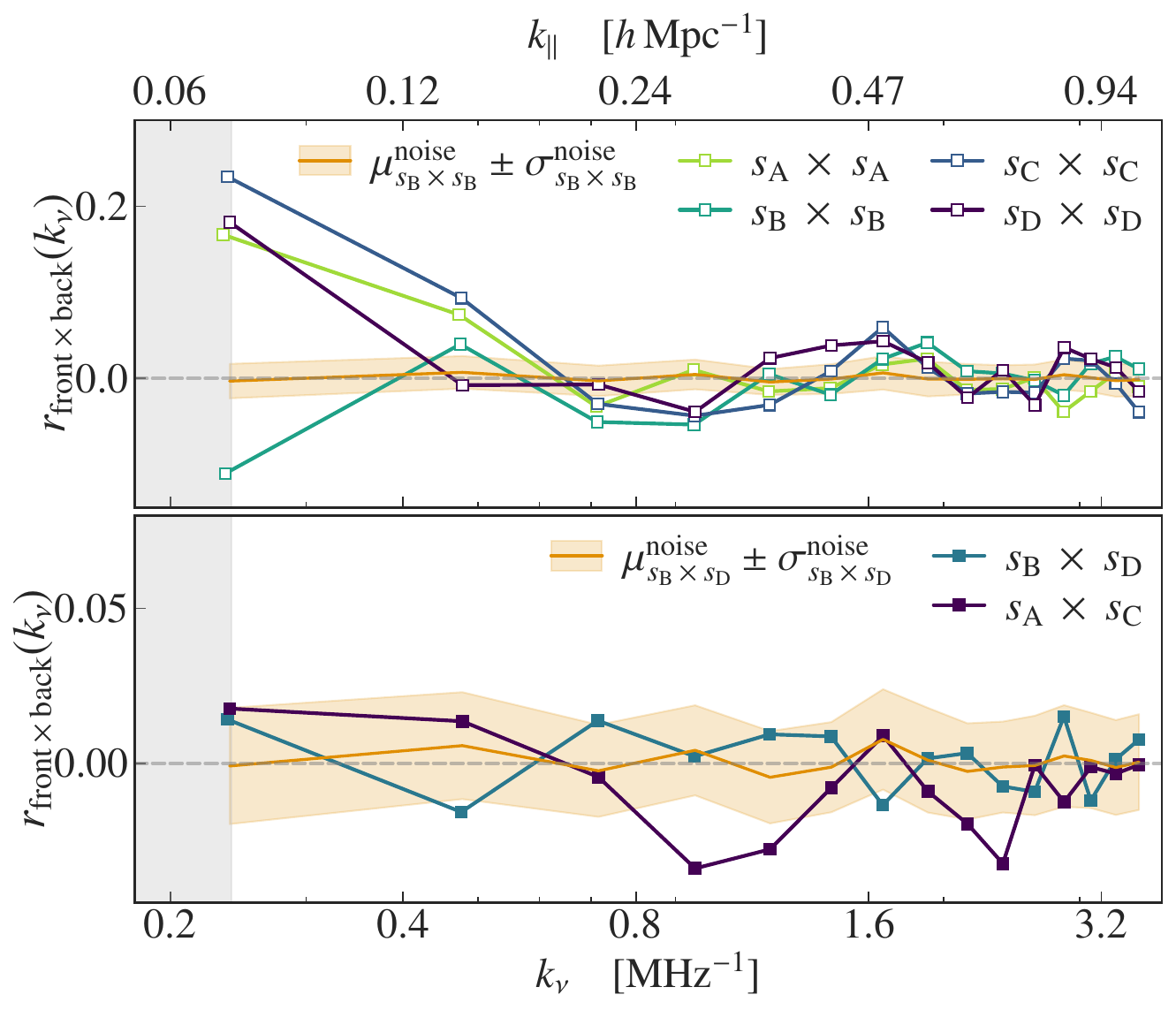}
        \caption{
        Cross-correlation coefficients of the front and back batches of each subset for the \chess\ division, i.e., computed cross-correlating disjoint frequency ranges within the same subset (empty points in the top panel) and for the two super-cross, i.e., cross-correlating the front and back batches of different subsets (bottom panel). Any significant correlation could be traced back to residual astrophysical foreground contamination or systematics such as calibration errors. The vertical grey shaded region indicates the range excluded in the primary analysis by the $k_{\|{\rm min}}$ cut. The orange shaded areas are the $1\sigma$-scatter from independent thermal-noise realisations that underwent the same $r_{\rm front\times back}$ computations; we show one case only as they are equivalent.
        }
        \label{fig:radial_Pk_frbk}
\end{figure}
\begin{figure*}
        \centering
        \includegraphics[width=2\columnwidth]{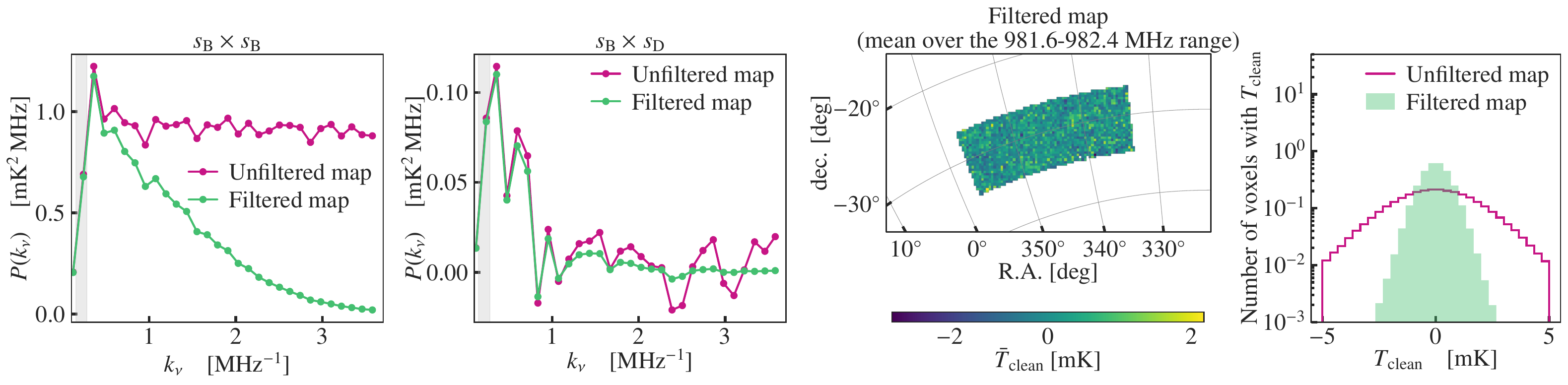}
        \caption{Example of a radial low-pass filtering test described in \cref{sec:lowpass}. The $s_{\rm B}$ subset is filtered along the frequency direction with a Gaussian kernel of $\sigma=1.2$ MHz (first panel) to suppress the information that does not cross-correlate with the other subsets (second panel). Visually, the maps of the filtered $s_{\rm B}$ do not show any unexpected structure, as, for example, shown in the third panel; indeed, the temperature distribution of their voxels still resembles the same symmetric shape with less variance, i.e., noise (fourth panel).}
        \label{fig:lowpass}
\end{figure*}

\Cref{fig:radial_Pk} shows the cross-$\smash{\widehat P(k_\nu)}$ computed for the super-cross combinations of the \chess\ division. The error bars are computed assuming Gaussian statistics and adapting \cref{eq:PvarGauss} to this specific case.
We observe positive $10^{-2}$--$10^{-1}$ mK$^2$ MHz power at the largest radial scales, quickly approaching zero as $k_\nu$ increases (roughly after $k_\nu\gtrsim0.8\,\fum^{-1}$), matching the expectations from \hi-only simulations (see top panel of \cref{fig:Pknu_sims} in \cref{app:radial_valid}), rescaled to the inferred best-fitted $\Ohi\bhi$ value and undergone through the same windowing scheme as data (solid magenta line in \cref{fig:radial_Pk}). The smallest two $k_\nu$-bins show highly suppressed clustering power, which is unexpected and likely due to signal loss due to the cleaning; indeed, it corresponds to the region in $k_{\|}$ that we exclude from the main analysis through the $k_{\|{\rm min}}$ cut, highlighted by the grey shaded region.

\subsubsection{Frequency chunks: front and back splitting} \label{sec:fXb}

Another way to look at the line-of-sight clustering properties of the cleaned subsets is to cross-correlate their `front' and `back' partitions. 
Moreover, this also tests the presence of multiplicative systematics that may decorrelate when reshuffling, as in the first null tests in \cref{sec:null}, but can be preserved in this splitting, such as bandpass calibration errors and standing waves generated by the MeerKAT dishes \citep{2021MNRAS.506.5075M,Zhaoting2025}.

In \cref{fig:radial_Pk_frbk} we show the correlation coefficient of the front and back batches of each subset,
\begin{multline} \label{eq:r_frontback}
    r_{{\rm front\times back,}ij}=\frac1{2}\left[\frac{\widehat P^{\rm front\times back}_{ij}(k_\nu)}{\sqrt{\widehat P^{\rm front\times front}_{ii}(k_\nu)\,\widehat P^{\rm back\times back}_{jj}(k_\nu)}} \right.\\+ \left.\frac{\widehat P^{\rm front\times back}_{ji}(k_\nu)}{\sqrt{\widehat P^{\rm front\times front}_{jj}(k_\nu)\,\widehat P^{\rm back\times back}_{ii}(k_\nu)}}\right] \; .
\end{multline}

The overall results are encouraging: for most scales and for all auto-correlations the ratio oscillates around zero. Although of small amplitude, some deviations from zero are present at the smallest $k_\nu$ bin, indicating the possible influence of residual spurious, non-cosmological signals at those scales. Nevertheless, these deviations are no longer observed when moving to the cross-correlations, corroborating the robustness of these combinations. We anyway exclude that first bin from the main analysis through the $k_{\|{\rm min}}$ cut, as shown by the grey region in \cref{fig:radial_Pk_frbk}. The behaviours shown in \cref{fig:radial_Pk} and \cref{fig:radial_Pk_frbk} suggest that the adopted $k_\|$ cut is a robust choice. Next in \cref{sec:scale_indep}, we also test what happens to our internal cross-correlation fits with more conservative cuts and find no notable change in the results.

Finally, to put the $r_{{\rm front\times back,}ij}$ oscillations around zero into context, we compared them against the scatter expected from idealised subsets composed only of thermal noise. For each subset, we generated 50 independent noise realisations using \cref{eq:radiometer}, and computed the corresponding $r_{{\rm front\times back,}ij}$. As shown in \cref{fig:radial_Pk_frbk}, the resulting fluctuation amplitude matches the oscillations observed in the data within $1\sigma$ for the super-cross cases, $2-3\sigma$ for the auto-correlations. The latter exercise is reassuring for all spectrum combinations and once again shows the better behaviour of the super-cross combinations.

%%%%%%%%%%%%%%%%%%%%%%%%%%%%%%%%%%%%%%%%%%%%%%%%%%
%%%%%%%%%%%%%%%%%%%%%%%%%%%%%%%%%%%%%%%%%%%%%%%%%%

\subsection{Inspecting maps after a radial low-pass filtering} \label{sec:lowpass}

When considering the one-dimensional cross-$\smash{P(k_\nu)}$ shown in \cref{fig:radial_Pk} and discussed in \cref{sec:Pknu}, we have demonstrated how, as expected, most of the clustering power is from the largest sampled parallel modes (excluding the $k_\parallel$-cut). We want to inquire further into this clustering signal revealed by the cross-$\smash{P(k_\nu)}$ and check that also in the angular direction (in pixel space) it visually corresponds to something compatible with the sum of the cosmological signal and uncorrelated noise ($\mathcal{H}_{8}$). 

We proceed as follows. We low-pass filter the cleaned subsets along the radial direction (Gaussian smoothing with a standard deviation of about 12 channels, that is $\approx1.2$ MHz) to isolate the high-clustering portion of the signal. We visually inspect the corresponding maps: are they still noisy, as was shown in \cref{fig:cleanedmaps}, or does some evident structure of non-cosmological origin arise? For example, in \citet{Carucci2024}, we have shown that evident stripes and regular patterns are present in the MeerKAT unclean intensity maps, which we conclude (and are probing and characterising in ongoing analyses) originate from systematics. Results are reassuring, as shown for the $s_{\rm B}$ case in \cref{fig:lowpass}. 

Both the unfiltered and the low-pass-filtered maps are compatible with a \hi\ plus noise signal, as also quantitatively shown by the histograms of the voxel temperature in the rightmost panel of \cref{fig:lowpass}. To validate quantitatively this agreement, we compared the skewness and kurtosis of the cleaned data with those of simulated thermal noise maps generated using the actual survey hit map; the skewness of the cleaned subsets is reduced to near-vanishing values ($\lesssim 10^{-2}$), the residual excess kurtosis ($\lesssim 0.8$) is found to be in excellent agreement with the noise simulations and is a deterministic consequence of the non-uniform integration time across the observed region.

The results indicate that our foreground removal procedure effectively removed the dominant, coherent foreground components. The remaining residuals are stochastically distributed at the pixel level, with no significant non-Gaussian signatures exceeding the noise floor. This outcome is consistent with expectations, as the cleaned subsets are individually dominated by noise. 

%%%%%%%%%%%%%%%%%%%%%%%%%%%%%%%%%%%%%%%%%%%%%%%%%%
%%%%%%%%%%%%%%%%%%%%%%%%%%%%%%%%%%%%%%%%%%%%%%%%%%

\subsection{Signal injection test} \label{sec:injection}

\begin{figure}
    \centering
    \includegraphics[width=\columnwidth]{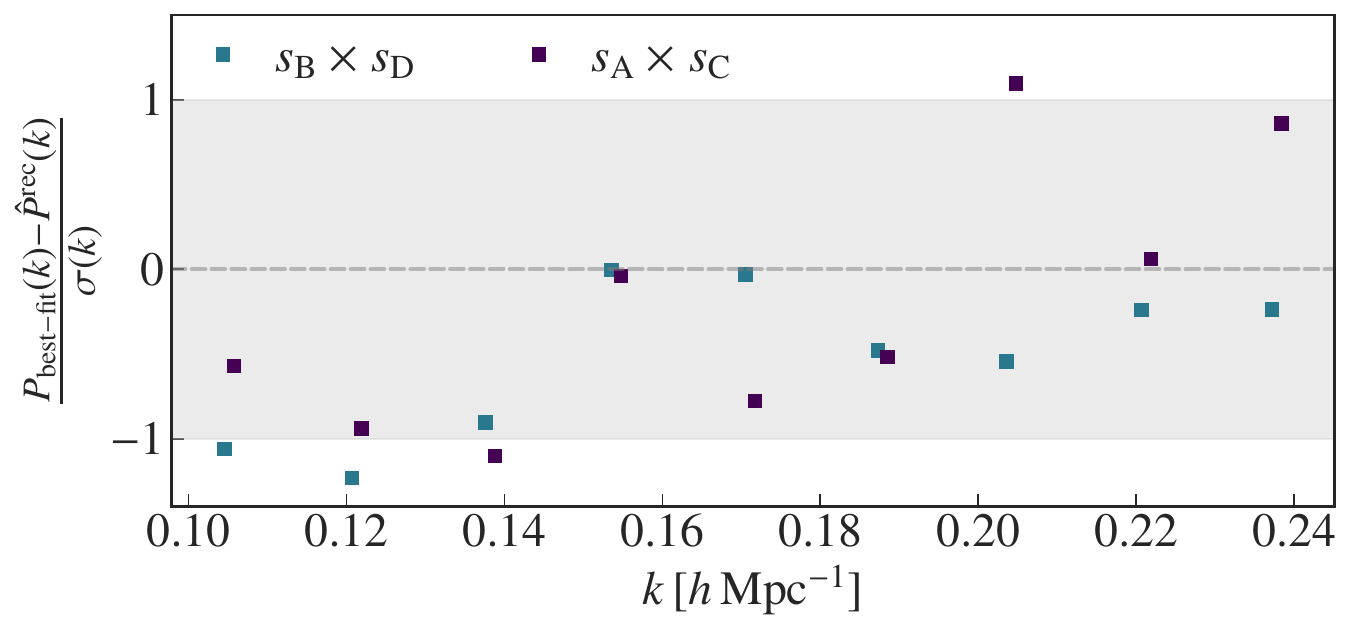}
    \caption{Results of the signal injection test performed on the power spectra estimated from the super-cross combinations. In both cases, the departure between the inferred and reconstructed power spectra remains within the $68\%$ confidence level region (highlighted in by the grey region) for most $k$-bins.
    }
    \label{fig:signal_inj}
\end{figure}

As stressed before, part of the \hi\ signal is inevitably eroded by blind foreground cleaning techniques such as the adopted mPCA. During inference, we characterise this effect with the kernel $\mathcal{D}(k)$ of our model (\cref{eq:Dsl}). To estimate the impact of signal loss on the measured power spectra and to test that $\mathcal{D}(k)$ indeed captures it ($\mathcal{H}_{9}$), we perform a signal injection test.

We inject synthetic \hi\ signal in the observed maps just before the foreground subtraction step, leaving the rest of the analysis unchanged, and compare the resulting power spectrum with the input one. In particular, we adapt the formalism described in \citet{Cunnington2023}, consisting in deriving a scale-dependent transfer function (TF) $\pazocal{T}(k)$ to reconstruct the signal lost because of the foreground removal process at the power spectrum level. The reconstructed power spectrum reads

\begin{equation} \label{eq:TF_std}
    \widehat P^{\rm rec}_{ij}(k)=\frac{\widehat P_{ij}(k)}{\pazocal{T}^\times_{ij}(k)} \; ,
\end{equation}
where $\smash{\pazocal{T}^\times_{ij}(k)}$ refers to the transfer function for a specific cross-spectrum, and it is computed averaging over $N_{\rm mocks}=200$ realisations (see \cref{app:tf_valid} for further details). 

Having the reconstructed power spectra, we can measure its agreement in each $k$-bin with the final, inferred $P_{\hi}(k)$ model of our primary analysis. This comparison is shown in \cref{fig:signal_inj}, where the difference between the best-fit power spectra and the reconstructed ones is represented in units of the (asymmetric) $68\%$ confidence level associated to the fitted value of $\Ohi\bhi$. The uncertainty on the two $\beta$ parameters is not considered in this test, which makes the comparison more conservative. There is broad agreement between the simulation-based way to quantify signal loss and the inferred one through the $\mathcal{D}(k)$ kernel. We quantify the latter agreement by computing the $\smash{\chi^2_{\rm inj}}$ statistics,
\begin{equation}
    \chi^2_{\rm inj}=\left(\bm P_{\rm best-fit} - \widehat{\bm P}^{\rm rec}\right)^{\sf T}\,{\sf\Sigma}_{\rm inj}^{-1}\,\left(\bm P_{\rm best-fit} - \widehat{\bm P}^{\rm rec}\right)\;,
\end{equation}
where ${\sf\Sigma}_{\rm inj}$ is a diagonal covariance matrix with entries $\sigma_i = (\sigma^+_i + \sigma^-_i)/2$, where $\sigma^\pm_i$ are the asymmetric $68\%$ C.L.\ uncertainties on $\smash{\bm P_{\rm best-fit}(k_i)}$ introduced above, propagated from the uncertainty on $\Ohi\bhi$. As a cross-check, we repeated the test using the asymmetric prescription $\sigma_i = \sigma^+_i$ if $\smash{\widehat{\bm P}^{\rm rec}_i > \bm P_{{\rm best\text{-}fit},i}}$ and $\sigma_i = \sigma^-_i$ otherwise. For $s_{\rm B}\times s_{\rm D}$ the symmetric case yields $\smash{\chi^2_{\rm inj} = 9.74}$ ($p=0.37$) and for $s_{\rm A}\times s_{\rm C}$ we find $\smash{\chi^2_{\rm inj} = 8.25}$ ($p = 0.51$). When using the asymmetric prescription, we get $\smash{\chi^2_{\rm inj} = 4.10}$ ($p=0.90$) and $\smash{\chi^2_{\rm inj} = 5.24}$ ($p=0.81$) for $s_{\rm B}\times s_{\rm D}$ and $s_{\rm A}\times s_{\rm C}$ respectively, with the lowering of $\smash{\chi^2_{\rm inj}}$ reflecting the effect of the posterior asymmetry on the per-bin uncertainties.

Including the damping term in the model captures and compensates for the signal-loss effect measured by the signal injection test.

%%%%%%%%%%%%%%%%%%%%%%%%%%%%%%%%%%%%%%%%%%%%%%%%%%
%%%%%%%%%%%%%%%%%%%%%%%%%%%%%%%%%%%%%%%%%%%%%%%%%%

\subsection{Scale-independence test: robustness of the fits} \label{sec:scale_indep}

\begin{figure}
    \centering
\includegraphics[width=\columnwidth]{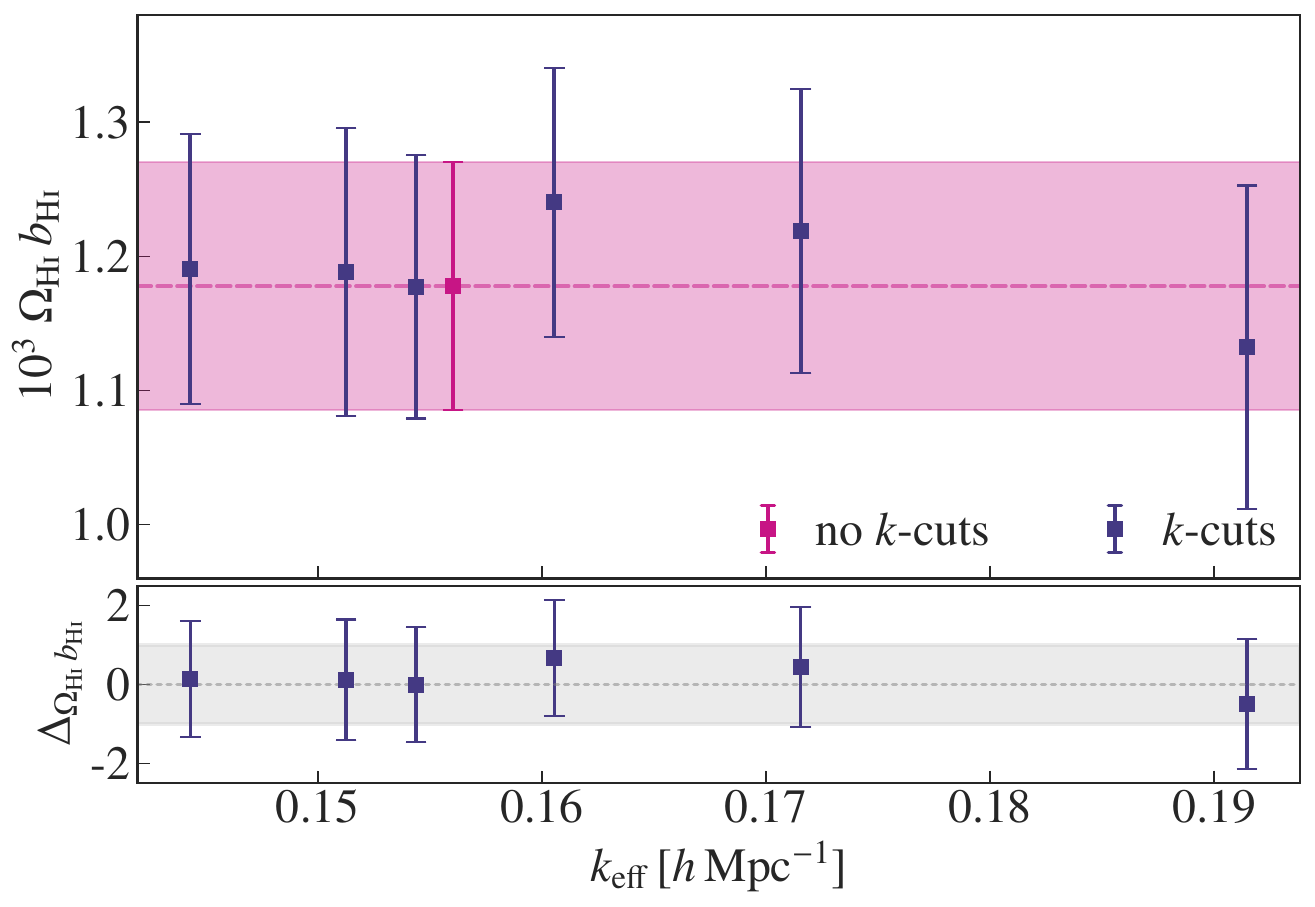}
    \caption{
    \textit{Top panel:} Best-fit values of $\Ohi\bhi$ resulting from the least-square error fits with scale cut. The purple square displays the benchmark value with all the $k$-bins available; all the results were obtaining using the ${\bm P}_{\rm cb}$ data vector with the corresponding full jackknife covariance and setting $\beta_{ij}=0$. 
    \textit{Bottom panel:} Values of $\Delta_{\Ohi\bhi}$ corresponding to the different choices of bin cuts; the $\pm1\,\sigma$ region is depicted with the shaded grey region. The error bars on $\Delta_{\Ohi\bhi}$ are obtained by propagating the uncertainties of $\Ohi\bhi^{\;{\rm no}\,k-{\rm cuts}}$ and $\Ohi\bhi^{\;k-{\rm cuts}}$.
    }
    \label{fig:scale_cuts}
\end{figure}

Our measurements are prone to various systematic errors, contributing differently at different scales. For example, signal loss affects mainly the low-$k$ region \citep[see, e.g.,][]{Cunnington2023}. To assess the stability across scale of our results ($\mathcal{H}_{10}$), we check whether performing the analysis with the large- or small-scale-only spectra information would lead to different amplitude fits. For this test, we opt for the conservative choice to perform least-squares fits with no free signal-loss damping kernel (i.e., we set all parameters $\beta=0$). The latter is to avoid the kernel absorbing the eventual shifts in amplitude that we want to highlight, if present.

We perform six new fits for each $i$--$j$ subset pair, removing one to three $k$-bins at large or small scales. We define the effective scale of each $k$-limited analysis as 
\begin{equation}
    k_{\rm eff}\coloneqq\sqrt{\frac{\left({\bm K}\odot{\bm P}_{\rm cb}\right)^{\sf T}\,\mathsf{\Sigma}^{-1}({\bm P}_{\rm cb},{\bm P}_{\rm cb})\,\left({\bm K}\odot{\bm P}_{\rm cb}\right)}{{\bm P}_{\rm cb}^{\sf T} \, \mathsf{\Sigma}^{-1}({\bm P}_{\rm cb},{\bm P}_{\rm cb}) \,{\bm{P}_{\rm cb}}}} \;,
\end{equation}
where $\odot$ stands for the Hadamard (element-wise) product and ${\bm K}=(\bm k^{\sf T}\cup\ldots\cup\bm k^{\sf T})^{\sf T}$, i.e., it is the concatenation (column stacking) of all the $k$-bins of each of the cross-spectra in ${\bm P}_{\rm cb}$. % (namely, the length of $\bm K$ is 4 $k$-bins for ${\bm P}_{\rm cb}$). 
In practice, when removing the large (small) scale information, $k_{\rm eff}$ increases (decreases). We check if there is any running of the best-fit value of $\Ohi\bhi$ with respect to $k_{\rm eff}$. We express the departure from the reference value (no scale cuts, fixing $\beta_{ij}=0$) $\Ohi\bhi^{\;{\rm no}\,k-{\rm cuts}}$ in units of its $1\,\sigma$ uncertainty with the quantity $\Delta_{\Ohi\bhi}$,
\begin{equation} \label{eq:DeltaAmpl}
    \Delta_{\Ohi\bhi}\coloneqq\frac{\Ohi\bhi^{\;k-{\rm cuts}} - \Ohi\bhi^{\;{\rm no}\,k-{\rm cuts}}}{\sigma_{\Ohi\bhi}^{{\rm no}\,k-{\rm cuts}}} \;.
\end{equation}

We show the results in \cref{fig:scale_cuts}. The best-fit values of $\Ohi\bhi$ exhibit remarkable stability against the variation of the scale range used for the fit; also, although not shown for brevity, the corresponding $\chi_{\rm red}^2$ remain optimal despite the $k$-bin cuts. 

Since we have neglected signal loss for this test, we can claim that its effect (which we expect to be $k$-dependent) is minor and under control compared to the uncertainties that characterise our measurements. The outcome of this test further supports the Bayesian analysis in \cref{sec:MCMCfit} that has shown that, when letting the nuisance $\beta_{ij}$ parameters vary, the resulting amplitude agrees with the $\beta_{ij}=0$ derived amplitudes, although with an upward-inflated error bar. 

Following the same approach of \cref{sec:splitting_test}, we evaluated the consistency of the results from these fits using the \texttt{tensiometer} package, computing the shift probabilities $P_{\rm shift}$ and corresponding effective significance $\sigma_{\rm eff}$. To perform this test, the best-fit values of the amplitude parameter and associated uncertainties from the LSE fits were used to construct Gaussian posteriors for $\Ohi\bhi$. All cases considered (with and without $k$-bin cuts) were compared against each other and against the reference. When comparing against the main analysis result (free $\beta_{ij}$ and full $k$-range), the shift probability remains in the range $0.01$--$0.09$ (corresponding to $0.02$--$0.17\,\sigma_{\rm eff}$). Comparing results of the LSE fits with and without $k$-bin cuts yields higher shift probabilities, namely between $0.07$ and $0.36$, corresponding to $0.03$--$0.47\,\sigma_{\rm eff}$; the latter is driven by the smaller error bars.

%%%%%%%%%%%%%%%%%%%%%%%%%%%%%%%%%%%%%%%%%%%%%%%%%%
%%%%%%%%%%%%%%%%%%%%%%%%%%%%%%%%%%%%%%%%%%%%%%%%%%

\subsection{Stability of the $\Nfg$ choice in the component separation process} \label{sec:Nfg_test}

\begin{figure}
    \centering    \includegraphics[width=1\columnwidth]{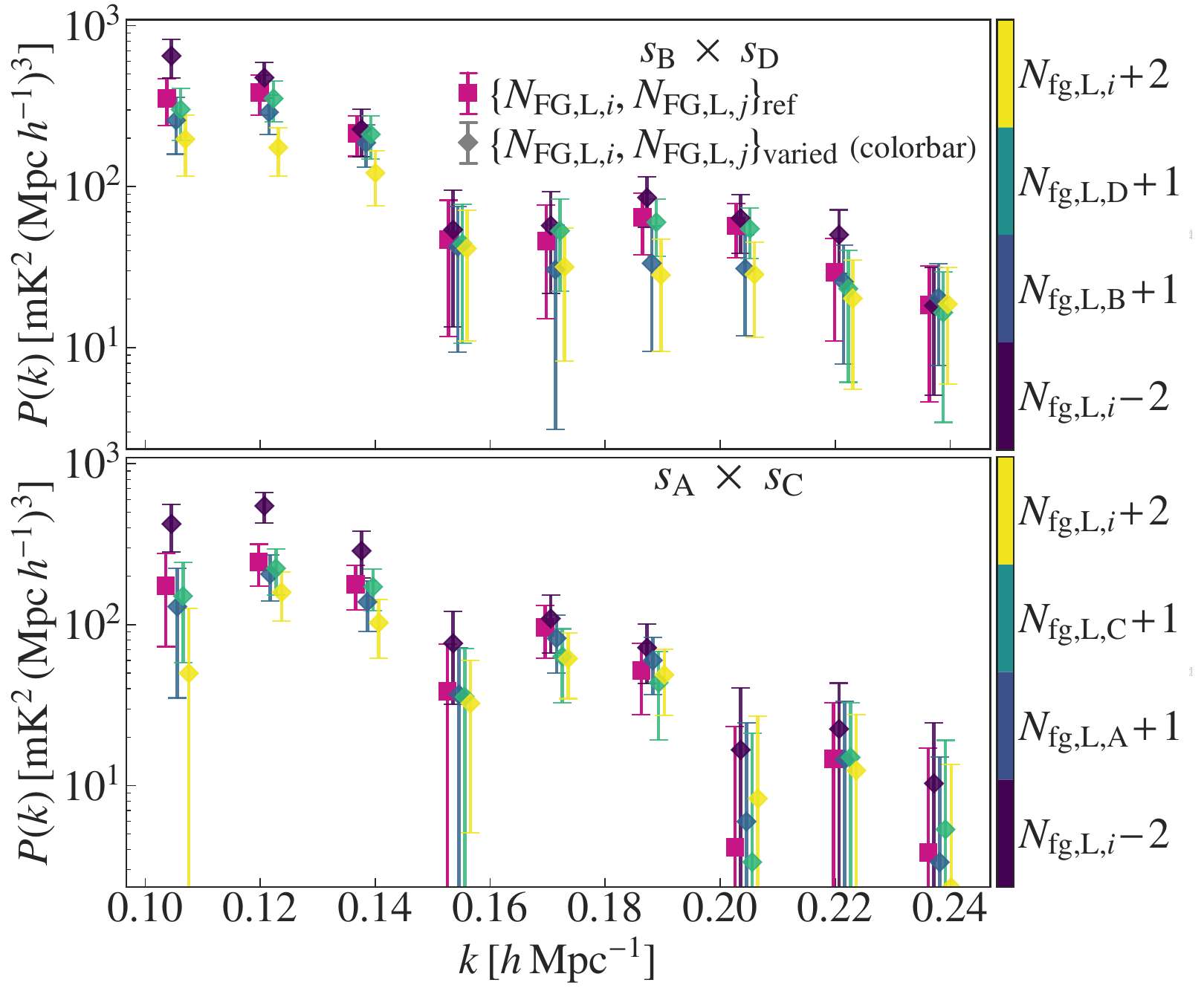}
    \caption{Power spectra of the two super-cross (\textit{top panel}: $P_{\rm BD}$; \textit{bottom panel}: $P_{\rm AC}$) for the reference cleaning level (magenta points with error bars), compared with the other cases tested, identified by different colours in the colour bar. The two cases with the largest difference in cleaning (two fewer or two additional modes removed) are shown in the darkest and lightest shades of the colour bar, respectively, while cases in which one of the involved subsets has been cleaned with one additional mode at large scales are shown in intermediate colours.}
    \label{fig:Nfg_variations_pk}
\end{figure}

Although the measured spectra can suffer from signal loss, this can be compensated for by low-$k$-enhancing systematics. The latter situation seems unlikely given that, e.g., the signal injection tests have shown that we do not expect a radically different signal loss than that we eventually infer and that the radial low-pass filtering analysis has not revealed any striking systematics in the maps acting at the scales of interest.

Here, we further test that, in the fixed-cosmology approach and by `playing' with the component separation freedom given by the choice of the $\Nfg$ parameters, we are not incurring into the situation described above (competing systematic effects that cancel each other, leading to an incorrect spectrum amplitude). We thus assess the independence of our inferred power spectrum amplitude with respect to the $\Nfg$ choice ($\mathcal{H}_{11}$).

We first proceed by varying, one at a time, the $\Nfgl$ values of the four subsets, increasing the value by one or two units, while keeping the small-scale cleaning fixed, as reported in \cref{table:cleaning}, since \textit{(i)} the level of cleaning at large scales has the strongest impact on the spectrum amplitude, and \textit{(ii)} for cleaning setups more different than these the spectrum fits deteriorate, becoming less meaningful. The rationale behind this test is that we expect that in the presence of a more significant signal loss (i.e., for higher values of $\Nfgl$), the $\mathcal{D}(k)$ kernel should be able to compensate for it (pointing to higher $\beta_{ij}$ parameters), while they would be less effective at correcting for possible additive contributions from residual systematics. Overall, we want to address whether the inferred $\Ohi\bhi$ stays in agreement among these scenarios, while the $\beta_{ij}$ might vary. 

\begin{figure}
    \centering    \includegraphics[width=1\columnwidth]{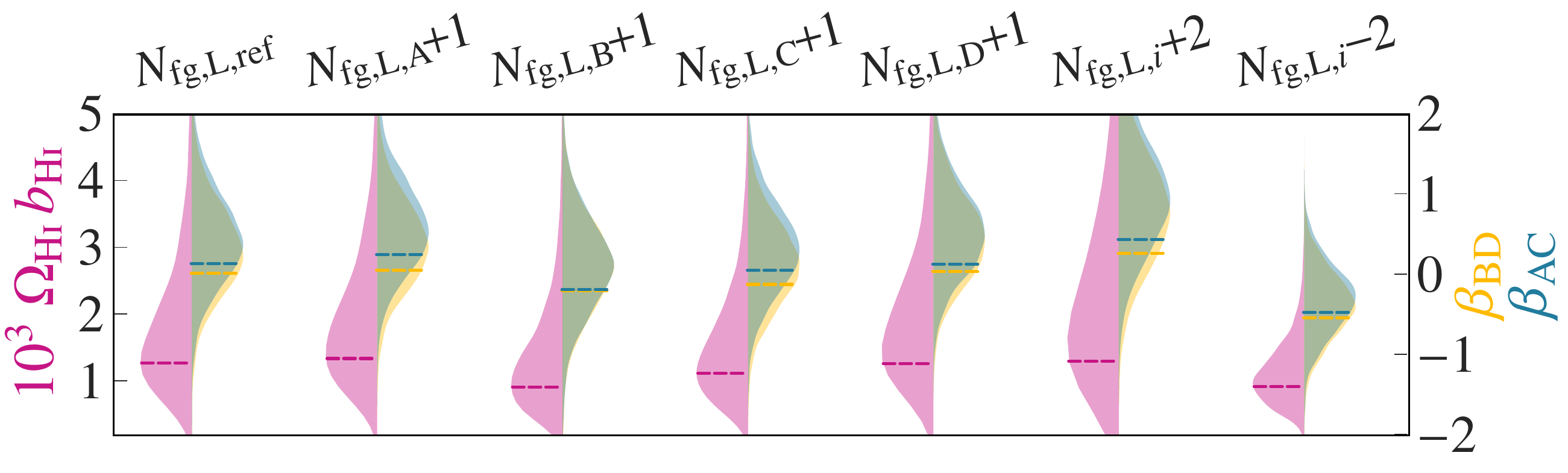}
    \caption{One-dimensional posterior distribution of the free parameters of the MCMC fits. The left side of each violin plot (pink area) refers to the amplitude $\Ohi\bhi$, while on the right side the posterior distributions of the signal-loss parameters $\beta_{\rm BD}$ and $\beta_{\rm AC}$ are shown in yellow and blue respectively. Each element of the plot corresponds to a different cleaning level, as indicated in the title, that reports which subset has an varied $\Nfgl$ value (or all subsets in the case of the generic subscript $i$), and by how many units it is varied relative to the reference value depicted in the first column.}
    \label{fig:Nfg_variations}
\end{figure}

We show examples of these scenarios in \cref{fig:Nfg_variations_pk}, where we compare the power spectra measured with different cleanings of the subsets with respect to the reference case (highlighted with magenta squares). As expected, we observe that the largest discrepancies, though well within $1\,\sigma$ except for a couple of bins at the largest scales, occur when the cleaning is reduced; conversely, removing more modes results to have a milder effect. This suggests that the former case likely suffers from residual spurious contributions that are subsequently removed when $\Nfgl$ increases, whereas once the optimal cleaning level is achieved, the subtraction of further components does not significantly affect the spectra. The latter is a particularly positive finding, as it implies that signal loss does not play a dominant role in our analysis.

Being more quantitative, we run the inference analysis for each of these scenarios. Results are in \cref{fig:Nfg_variations}, where each element refers to a different MCMC fit, showing the one-dimensional posterior distribution of the cross-spectrum amplitude $\Ohi\bhi$ on the left and on the two involved $\beta_{ij}$ parameters on the right (often overlapping yellow and blue regions). The first element on the left refers to the primary analysis choice of fixing $\{N_{\rm fg,L,A},N_{\rm fg,L,B},N_{\rm fg,L,C},N_{\rm fg,L,D}\} = \{11,11,10,8\}$; in the other cases, we let one of those cleaning parameters vary. Inferred $\Ohi\bhi$ %and $\beta_{ij}$ 
values are robust and coherent among different (yet reasonable) cleaning setups. Indeed, when comparing the posterior distributions of $\Ohi\bhi$ obtained at different cleaning levels against the reference result, i.e., the magenta areas of \cref{fig:Nfg_variations}, all are found to be consistent to the reference within $0.11\,\sigma_{\rm eff}$, with the sole exception of the moderate-cleaning case $N_{\rm fg,L}-2$, which reaches $0.34\,\sigma_{\rm eff}$.

In particular, a significantly more aggressive cleaning (the $N_{\rm fg,L}+2$ scenario) leads to a broadening of the posterior distribution, as a consequence of the reduced SNR when going towards a (moderately) over-cleaned regime; at the same time there is a slight indication of more significant signal loss, as evidenced by the $\beta_{ij}$ parameters shifting toward larger values. In the opposite scenario, where cleaning is milder ($N_{\rm fg,L}-2$), we observe a tightening of the constraints, likely driven by the increased amplitude of the spectra due to foreground residuals; this interpretation is supported by the $\beta_{ij}$ parameters reaching negative values (although still compatible with 0), which suggests the presence of potential additive contaminations. 

%%%%%%%%%%%%%%%%%%%%%%%%%%%%%%%%%%%%%%%%%%%%%%%%%%
%%%%%%%%%%%%%%%%%%%%%%%%%%%%%%%%%%%%%%%%%%%%%%%%%%

\subsection{Cross-correlation with a galaxy catalogue} \label{sec:xGAMA}

\begin{figure}
    \centering
    \includegraphics[width=\columnwidth]{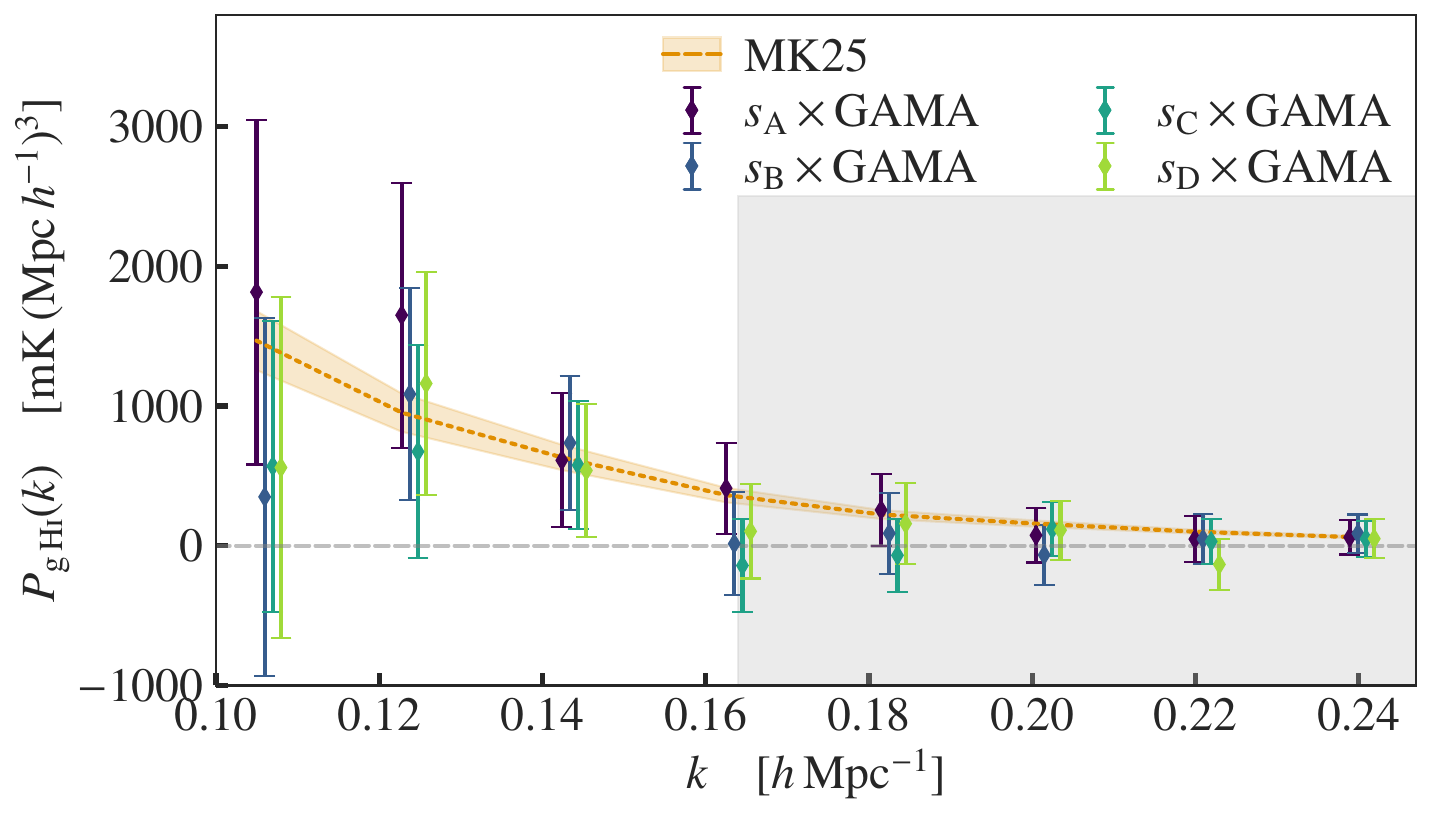} 
    \caption{\hi-galaxy power spectra estimated cross-correlating each subset with the GAMA galaxy catalogue (points with error bars). The $68\%$ confidence level error bars are computed under the assumption of Gaussian variance.
    The theoretical line, plotted as reference, is obtained assuming the best-fit value of the power spectrum amplitude derived in \citetalias{MKxGAMA} applied to the theoretical power spectrum template of the present work, while the shaded area represents the $1\,\sigma$ uncertainty. Note that the measurements have not been corrected for signal loss, unlike the analysis in \citetalias{MKxGAMA}, from which the theoretical prediction is derived. The grey area marks the region where the beam damping suppresses at least $50\%$ of the \hi\ signal and, due to the lower signal power, oscillations and negative points are more likely to occur.
    }
    \label{fig:crossGama}
\end{figure}

The analysis in \citetalias{MKxGAMA} reported a successful detection of the \hi\ cosmological signal from the cross-correlation of the data set under scrutiny in cross-correlation with the spectroscopic GAMA galaxies. 
We repeat this cross-correlation analysis with each subset to demonstrate that, individually, each subset contains cosmological signal ($\mathcal{H}_{12}$).

As in \citetalias{MKxGAMA}, we can use only a fraction ($\approx25\%$) of the observed sky area, due to the limited overlap region with the GAMA survey footprint. Concerning the processing of the galaxy catalogue and the estimator, we refer to \citetalias{MKxGAMA} for details. In practice, besides the subsets' cubes being more noise-dominated than the whole data set, the estimation can straightforwardly follow that in \citetalias{MKxGAMA}.

The measured \hi-galaxy power spectra are displayed in \cref{fig:crossGama}, where we also plot, for comparison, the best-fit resulting from the analysis in \citetalias{MKxGAMA} and its $68\%$ confidence level. For the latter, since in that work maps were resmoothed to a lower resolution, we use the reported best-fit $\Ohi\bhi$ value and its $1\,\sigma$ uncertainty and scale our theoretical model accordingly; concerning the other parameters that enter the model, we fix the \hi-galaxy correlation coefficient $r=0.9$, the galaxy bias $b_{\rm g}=1.9$ and the pairwise velocity dispersion of both galaxies and \hi\ $\sigma_{\rm v}=400\,{\rm km/s}$, still following \citetalias{MKxGAMA}. We do not perform any parameter estimation as the measurements suffer from a very low signal-to-noise ratio due to the reduced sky coverage, and given that each subset carries $\approx1/4$ of the signal present in the complete data set. 

Considering that additional spurious contaminants, if present, should not correlate with the galaxy positions, we appreciate the consistency among the four different galaxy-\hi\ cross-power spectra, as it hints to comparable cosmological information carried by the different subsets, and also the good agreement with the results of the previous analysis from \citetalias{MKxGAMA}. The \citetalias{MKxGAMA} estimation (dashed line in \cref{fig:crossGama}) reflects a corrected (boosted) power spectrum via the transfer-function approach, while the power spectrum measurements reported for the subsets do not account for signal loss. However, the fair agreement among the different analyses suggests that the low, zero-consistent values that we infer for the $\beta_{ij}$ parameters provide a proper representation of the low level of signal loss.

Additionally, we use this cross-correlation with the galaxy catalogue to reiterate the optimality of the chosen $N_{\rm fg}$ cleaning parameters. We report that overcleaning (using $N_{\rm fg,L}+2$) results in cosmological signal loss; the first three bins of the spectra negatively shift in amplitude by about half of their $1\sigma$ uncertainty. Conversely, when adopting $N_{\rm fg,L}-2$, the amplitudes of the resulting spectra remain roughly the same as expected because contaminants’ residuals do not correlate with galaxy positions; however, error bars increase by about 1.5 due to noisier measurements.

\begin{table*}
\centering
 \caption{Summary of tests performed to reinforce the reliability of our findings.
}
 \label{table:tests}
 % \begin{tabularx}{ccc}
 \renewcommand{\tabularxcolumn}[1]{p{#1}}
 \begin{tabularx}{1\textwidth}{
  >{\raggedright\arraybackslash}X 
  >{\raggedright\arraybackslash}X 
  >{\raggedright\arraybackslash}X
  }
 \toprule
 Hypothesis & Test & Outcome \\ 
  \midrule
$\bm {\mathcal{H}_{1}}$: The adopted data splitting strategy identify subsets in which systematics do not behave consistently. Within each strategy, (1) results do not depend on the specific random shuffling used; and (2) the super-cross decorrelates residual systematics. & Re-analyses performed varying the splitting strategy to intentionally maximise or minimise the coherence of systematics in the Level-6 calibrated subsets; comparison of the inferred value and posterior distribution of $\Ohi\bhi$. \cref{sec:splitting_test}, \cref{fig:deltaOb_post}. & 
When different splitting strategies are adopted, the super-cross $P(k)$ yield consistent inferred $\Ohi\bhi$ values, with shift probabilities $P_{\rm shift}$ ranging from $0.02$ to $0.19$. The constraining power is nonetheless affected by non-optimal splitting choices. \\
    &  &  \\
$\mathcal{H}_{2}$: The detections of the internal cross-correlations are robust.  & Null tests, either shuffling channels or pixels of one of the two subsets.  \cref{sec:null}, \cref{fig:nulltest}. & 
Spectra with a shuffled subset are fully consistent with a null spectrum; the data reject consistency with these null realisations ($p$-value $< 5 \times 10^{-3}$). \\
    &  &  \\
$\mathcal{H}_{3}$: The clustering signal measured by the internal cross-correlations is sky-driven, hence subset-independent. & Patchwork tests. We cross-correlate each subset with data cubes where each voxel is randomly selected from one of the other subsets, to preserve angular and frequency order while breaking eventual subset-dependent systematics. \cref{sec:frankenstein}, \cref{fig:franktest}. & The cross-correlations with the patchwork subsets are in agreement with the two super-cross (consistency with $p$-value$>0.4$ except one case of $p$-value$\approx0.13$).\\
      &  &  \\ 
$\mathcal{H}_{4}$: The cleaning has succeeded in decomposing each subset in two spatially uncorrelated parts.
& Measuring the cross-correlation coefficient between the signal and contaminants parts of the mPCA separation, $r_{{\rm LK}}$. \cref{sec:rleakage_lk}, \cref{fig:rleak}. & $r_{{\rm LK}}$ oscillates around zero with a scatter of $\approx10\%$ and no appreciable $k$-dependence. \\
      &  &  \\ 
$\mathcal{H}_{5}$: Residual astrophysical foregrounds in each subset are negligible. & Measuring the cross-correlation coefficient between our signal and uncleaned data from non-cosmological frequencies where foregrounds live as well. \cref{sec:rleakage_fg}, \cref{fig:rnoncosmo}. & $r_{{\rm FG}}$ oscillates around zero with a scatter of $\approx10\%$ and no appreciable $k$-dependence.\\
      &  &  \\ 
$\mathcal{H}_{6}$: Residual astrophysical foregrounds are negligible. The internal cross-correlation signal is compatible with a cosmological one.
& Radial $\widehat P(k_\nu)$ tests. The cosmological signal is expected to have negligible coherence along frequency, as opposed to the astrophysical foregrounds. 
\cref{sec:Pknu}, \cref{fig:radial_Pk}. & There is a clustering signal at the largest radial scale; then it flattens to no correlation. This agrees with the expected behaviour of the \hi\ cosmological signal.\\
      &  &  \\ 
$\mathcal{H}_{7}$: Residual astrophysical foregrounds in each subset are negligible. & `front\texttimes back' tests. Foregrounds are highly correlated in frequency: their low-frequency part correlates with the high-frequency part (see \cref{fig:Pknu_sims}). The cosmological signal, however, should be uncorrelated between these two partitions. \cref{sec:fXb}, \cref{fig:radial_Pk_frbk}. &  The cross-correlation coefficient $r_{\rm front\times back}$ oscillates around zero with $\lesssim 7\% \, (4\%)$ scatter when considering each subset's `front' and `back' partitions (or combined with their super-cross associate). The oscillation amplitude matches that of independent thermal-noise realisations. \\
      &  &  \\ 
$\mathcal{H}_{8}$: The clustering signal in the cross-correlations is visually compatible with a cosmological one (at voxel level). & Low-pass filtering in the parallel direction to highlight where most of the information is held (see the $\widehat P(k_\nu)$ test above). Visually inspecting the corresponding maps. \cref{sec:lowpass}, \cref{fig:lowpass}. & No unexpected `structures' are visible in the maps. At voxel level, what drives the bulk of the clustering behaves as cosmological signal (and uncorrelated noise). \\
     &  &  \\ 
$\mathcal{H}_{9}$: The $\mathcal{D}(k)$ kernel is a good description of the signal loss due to the component separation. &  
Injecting \hi\ signal mocks in the pristine, dirty subsets and computing their loss in clustering due to the cleaning. \cref{sec:injection}, \cref{fig:signal_inj}. & The estimated loss through simulations is in agreement with the inferred one ($p$-value$>0.37$). \\
      &  &  \\ 
$\mathcal{H}_{10}$: Cleaning-induced signal loss is not destructive, in agreement with the estimated small $\beta_{ij}$ parameters. There are no strong systematic effects afflicting a particular spatial scale of the measurements. The estimated spectrum amplitudes are not driven by a specific spatial scale of each spectrum measurement. & Scale-independence test. Ignoring the $\mathcal{D}(k)$ damping kernel by setting $\beta_{ij}=0$, we check whether the amplitudes of the internal cross-correlations change when using only large- or small-scale information. E.g., strong signal loss should introduce artificial $k$-dependence in the spectra. \cref{sec:scale_indep}, \cref{fig:scale_cuts}. &  
The LSE fitted amplitude (fixing $\beta_{ij}=0$) of each internal cross-correlation is insensitive to whether we use only the small- or large-scale information of the measurement, with $P_{\rm shift}=0.01-0.09$ relative to the main analysis (with free $\beta_{ij}$) and $P_{\rm shift}=0.07-0.37$ relative to the LSE fit without $k$-cuts and $\beta_{ij}=0$.
\\
      &  &  \\ 
$\mathcal{H}_{11}$: With fixed cosmology, our estimate of $\Nfg$ is robust; the choice of $\Nfg$ does not affect the inferred power spectrum amplitude. & Testing whether slightly changing $\Nfgl$ changes the final, inferred amplitude parameters. \cref{sec:Nfg_test}, \cref{fig:Nfg_variations}.
& For reasonable cleaning settings ($\Nfg'=\Nfg\pm2$), the inferred amplitude stays within the $68\%$ C.L. threshold, maintaining competitive $\chi^2_{\rm red}$. \\
      &  &  \\ 
$\mathcal{H}_{12}$: Each subset carries cosmological signal, which is comparable among subsets. & Measuring the cross-correlation signal of each subset with the overlapping GAMA galaxy catalogue. \cref{sec:xGAMA}, \cref{fig:crossGama}. & Subset\texttimes galaxy cross-correlations are in agreement among themselves and with a previous measurement from an independent analysis, even neglecting the signal-loss effect. \\
  \bottomrule
 \end{tabularx}
\end{table*}

%%%%%%%%%%%%%%%%%%%%%%%%%%%%%%%%%%%%%%%%%%%%%%%%%%
%%%%%%%%%%%%%%%%%%%%%%%%%%%%%%%%%%%%%%%%%%%%%%%%%%

\subsection{Additional tests}

We assessed our measurements using additional tests beyond those shown in this manuscript. In particular, we highlight a couple of them that, in the end, proved not informative due to the high uncertainties that still characterise the measurements, but that might be deemed valid for subsequent analyses of this kind.

We tried to quantify the `excess' contaminants present in the subsets and whether the internal cross-correlations indeed suppress their contribution to the clustering signal. By assuming negligible multiplicative systematics (or re-absorbable in the additive systematics term), we can express each cleaned subset as $s_i = \hi + e_i +n_i$, with \hi\ the cosmological signal, $e_i$ the residual contaminants (the `excess'), and $n_i$ the uncorrelated noise. We build `subtracted' or `summed' subsets and compute their spectra, $P_{\Delta_{ij}} \coloneqq \cP(s_i-s_j, s_i-s_j)$, $P_{\Sigma{ij}} \coloneqq \cP(s_i+s_j, s_i+s_j)$. It follows that:
\begin{align}
    P_{\Delta_{ij}} &= P_{e_{ii}}+P_{e_{jj}}-2P_{e_{ij}}+P_{n_{ii}}+P_{n_{jj}} \\
    P_{\Sigma_{ij}} & = P_{e_{ii}}+P_{e_{jj}}+2P_{e_{ij}}+P_{n_{ii}}+P_{n_{jj}}+2P_{\hi}\,. 
\end{align}
We have verified that the assumptions implied in the way we expressed the subsets reasonably hold by, for instance, asserting that $\smash{P_{\Delta_{ij}} < P_{ii} + P_{jj} < P_{\Sigma{ij}}}$ for all our \chess\ subsets' spectra measurements. 
Since we have estimates of the uncorrelated noise carried in each subset and of the cosmological signal (through the best-fit model of the reference analysis), we are able to isolate the excess contaminants' terms of interest, $\smash{P_{e_{ij}}}$. For instance, we can derive the expression:
\begin{equation}
    \frac{P_{e_{ij}}}{P_{\hi}} = \frac{P_{ii} + P_{jj}-P_{\Delta_{ij}} - P_{ij}}{P_{\hi}} - 1 \,.
\end{equation}
If $P_{ii} + P_{jj} -P_{\Delta_{ij}} - P_{ij} \to P_{\hi} $, then it follows that $P_{e_{ij}}\to 0$. We have verified this; however, the corresponding uncertainties are such that this test becomes inconclusive in our case.

\hfill \break
We summarise in \cref{table:tests} the validation tests performed in this section. To each hypothesis we list the tests we performed and how their outcome backs our detection.

%%%%%%%%%%%%%%%%%%%%%%%%%%%%%%%%%%%%%%%%%%%%%%%%%%
%================== DISCUSSION ==================%
%%%%%%%%%%%%%%%%%%%%%%%%%%%%%%%%%%%%%%%%%%%%%%%%%%

\section{Discussion and perspectives}
\label{sec:discussion}

\begin{figure}
        \centering        \includegraphics[width=\columnwidth]{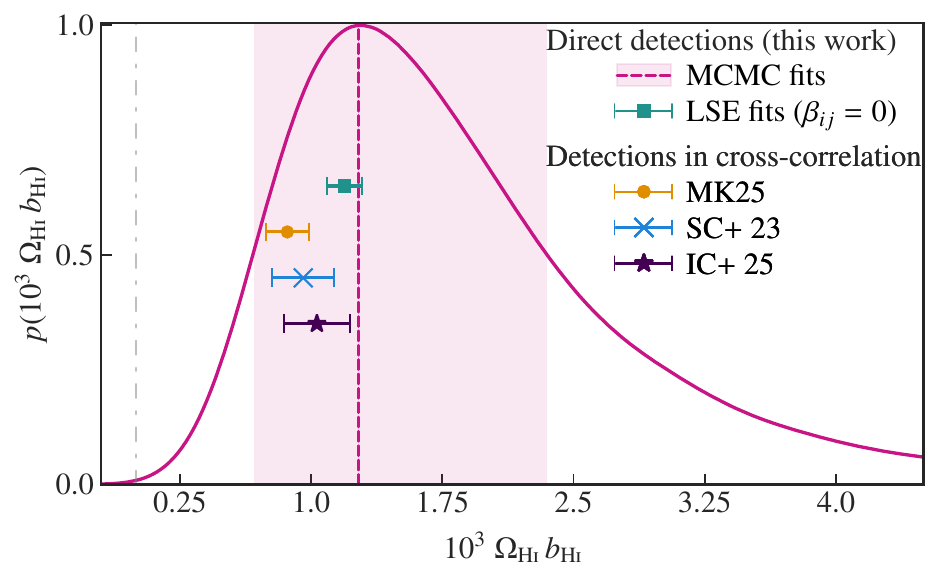} \caption{Normalised marginalised posterior distributions of $\Ohi\bhi$, obtained from the multi-subset analyses performed with the super-cross data vector ${\bm P}_{\rm cb}$. The MAP value and 68\% confidence level region are represented with the purple solid line and the shaded region. We show with the teal square the best-fit result obtained with the simple least-square-estimation (LSE) fits with fixed $\beta_{ij}=0$. For comparison, we include the results (all LSE fits) from the previous MeerKLASS analyses in cross-correlation with galaxy catalogues (in those cases the inferred parameter is $\Ohi\,\bhi\,r$, with $r$ the cross-correlation coefficient, which we assume to be $r=0.9$). We note that, while \citetalias{MKxGAMA} used the same data as the analysis presented in this work (albeit restricted to the region overlapping with the GAMA galaxies), SC+ 23 \citep{MKxWZ} and IC+ 25 \citep{Carucci2024} are based on the data from the previous MeerKLASS 2019 L-band pilot survey. The vertical dash-dotted line corresponds to the null amplitude. }
    \label{fig:ohibhi_posterior}
\end{figure}

In the previous sections, we presented constraints on the amplitude of the signal power spectrum at cosmological scales using the measurements of spectra from cleaned maps and their multi-subset analysis, without an external tracer for mitigating systematics.

In the following, we discuss our results, highlighting strengths and weaknesses of our analysis.

%%%%%%%%%%%%%%%%%%%%%%%%%%%%%%%%%%%%%%%%%%%%%%%%%%
%%%%%%%%%%%%%%%%%%%%%%%%%%%%%%%%%%%%%%%%%%%%%%%%%%

\subsection{Detection significance and constraints} \label{sec:constraints}

We compute the measurements signal-to-noise ratio as
\begin{equation} \label{eq:SNRmeas}
    {\rm SNR}_{\rm meas}\coloneqq{\sqrt{{\bm P}_{\rm data}^{\sf T}\,\mathsf{\Sigma}^{-1}\,{\bm P}_{\rm data}}} \; ,
\end{equation}
where ${\bm P}_{\rm data}$ is the data vector and $\mathsf{\Sigma}$ is the covariance matrix. For the super-cross data vector ${\bm P}_{\rm cb}$ defined in \cref{eq:Pcb}, analysed with the jackknife covariance and varying $\beta_{ij}$ (i.e., our reference analysis), we find ${\rm SNR}_{\rm meas}=7.3$. In order to measure the level of coherence between the data and the model, we also compute the model-dependent signal-to noise ratio of the fit:

\begin{equation} \label{eq:SNRfit}
    {\rm SNR}_{\rm fit}\left(\thv_{\rm bf}\right)\coloneqq\frac{{\bm P}_{\rm data}\,\mathsf{\Sigma}^{-1}\,{\bm P}_{\rm th}(\thv_{\rm bf})}{\sqrt{{\bm P}_{\rm th}(\thv_{\rm bf})\,\mathsf{\Sigma}^{-1}\,{\bm P}_{\rm th}(\thv_{\rm bf})}} \; ,
\end{equation}
with $\thv_{\rm bf}$ the array of the MAP best-fit values of the free parameters at which we evaluate the theoretical model ${\bm P}_{\rm th}$. In this case, we obtain ${\rm SNR}_{\rm fit}=6.2$. By construction ${\rm SNR}_{\rm fit} \leq {\rm SNR}_{\rm meas}$, with equality only when the data are perfectly described by the model and in the absence of scatter due to the covariance. Hence, ${\rm SNR}_{\rm fit}=6.2$ highlights a strong consistency between the measurements and the model we employ. 

To quantify the consistency between ${\rm SNR}_{\rm fit}$ and ${\rm SNR}_{\rm meas}$, we performed the following test. We generated 500 different realisations of the \hi\ auto-spectrum using the theoretical model ${\bm P}_{\rm th}$ of \cref{eq:Pmodel_DB} (without thermal noise as expected for the internal cross-correlations), the best-fit parameters from \cref{table:fits}, and the complete, true data covariance as in \cref{eq:multi_xchess}. For each realisation, we computed ${\rm SNR}_{\rm fit}$ and ${\rm SNR}_{\rm meas}$. This yielded median values of $6.1$ for the former and $7.4$ for the latter, demonstrating that the measured difference between the two signal-to-noise estimates is consistent with Gaussian statistical fluctuations, and confirming that our estimate of ${\rm SNR}_{\rm meas}$ is not artificially boosted by the presence of additive systematics in the measurements.

We have already stressed the preference for the more conservative super-cross ${\bm P}_{\rm cb}$ data vector. However, to show the depth and constraining power of the survey, we quote that the ${\rm SNR}_{\rm meas}$ corresponding to the fitting of the data vector built with all the six cross-$P(k)$ of the \chess\ division is ${\rm SNR}_{\rm meas}=12.1$, although characterised by a worse fit quality (e.g., in terms of $\smash{\chi^2_{\rm red}}$), indeed we measure ${\rm SNR}_{\rm fit}=9.7$, that is $\approx20\%$ smaller than ${\rm SNR}_{\rm meas}$. We provide below more details about our choice not to use all six possible cross-correlations from the \chess\ division but instead rely on the two fully independent ones. 

One possible way our fitting procedure can accommodate extra additive or multiplicative systematics with scale dependencies more complex than the assumed functional form of $\mathcal{D}(k)$ is to point to the highest possible cosmological signal model subject to a strong $\mathcal{D}(k)$ damping. We see this effect when we work with the complete set of six cross-spectra from the \chess\ division. Specifically, \textit{i)} the MAP value of $\Ohi\bhi$ shows a $\approx45\%$ positive shift; \textit{ii)} the corresponding values of the $\beta_{ij}$ parameters, all reaching $\approx0.4$, point to a more significant signal loss; and \textit{iii)} the parameter error bars are boosted by $40\%$, even though the longer length of the data vector should, in principle, enhance its constraining power. The latter result indicates that adding the information from the remaining four cross-spectra of the \chess\ division does not compensate for the increased risk of incorporating additional (and not modelled) systematics alongside the signal. 

At the beginning of the analysis, we motivated our choice of the super-cross to minimise both time- and dish-dependent systematics. A posteriori, we can endorse that choice based on the results reported in the previous paragraph.

In \cref{fig:ohibhi_posterior}, we plot with a purple curve the posterior distribution of the $\Ohi\bhi$ amplitude, with the corresponding vertical line highlighting the best-fit, MAP value compared to the LSE-fit case where we do not permit the phenomenological signal-loss kernel to be free (i.e., imposing $\beta_{ij} = 0$) with a teal error bar (values reported also in \cref{table:fits}). The multivariate fit $68\%$ confidence region (shaded area) is broad and the $\beta_{ij} = 0$ LSE-fit result falls in it. Consistent with the fact that signal loss is not accounted for, the LSE fit is slightly shifted towards lower values with respect to the main result. When accounting for the effect of signal loss, we find that opening up the parameter space, including the signal-loss damping kernel and allowing the slope parameters to vary, leads to a significant loosening of the constraining power on the power spectrum amplitude.

Moreover, in \cref{fig:ohibhi_posterior} we show: \textit{i)} the result obtained from a different analysis of a subset of same data set in cross-correlation with a spectroscopic galaxy survey (orange filled circle and error bar, \citetalias{MKxGAMA}); and \textit{ii)} the results obtained from two different analyses that dealt with a different MeerKLASS data set (a different sky patch and shallower survey). The latter is at the same redshift, again in cross-correlation with an ancillary galaxy data, light blue X and bar for `SC+ 23' \citep{MKxWZ} and dark blue star and bar for `IC+ 25' \citep{Carucci2024}. The overall consistency among all these differently derived constraints, with the residual difference remaining within the current uncertainties, is reassuring. 

The differences between the estimates of $\Ohi\bhi$ from the various cross-correlation analyses are discussed in \citet{Carucci2024} together with their possible sources, which include some of the preprocessing steps performed on the intensity maps, the different foreground removal techniques and strength of the cleaning, and the unaccounted impact of mode-mixing in the computation of the TF. The case of MK25 is not addressed specifically in \citet{Carucci2024}, but the same considerations hold; additionally, in this case, the way covariance was estimated through the TF computation, especially at large scales, may play a role, as discussed in \citet{Zhaoting2025QE}. Similar methodological differences are also relevant for the comparison between the results from cross-correlation analyses and the one presented here (which shares many of the choices adopted in \citet{Carucci2024}. Beyond this, comparing the best-fit values of $\Ohi\bhi$ from \hi-galaxy cross-correlations or from the \hi\ auto power spectrum requires assuming a fiducial value of the correlation coefficient $r$, since the former constrain the combination $\Ohi\bhi r$ rather than $\Ohi\bhi$ alone. While the choice $r=0.9$ is consistent with previous literature \citep[see, e.g.,][\citetalias{MKxGAMA}]{Chang2010,2011MNRAS.415.2580K,2016MNRAS.458.3399W,Wolz2022,MKxWZ}, the value is not tightly constrained, and this allows only for an approximate comparison between the exact best-fit values, even though the overall agreement remains meaningful.

We note that SC+23 and MK25 dealt with signal loss via the TF technique rather than as an additional free parameter in the fitting procedure; this in turn enabled simpler fits via the LSE technique, in which the single free parameter $\Ohi\bhi r$ is thus constrained to tighter uncertainties. IC+25 achieves the same constraining power as it adopts the same LSE fitting strategy, although it did not apply any correction for signal loss\footnote{The IC+25 result is intentionally biased low as the goal was to show that mPCA minimises signal loss compared to the other works that used a simpler PCA.}. Our current results from the LSE fits with no damping kernel hence no signal-loss correction yield tighter error bars on $\Ohi\bhi$; indeed, measuring a signal exclusively from \hi\ data increases the constraining power on this parameter combination since $\Ohi\bhi$ enters quadratically in our model, as opposed to linearly, which is the case in a \hi-galaxy cross-spectrum analysis.

The posterior distribution of the slope parameters peaks in the positive region, with the $\beta_{ij}$ still consistent with zero at $68\%$ confidence level.
Looking back at the subsets auto-spectra (bottom left panel of \cref{fig:allspectra}), we expect systematic effects with a positive contribution to the clustering power budget. If some of those were still present in the cross-spectra, they might have compensated for the signal loss to some extent, since our $\beta_{ij}$ parameters are consistent with zero.

However, the battery of tests performed in \cref{sec:robustness} and summarised in \cref{table:tests} provide support for our conclusion that residual contaminants are minor in the cross-correlation among subsets, and our choice not to model them. In the last test, we corroborated this by measuring the cross-correlation signal of each subset with an independent external data set (see \cref{sec:xGAMA}). That exercise is a demonstration of the cosmological origin of the signal we have directly measured through the internal, \hi-only cross-correlations. 

%%%%%%%%%%%%%%%%%%%%%%%%%%%%%%%%%%%%%%%%%%%%%%%%%%
%%%%%%%%%%%%%%%%%%%%%%%%%%%%%%%%%%%%%%%%%%%%%%%%%%

\subsection{Contaminant separation}
\label{sec:cleaning_discussion}

In this work, we address contaminant separation using wavelet-based mPCA, which has successfully been established using previous MeerKLASS data \citep{Carucci2024}. By combining the decomposition into spatial scales with the standard PCA, mPCA complements the latter by leveraging the contaminant characterisation in the angular direction, in addition to their frequency-directional expected behaviour. In brief, after cleaning the data cubes using the mPCA approach, the correlation between frequency channels is lower than when the standard PCA is applied. Moreover, when the same number of components $\Nfg$ is used, mPCA results in cleaned data cubes with a higher signal-to-noise ratio (hence, less signal loss) while achieving less overall variance (linked to lower residual systematics).

The significant drawbacks that still characterise this contaminant separation approach are the choice of the $\Nfg$ parameters and the control of separation leakages, i.e., residual contaminants and signal loss. By working within a fixed cosmology, here we could overlook the first problem by defining the $\Nfg$ parameters using the LSE procedure in \cref{sec:tablecloth}; we further tested in \cref{sec:Nfg_test} that this choice did not bias our final results. In perspective, we foresee incorporating the cleaning setting into a larger inference framework when we begin constraining cosmology as well. 

For the latter challenge---controlling leakages---and how it affected the claimed results, we articulate below our lessons learnt and perspectives.

%%%%%%%%%%%%%%%%%%%%%%%%%%%%%%%%%%%%%%%%%%%%%%%%%%
%%%%%%%%%%%%%%%%%%%%%%%%%%%%%%%%%%%%%%%%%%%%%%%%%%

\subsection{Minimising residual foregrounds and systematics}

\hi\ auto-spectrum measurements from the internal cross-correlations, not relying on cross-correlation with external data sets, are sensitive to residual contaminants common in the cleaned subsets, instrumental ones such as correlated calibration errors, and residual astrophysical foregrounds. Indeed, if those residual contaminants are shared across all subsets, their cross-correlations capture this spurious clustering signal.

In our analysis, we addressed this challenge in three ways. \textit{(i)} We performed contaminant separation independently for each subset. Since we expect different systematics acting in the subsets, the mode-mixing at play in each of them makes the separation itself different; for example, the $\Nfg$ parameters we use differ among $s_i$, see \cref{table:cleaning}, but also the filtering mixing matrices differ from one subset to the other. With the latter, we also foresee the residual foregrounds (the common sky) to be different from subset to subset.
To verify the latter holds, \textit{(ii)} we performed dedicated validation tests (especially the ones focusing on the radial direction, where foregrounds leave distinct features, \cref{sec:rleakage} and \cref{sec:pknu_residual}) and eventually set limits in the $k$-range of the analysis to exclude the identified dubious regions. Finally, \textit{(iii)} we demonstrated that the clustering signal present in the subsets is, first of all, predominantly sky-driven performing the patchwork tests (where subset-specific systematics are broken in all directions, and yet the overall clustering signal with the other subsets is unchanged, \cref{sec:frankenstein}), and, secondly, likely of cosmic \hi\ origin since we measure---individually per-subset---a comparable cross-correlation signal with an ancillary external data set (\cref{sec:xGAMA}), the latter free by construction from cleaning-induced signal loss, Milky Way emissions' residuals and leftover contamination from the MeerKAT observations and our data reduction.

For future analyses, leveraging larger data sets, a refinement of our understanding of systematics and noise sources will be a key priority. The continuous improvement of the calibration and map-making pipelines will play a crucial role in producing data cubes where instrumental effects and RFI leakages are increasingly under control. Accessing a wider frequency range will also enhance the effectiveness of component separation, thereby reducing leakages between foregrounds and the cosmological signal. The advances described above will also enable the characterisation of the subsets' auto-spectra, allowing their inclusion in the full multi-subset analysis. 

%%%%%%%%%%%%%%%%%%%%%%%%%%%%%%%%%%%%%%%%%%%%%%%%%%
%%%%%%%%%%%%%%%%%%%%%%%%%%%%%%%%%%%%%%%%%%%%%%%%%%

\subsection{Addressing signal loss} \label{sec:discussion_loss}

Currently, one of the biggest drawback of blind contaminant separation methods, such as mPCA, is the loss of cosmological signal. So far, the community has opted to reconstruct the power-spectrum a posteriori with signal injection techniques through a TF approach \citep{Switzer2015,Cunnington2023}. This method provides a quantitative analysis of signal loss in a data-driven way. Indeed, the TF can be used to correct for signal loss in the measured power spectrum, and it is numerically equivalent to a renormalisation of the power spectrum estimator. However, as shown in \citet{Zhaoting2025QE}, foreground subtraction also induces mode-mixing, and treating it as signal loss may result in potential biases in the estimation of the correlation between $k$-bins at large scales. The latter can be noticed in the observational analysis conducted in \citet{Carucci2024}. 

Instead, in this analysis, we opted to forward-model the expected loss with the kernel $\mathcal{D}(k)$ described in \cref{sec:signal_loss}, adding a nuisance parameter in the model. The functional shape of the kernel is motivated by previous observational works, the present one (see \cref{fig:signal_inj} and \cref{fig:TF_sims}) and simulations (again, see the findings in \cref{app:tf_valid}). 
Validation tests were also performed to test the optimality of the $\mathcal{D}(k)$ kernel in our setup through signal-injection tests (\cref{sec:injection}) and by checking the dependence of our fits on the particular $k$-scale information analysed (\cref{sec:scale_indep}). 

As shown in \cref{fig:ohibhi_posterior}, the value we infer for $\Ohi\bhi$ when adopting this parameterisation is consistent with the results from previous cross-correlation analyses, whether or not the TF correction was in place in those analyses. A detailed comparison is provided in \cref{sec:constraints}. The kernel allows for a conservative estimation of the parameter posteriors, which gives a robust and still significant detection of the \hi\ auto spectrum amplitude. Although the additional nuisance parameters reduce the constraining power on the amplitude, as discussed in \cref{sec:constraints}, this approach represents the most reliable and conservative approach for our analysis setup. Moreover, a systematic and detailed study of the TF technique applied to our specific case of internal cross-correlations is still lacking (\cref{app:tf_valid} represents a first step in this direction), and further shortcomings need to be addressed \citep{Zhaoting2025QE}.

One could refine the modelling of signal loss, introducing a more complicated form for $\mathcal{D}(k)$. For example, a reasonable extension would be separating the radial and the perpendicular directions to capture the known different impact that the foreground cleaning pipeline induces at the level of the power spectrum. At the moment, further parameters in $\mathcal{D}(k)$ would remain unconstrained due to the limited data volume and the strong degeneracies in the prior parameter space, hence we opted for the simplest yet reasonable model (see again the test in \cref{sec:injection}). In the future, a more robust TF method or a simulation-informed prior on a more detailed $\mathcal{D}(k)$ will be necessary to constrain cosmology with the largest scales and broadband of the \hi\ power spectrum \citep[see, e.g.,][]{2020MNRAS.499.4054C}.

%%%%%%%%%%%%%%%%%%%%%%%%%%%%%%%%%%%%%%%%%%%%%%%%%%
%%%%%%%%%%%%%%%%%%%%%%%%%%%%%%%%%%%%%%%%%%%%%%%%%%

\subsection{Limitations of the modelling framework}
\label{sec:discussion_modelling}

To model the cosmological \hi\ signal, we applied certain approximations that are justified in the current setup, but they will begin to reveal their limitations as we enter a regime of higher precision.

We assume a linear matter power spectrum, linear bias, and linear (matter-only) RSD (see \cref{eq:pmodel}). Indeed, since we measure only the power spectrum Legendre monopole, we model matter-only RSD for the linear term (rather than related to the \hi\ field) to break the degeneracy between $f\sigma_8$ and $\Ohi\bhi$. Improved RSD models, supported by a better understanding of their impact on the \hi\ power spectrum \citep[see, e.g.,][]{2025arXiv250301956O}, will be essential for future analyses. Non-linearities in matter clustering and bias will be addressed in future work, for example, through perturbative expansions \citep[see, e.g.,][]{2022JCAP...02..026M, 2023PhRvD.108h3528O}. We also neglect contributions such as shot noise and \hi\ stochasticity, which could become significant in the future due to improved measurement sensitivity \citep{2024PhRvD.110f3555F}.

For example, extending to non-linear RSD contributions would require modelling Finger-of-God (FoG) effects \citep{1972MNRAS.156P...1J,2018MNRAS.476...96S}, which are described as the suppression of power at small scales along the line of sight. Its strength is set by the \hi\ velocity dispersion $\sigma_{\rm v}$ \citep{2001MNRAS.325.1288S}, a parameter with no precise observational constraints and strongly degenerate with the signal amplitude \citepalias{MKxGAMA}. Our data lack the sensitivity to meaningfully constrain these additional degrees of freedom. However, non-linearities in clustering and RSD are most prominent at small scales. By suppressing small-scale power, beam smearing lowers the SNR of these modes, making the contribution of such non-linear effects subdominant to the statistical errors in the inferred power spectrum.

Ultimately, this work aims to robustly detect the \hi\ signal and characterise it in terms of its amplitude, set by \hi\ abundance and linear bias. The current constraining power motivates the choice of the simplified modelling framework described above. Our model therefore adopts the combination $\Ohi\bhi$ and the damping kernel $\smash{\mathcal{D}(k)\propto k^{\beta}}$ as free parameters, allowing for a conservative treatment of signal loss and providing flexibility to accommodate potential incompleteness in the modelling. In the following we demonstrate that the impact of the simplifying assumptions is negligible both compared to the error bars on the measurements and to the final posterior uncertainty on $\Ohi\bhi$.

\begin{figure}
    \centering
    \includegraphics[width=\columnwidth]{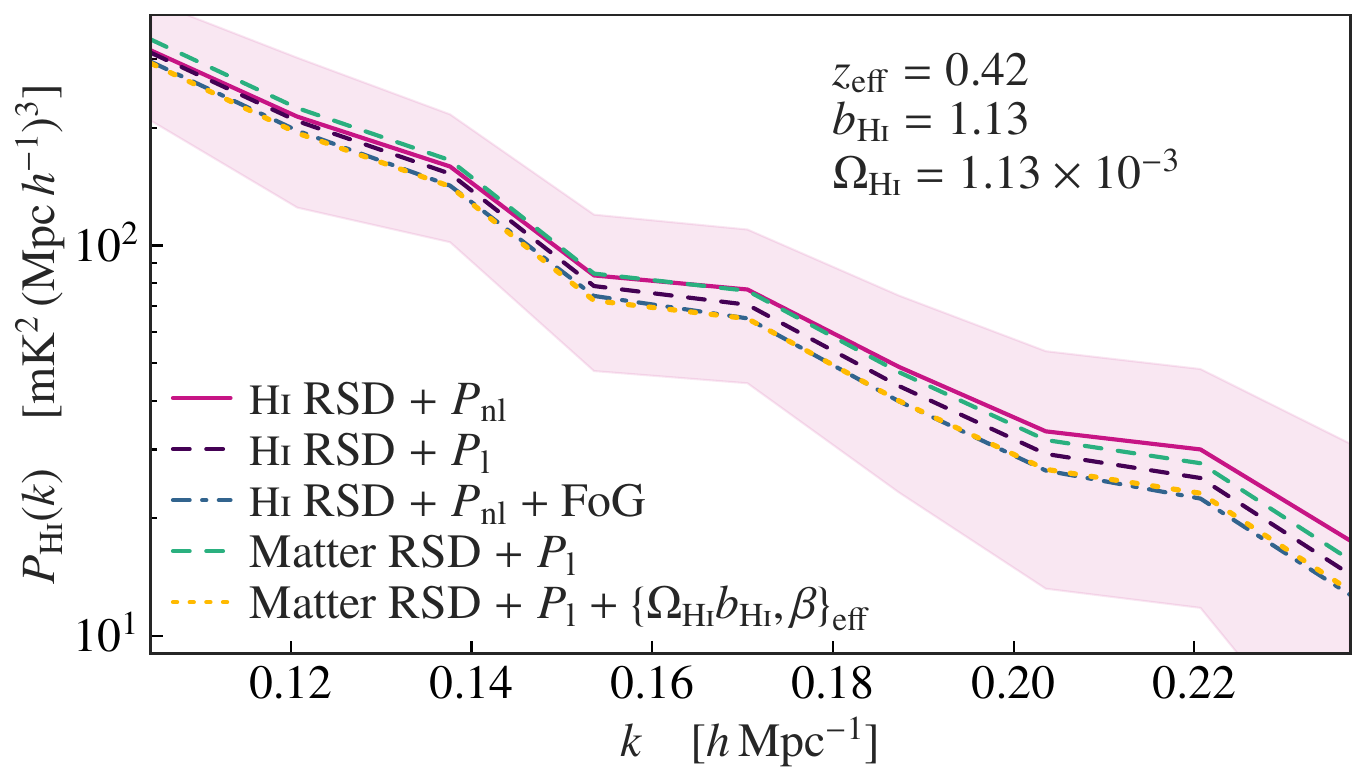}
    \caption{Theoretical \hi\ power spectra computed under different prescriptions. Specifically we compare cases where RSD are applied to the \hi\ field versus matter-only RSD, cases adopting either the linear ($P_{\rm l}$) or the non-linear ($P_{\rm nl}$) matter power spectrum, the inclusion of FoG effects and an illustrative case showing how the damping kernel $\mathcal{D(K)}$ enables an effective treatment of various clustering/RSD effect that we simplified in the main analysis. All predictions, except for the latter, are computed using the parameters displayed in the top-right corner of the figure. The linear bias was computed using \cref{eq:bhiFid} and we set $\Ohi$ in order to match the best fit value that we obtain for $\Ohi\bhi$. For the prediction with FoG effects, we set $\sigma_{\rm v}=200\,{\rm km/s}$. The case with effective $\Ohi\bhi$ and $\beta$ is computed with a $10\%$ lower amplitude compared to the other cases (a variation well within our uncertainties) and $\beta=-0.05$; this configuration matches the prediction of the more complex modelling also including FoG at percent level. The shaded region represents the average standard deviation of the two super-cross $P(k)$ used in the analysis: all variations arising from modelling choices are negligible compared to the size of the current error bars.}
    \label{fig:model_comparison}
\end{figure}

To be more quantitative, in \cref{fig:model_comparison} we show that several modelling prescriptions—including \hi\ or matter-only RSD with non-linear matter power spectrum ($P_{\rm nl}$), linear power spectrum ($P_{\rm l}$), FoG damping, and matter-only RSD with effective parameters—yield comparable predictions within current measurement uncertainties (shaded area).

To further demonstrate that our modelling choices do not introduce a significant bias on the inferred $\Ohi\bhi$ value, we performed a consistency check by relaxing baseline assumptions and adopting a more complex treatment of the power spectrum. In particular, we account for non-linearities in the matter power spectrum (i.e., in this case $P_{\rm m}$ in \cref{eq:pmodel0} is the non-linear matter power spectrum) and apply the RSD modelling directly to the \hi\ field. To lift resulting degeneracies, we fit for an overall amplitude parameter defined as \citep{2023ApJ...947...16A}
\begin{equation} \label{eq:AHI_ampl}
    \Ahi=\Thi^2\,\left(\bhi+f\,\langle\mu^2\rangle\right)^2 \; ,
\end{equation}
where $\langle\mu^2\rangle$ is the angle-average of $\mu^2$ on the grid and \(f\) is the value of the growth rate at the effective redshift \(z_{\rm eff}=0.42\). By adopting the same $\Ohi$--$\Thi$ relation used for the benchmark analysis (see \cref{eq:A_HI-def}), and assuming the $\bhi(z)$ relation in \cref{eq:bhiFid},
we can derive constraints on $\Ohi$ from the best-fit value of $\Ahi$. We find this estimate compatible with our main results, detailed in \cref{sec:omegahi_constraints}, within $0.15\sigma$.

In summary, given current measurement limitations, the impact of the contributions mentioned above would likely be largely unconstrained, leading to strong degeneracies with the nuisance parameters and our target amplitude, $\Ohi\bhi$. Therefore, we choose to neglect them to obtain a constraint on $\Ohi\bhi$ that can be compared with independent measurements, albeit with the necessary caveats. 

%%%%%%%%%%%%%%%%%%%%%%%%%%%%%%%%%%%%%%%%%%%%%%%%%%
%%%%%%%%%%%%%%%%%%%%%%%%%%%%%%%%%%%%%%%%%%%%%%%%%%

\subsection{Dealing with the MeerKAT beam}
\label{sec:beam_discus}

In the model of the \hi\ power spectrum, we include the damping in the direction transverse to the line of sight using a Gaussian profile evaluated at the median frequency of the observations (\cref{sec:beam}, reaching a $70\%$ of damping at $k \gtrsim 0.15 \,\kum$), in analogy with previous MeerKLASS analyses \citep[\citetalias{MKxGAMA};][]{MKxWZ,Carucci2024}. We therefore neglect the frequency dependence, side lobes, and the presence of slight departures from perfect symmetry in the pattern, which have, however, been observed in the measurements of MeerKAT's primary beam in the L-band \citep{6410347, 2021MNRAS.502.2970A, 2021MNRAS.506.5075M,2022AJ....163..135D}. 

In forthcoming analyses, it will be interesting to include a more accurate modelling of the beam damping in the power spectrum. 
For instance, it will be relevant to study the potential impact of the beam on the radial modes as well. Such effects may arise, for example, from beam chromaticity, as a frequency-dependent beam size induces coupling between transverse and radial modes: the resulting mode mixing may also affect the shape of the power spectrum along the $k_\|$ direction. In addition, when adopting the plane-parallel approximation in the power spectrum estimation, a beam which is purely transverse with respect to the true local line of sight is projected onto a single global line-of-sight direction, so that the impact of the beam also leaks into the radial modes.

An effort is therefore underway to move beyond the plane-parallel approximation both in the estimator and in the corresponding theoretical modelling of the \hi\ anisotropic power spectrum. Alternatively, work is ongoing to deal with beam effects at the map-making level, thus easing the modelling efforts. We will keep exploring both approaches.

Besides its contribution at the power spectrum level, the beam convolution can affect, e.g., the contaminant separation process \citep{2021MNRAS.506.5075M,2022MNRAS.509.2048S}.
Recently, \citet{Zhaoting2025} highlighted how the beam ripple induced by the interference between the primary and the secondary reflector of MeerKAT antennas gives rise to a systematic effect in the radial direction, at the scale corresponding to the characteristic frequency of this feature ($\approx18\,\fum$), that standard PCA is not able to isolate.
We do not see this effect in our analysis (see \cref{sec:Pknu}); as we expect standing waves to be slight dish-dependent and elevation-dependent (hence block- and again dish-dependent),\footnote{To scan the same sky patch, different dishes have to enforce different elevations during the observation.} we speculate that this might be due to the internal cross-correlations, after independent mPCA component separation. Indeed, even for milder (small $\Nfg$ values) cleaning, there is no evident ripple-induced peak in the radial, one-dimensional spectra at the expected $k_\nu \approx 0.05 {\rm MHz}^{-1}$ (see \cref{fig:radial_auto} for the auto-spectra, corresponding cross-spectra are even flatter).
Nonetheless, we tested whether discarding the beam-ripple $k$-region identified in \citet{Zhaoting2025} would lead to different outcomes. We repeated the measurements and fits using only 8 $k$-bins and applying different anisotropic cuts: $k_{\rm min}=0.113\,\kum$, $k_{\|{\rm min}}=0.1\,\kum$, $k_{\perp{\rm min}}=0.045\,\kum$. The results are consistent but with larger uncertainties (due to a higher degeneracy among the $\beta$ parameters and the spectrum amplitude) and an almost halved ${\rm SNR}_{\rm meas}$. This more conservative $k$-range analysis merely limited the amount of available information; hence, we preferred the analysis with the larger scale range. 

Unlike previous analysis by MeerKLASS, GBT and Parkes \citep[\citetalias{MKxGAMA};][]{Masui2013,Wolz2017,Anderson2018,Wolz2022,MKxWZ}, we do not apply a beam re-convolution to achieve a common angular (i.e., frequency independent) resolution of the maps. The idea behind beam re-convolution is that data variance and small-scale, unwanted artefacts (for example caused by polarization leakage) can be reduced by convolving all maps to the lowest-frequency map `gaussianised' beam (as in \cref{eq:Bbeam}). However, more recent literature is consistently finding beam reconvolution detrimental in the \hi\ intensity mapping observational context \citep{2021MNRAS.506.5075M,2022MNRAS.509.2048S,Carucci2024,Zhaoting2025}. The shared conclusion is that it modifies the data cube in a frequency-dependent way, hence increasing the mode-mixing between the foreground and signal components of the data linear decomposition assumed by, e.g., PCA, thus entangling such contaminant separation processes.

%%%%%%%%%%%%%%%%%%%%%%%%%%%%%%%%%%%%%%%%%%%%%%%%%%
%%%%%%%%%%%%%%%%%%%%%%%%%%%%%%%%%%%%%%%%%%%%%%%%%%

\subsection{Characterising the abundance of cosmic \hi} \label{sec:omegahi_constraints}

\begin{figure}
    \centering    \includegraphics[width=\columnwidth]{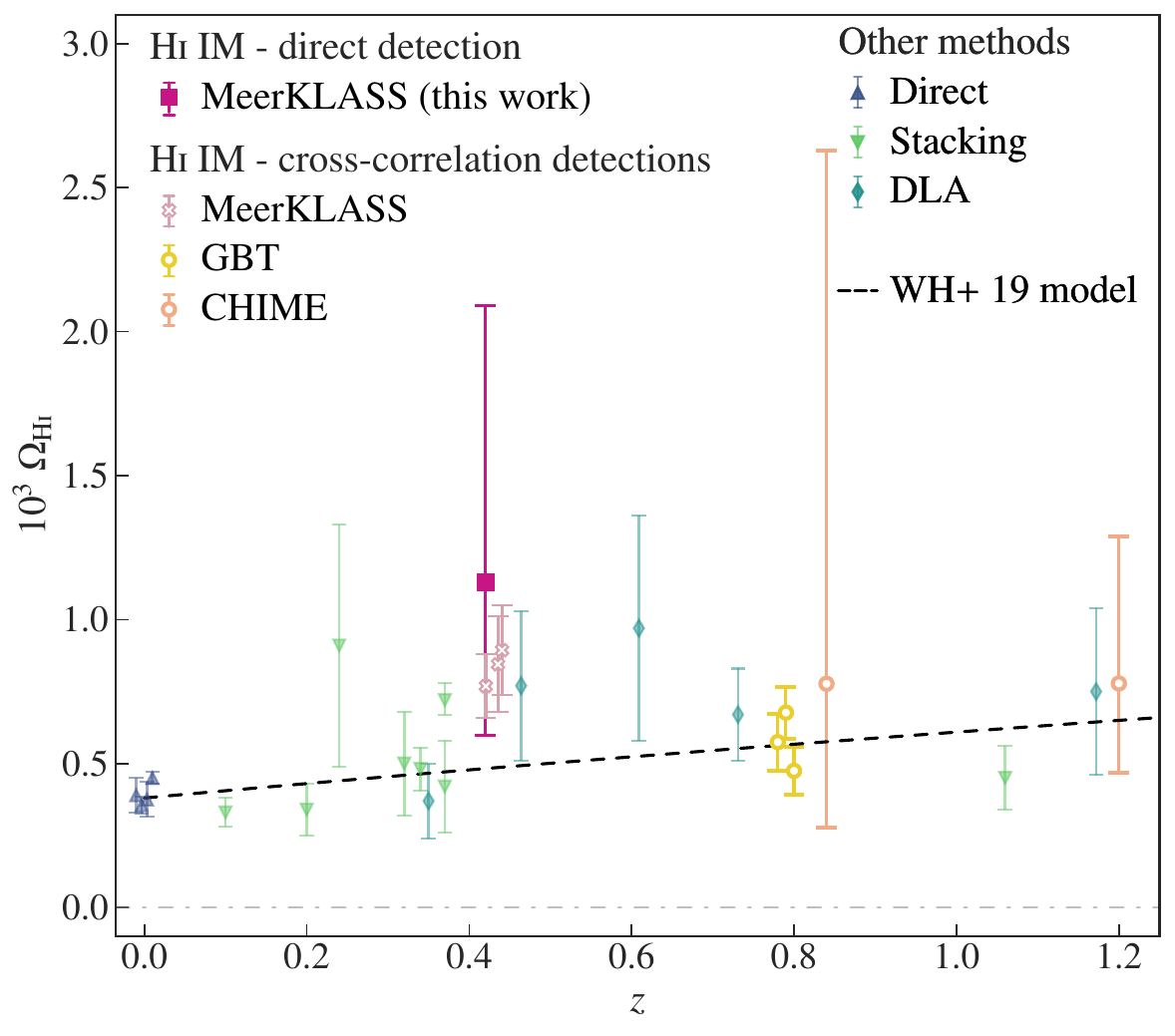}
    \caption{
    The cosmic abundance of \hi\ estimated with different observational techniques and at different redshifts. The result of this analysis is reported with the purple square. Other estimates of $\Ohi$ coming from intensity mapping measurements in cross correlation with external data sets are displayed with empty pink crosses for MeerKLASS and empty circles (yellow for GBT and orange for CHIME). Direct measurements of $\Ohi$ from \hi\ galaxy survey are reported with blue triangles, stacking measurements with reversed green triangles and constraints from observations of DLAs with teal diamonds. Corresponding publications are cited in the main text. We also include a low-$z$ inferred (and extrapolated) model (black dashed line) fitted using the stacking observations in \citet{2019MNRAS.489.1619H}, labelled as `WH+ 19 model'. The dash-dotted line corresponds to the null abundance.}
    \label{fig:ohi_comparison}
\end{figure}

By making an assumption on the linear bias $\bhi$ (\cref{eq:bhiFid}) we can translate our \hi\ power spectrum amplitude constraint into an estimate of $\Ohi$ only, and compare it with other observations.
Our results are not only in good agreement with the outcome of the analysis of MeerKLASS data in cross-correlation with galaxy surveys \citep[\citetalias{MKxGAMA};][]{MKxWZ,Carucci2024} as already shown in \cref{fig:ohibhi_posterior}, but also with other observational constraints. In \cref{fig:ohi_comparison}, we compare our $\Ohi$ estimate with that obtained from other intensity mapping surveys, although always in cross-correlation with an ancillary data set \citep{Wolz2022,2023ApJ...947...16A}.\footnote{In the case of the CHIME measurements, we follow their prescription to go from the ${\cal A}_\hi$ parameter they fit for to $\Ohi$. For results from cross-correlations between \hi\ intensity maps and galaxies, we fix the cross-correlation coefficient that is reported in the respective publications.} We also show the comparison with estimates obtained with other observational methods; in particular, we report results from standard \hi\ galaxy surveys \citep{2018MNRAS.477....2J,2021MNRAS.501.4550X,2005MNRAS.359L..30Z,2025A&A...695A.241M}, which we refer to `Direct' data points in the plot, and, to reach results at higher redshift, from stacking analyses \citep{2013MNRAS.435.2693R,2007MNRAS.376.1357L,2016MNRAS.460.2675R,2018MNRAS.473.1879R,2019ApJ...882L...7B,2020Natur.586..369C,2025arXiv250611280S}. At even higher redshift, constraints on $\Ohi$ come from the analysis of damped Lyman-$\alpha$ absorption (DLA) systems observed in the spectra of quasars \citep{2006ApJ...636..610R,2013A&A...556A.141Z,2020MNRAS.498..883G}. Finally, we compare all data points with the $\Ohi(z)$ curve proposed in \citet{2019MNRAS.489.1619H}, which was fitted using a subset of low-$z$ data only.

%%%%%%%%%%%%%%%%%%%%%%%%%%%%%%%%%%%%%%%%%%%%%%%%%%
%================== CONCLUSION ==================%
%%%%%%%%%%%%%%%%%%%%%%%%%%%%%%%%%%%%%%%%%%%%%%%%%%

\section{Conclusions} \label{sec:conclusion}

In this study, we present a detection of the \hi\ auto-power spectrum derived from data of the MeerKLASS 2021 L-band deep survey (first presented in \citetalias{MKxGAMA}). The data set covered approximately $236\,\deg^2$ in the frequency band of $971.2\,\fum<\nu<1023.6\,\fum$. Observations were performed using the 64 antennas of the MeerKAT radio telescope operating in single-dish mode, with a total observation time of approximately 60 hours (reduced to about 40 hours after RFI flagging). 
To enhance the cosmological signal while minimising systematics, we divided the data set into independent subsets that were processed and analysed separately, allowing us to measure `internal' cross-spectra between the subsets. We inferred the amplitude of the \hi\ power spectrum---parameterised in terms of the \hi\ abundance and linear clustering---using a multi-subset formalism and a Bayesian inference framework. This framework included nuisance parameters to account for potential signal loss due to the blind contaminant subtraction method. The reliability of our measurements and the robustness of our analysis were further supported by a series of null and consistency tests, which we summarised in \cref{table:tests}.

Despite the presence of strong contaminants (astrophysical foregrounds coupled to systematic effects), we directly and independently detect the \hi\ cosmological signal present in the intensity maps, which we characterised through its auto-spectrum, measured with a 6.2 
%7.3 
significance in the scale range of $0.095\,\kum<k<0.245\,\kum$. This analysis allowed us to put constraints on the abundance and clustering of \hi\ at $z_{\rm eff}\approx0.42$, namely: $10^3\,\Ohi\bhi=1.271^{+1.080}_{-0.598}$. Crucial for this achievement were an improved calibration pipeline \citepalias{MKxGAMA}, an optimised component separation technique \citep{Carucci2024}, and the exploitation of internal cross-correlations to minimise systematics in the measured signal from different data subsets \citep{Masui2013,Wolz2017,Anderson2018,Wolz2022,2024A&A...691A.336S}.

This detection provides robust evidence that \hi\ intensity mapping on cosmological scales can be effectively carried out with MeerKAT, marking a significant advance toward surveying with the SKA Observatory (SKAO), for which MeerKAT serves as a precursor, and the realisation of its full potential for precision cosmology and extra-galactic astrophysical studies. 

\medskip

%%%%%%%%%%%%%%%%%%%%%%%%%%%%%%%%%%%%%%%%%%%%%%%%%%
%%%%%%%%%%%%%%%%%%%%%%%%%%%%%%%%%%%%%%%%%%%%%%%%%%

\paragraph*{Author contribution:} This work is a Tier 1 MeerKLASS publication that presents a significant scientific result from the MeerKLASS Collaboration.\footnote{\texttt{\url{https://meerklass.org/}}.} All authors are listed alphabetically. The collaboration conceived the idea of this project. All authors have contributed to the writing of MeerKAT observing proposals, participated in discussions on the analysis, and provided feedback on the results and on the manuscript.
MBS, supervised by SCa, and IPC were responsible for executing the analysis here presented and all three for writing the paper. JW led the calibration pipeline development. SCu led the development of the power spectrum estimation pipeline. JW, MS, and MI were responsible for processing the data and providing the calibrated maps for this analysis. 
MGS is the PI of the MeerKLASS Collaboration. LW is Deputy PI of the MeerKLASS Collaboration.

%%%%%%%%%%%%%%%%%%%%%%%%%%%%%%%%%%%%%%%%%%%%%%%%%%
%=============== ACKNOWLEDGEMENTS ===============%
%%%%%%%%%%%%%%%%%%%%%%%%%%%%%%%%%%%%%%%%%%%%%%%%%%

\section*{Acknowledgements}

The authors warmly thank all members of the MeerKLASS collaboration for their valuable insights, feedback, and continued analysis efforts. They also thank the anonymous referees, whose comments helped improve the quality of this work.
JW acknowledges support from the National SKA Program of China (No.\ 2025SKA0160100), and the National Natural Science Foundation of China (NSFC, Grant No.\ 12573112).
MBS would like to sincerely thank Luigi Guzzo and Emanuele Castorina for their constant guidance throughout this work and for their insightful advice. SCa joins MBS in warmly thanking Luigi Guzzo for his availability and steady support.
MBS also expresses gratitude to the Observatory of Trieste and IFPU for their hospitality during the development of this project (supported by grant PNRR-4-2-1.2 project No.\ SOE\textunderscore0000136, RadioGaGa). MBS and IPC appreciate the feedback received from the IFPU Large Scale Structure group.
MBS and SCa acknowledge support from the Italian Ministry of Foreign Affairs and International
Cooperation (MAECI), Grant No.\ ZA23GR03.
IPC is supported by the European Union within the Next Generation EU programme [PNRR-4-2-1.2 project No.\ SOE\textunderscore0000136, RadioGaGa]. 
JLB acknowledges funding from the project UC-LIME (PID2022-140670NA-I00), financed by MCIN/AEI/ 10.13039/501100011033/FEDER, UE.
This result is part of a project that has received funding from the European Research Council (ERC) under the European Union's Horizon 2020 research and innovation programme (Grant agreement No.\ 948764; PB).
SCa acknowledges support from the Italian Ministry of University and Research (MUR), PRIN 2022 `EXSKALIBUR – Euclid-Cross-SKA: Likelihood Inference Building for Universe's Research', Grant No.\ 20222BBYB9, CUP D53D2300252 0006, and from the European Union -- Next Generation EU. 
ZC's research is supported by a UK Research and Innovation Future Leaders Fellowship [grant MR/X005399/1].
SCu acknowledges support from the UKRI Stephen Hawking Fellowship (grant reference EP/U536751/1) and a UK Research and Innovation Future Leaders Fellowship grant [MR/V026437/1]. 
JF acknowledges support of Funda\c{c}\~{a}o para a Ci\^{e}ncia e a Tecnologia through the Scientific Employment Incentive program reference No.\ 2020.02633.CEECIND/CP1631/CT0002, the FCT exploratory project 10.54499/2023.15069.PEX, and the research grant
UID/04434/2025. 
WH acknowledges support from the South African Radio Astronomy Observatory and National Research Foundation (Grant No.\ 97948).
MGS acknowledges support from the South African Radio Astronomy Observatory and National Research Foundation (Grant No.\ 84156).
MS is supported by the French government through the France 2030 investment plan managed by the National Research Agency (ANR), as part of the Initiative of Excellence Université Côte d’Azur under reference number ANR- 15-IDEX-01.
LW is a UK Research and Innovation Future Leaders Fellow [grant MR/V026437/1]. 

The MeerKAT telescope is operated by the South African Radio Astronomy Observatory, which is a facility of the National Research Foundation, an agency of the Department of Science and Innovation. %We acknowledge the use of the \texttt{ilifu} cloud computing facility, through the Inter-University Institute for Data Intensive Astronomy (IDIA).
We acknowledge the use of the \texttt{ilifu} cloud computing facility,\footnote{\url{www.ilifu.ac.za}.} a partnership between the University of Cape Town, the University of the Western Cape, Stellenbosch University, Sol Plaatje University, and the Cape Peninsula University of Technology. The \texttt{ilifu} facility is supported by contributions from the Inter-University Institute for Data Intensive Astronomy (IDIA, a partnership between the University of Cape Town, the University of Pretoria, and the University of the Western Cape), the Computational Biology division at UCT and the Data Intensive Research Initiative of South Africa (DIRISA).

For the purpose of open access, the author has applied a Creative Commons Attribution (CC BY) licence to any Author Accepted Manuscript version arising from this submission. 

%%%%%%%%%%%%%%%%%%%%%%%%%%%%%%%%%%%%%%%%%%%%%%%%%%
%=============== DATA AVAILABILITY ==============%
%%%%%%%%%%%%%%%%%%%%%%%%%%%%%%%%%%%%%%%%%%%%%%%%%%

\section*{Data availability}

The MeerKLASS collaboration is committed to releasing processed data from its surveys regularly, see our \href{https://meerklass.org/HIdata.html}{Public Data Release webpage}.
The data used here was observed through the MeerKAT proposal ID SCI-20210212-MS-01. 
Access to the raw data used in the analysis is public (for access information please contact archive@ska.ac.za). 

%%%%%%%%%%%%%%%%%%%%%%%%%%%%%%%%%%%%%%%%%%%%%%%%%%
%================== REFERENCES ==================%
%%%%%%%%%%%%%%%%%%%%%%%%%%%%%%%%%%%%%%%%%%%%%%%%%%

\bibliographystyle{mnras}
\bibliography{bibliography} 

%%%%%%%%%%%%%%%%%%%%%%%%%%%%%%%%%%%%%%%%%%%%%%%%%%
%================== APPENDICES ==================%
%%%%%%%%%%%%%%%%%%%%%%%%%%%%%%%%%%%%%%%%%%%%%%%%%%

\appendix

%%%%%%%%%%%%%%%%%%%%%%%%%%%%%%%%%%%%%%%%%%%%%%%%%%
%================== APPENDIX A ==================%
%%%%%%%%%%%%%%%%%%%%%%%%%%%%%%%%%%%%%%%%%%%%%%%%%%

\section{Comparing noise estimates} \label{app:noise}

In addition to the ABBA method (\cref{sec:abba}), we can also rely on other indicators of the noise level. Comparing these with the results reported in the \cref{table:ABBAnoise} corroborates our estimates and characterises better our data. 

The variance of the residuals after the cleaning is the first alternative that we explore. In the case of noise dominated data sets, we expect that the residuals would hit the noise level for aggressive enough cleanings \citep{Wolz2017, Switzer2015}. Therefore, their variance should match the temperature root mean square or be slightly biased-low, given that noise is lost during the cleaning process, as is the case for the cosmological signal. It is possible to compute both the global variance, to assess its agreement with the values in \cref{table:ABBAnoise}, and the variance along the line-of-sight, which offers a closer look at its frequency dependence.
\begin{figure*}
    \centering
    \includegraphics[width=0.98\textwidth]{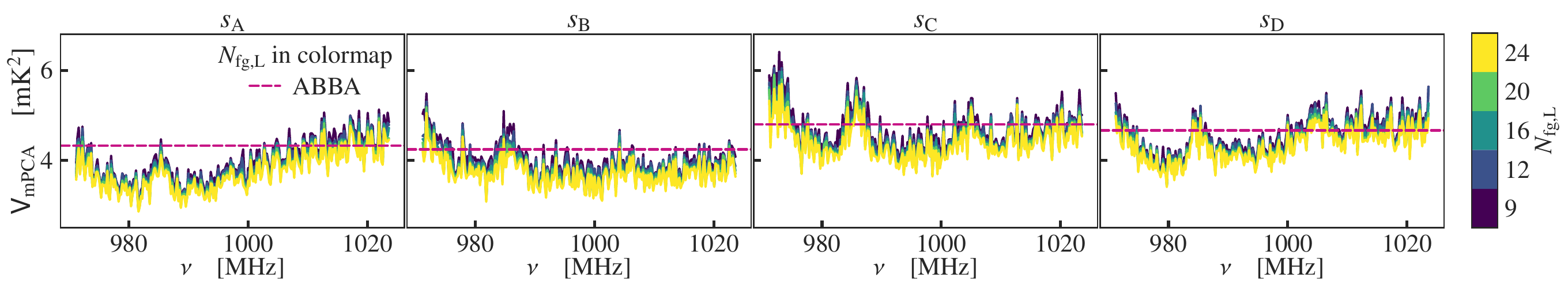}
    \caption{Variance of the residuals after performing the mPCA contaminant subtraction. The colour bar shows the number of modes removed from the large scales of the maps (from dark to light colours for increasing aggressiveness of the cleaning), while we fix the small scale cleaning to $\Nfgs=16$, as there are no major differences in the variance when varying it. For comparison, we plot the averaged ABBA noise level as estimated in \cref{sec:abba} with a dashed magenta line.}
    \label{fig:mPCA_variance}
\end{figure*}

\begin{figure*}
    \centering
    \includegraphics[width=0.98\textwidth]{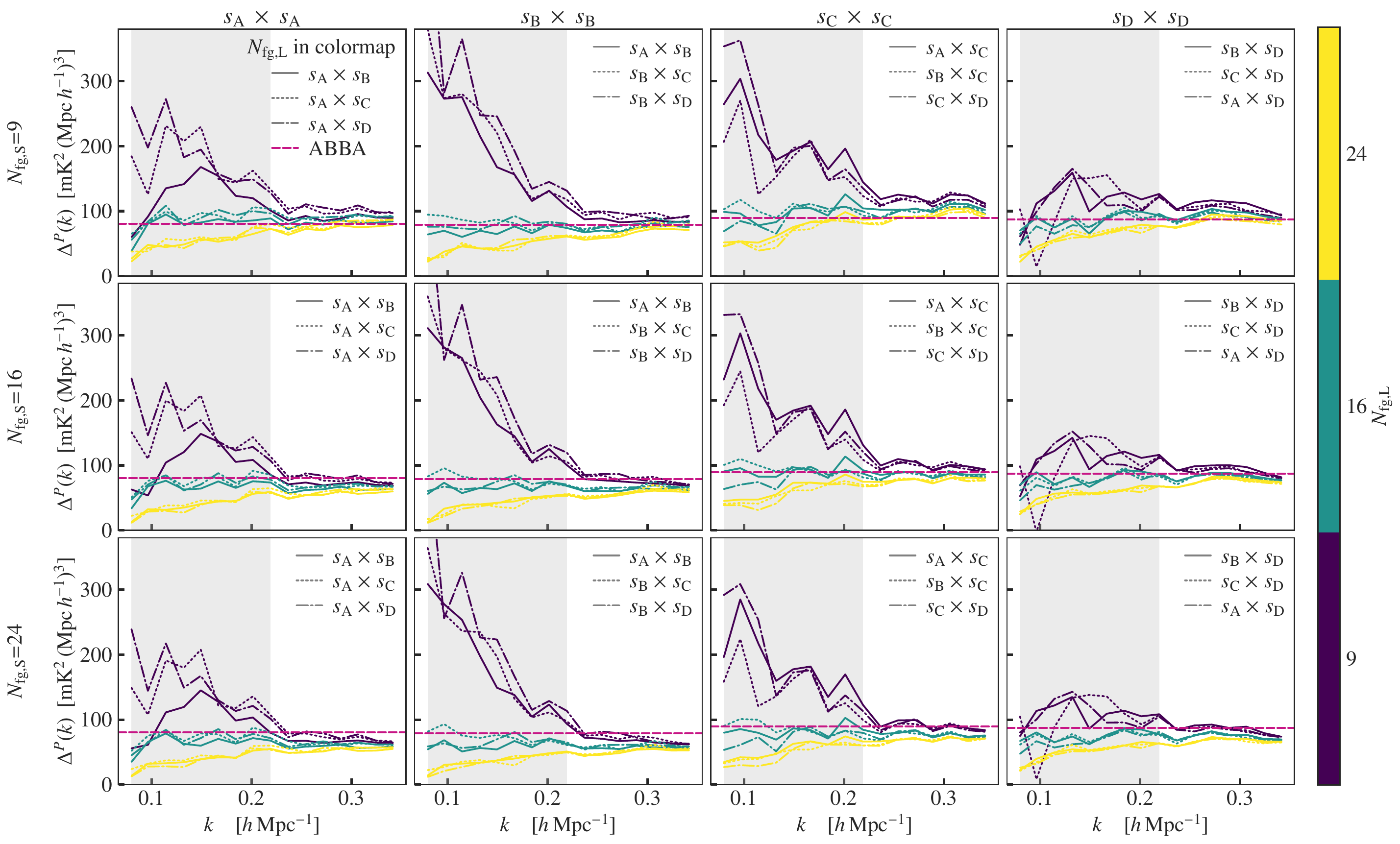}
    \caption{The difference between auto-$P(k)$ and the corresponding cross-$P(k)$. In the subsequent rows we increase the strength of cleaning at small scales ($\Nfgs=9,\,16,\,24$), while the cleaning at large scales is specified in the colour bar. Different line styles refer to the different possible cross-correlation that can be computed for each considered subset. We highlight the region where the telescope beam dampens at least $90\%$ of the signal as the non-shaded area. The thermal noise power spectrum estimated using the ABBA variance is represented with the magenta dashed line. }
    \label{fig:autodiff}
\end{figure*}
\begin{figure*}
        \centering
        \includegraphics[width=0.98\textwidth]{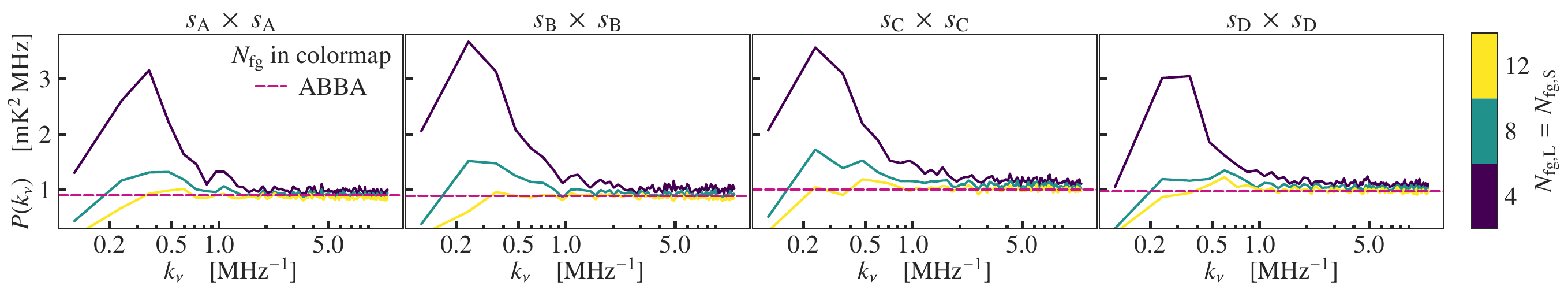}
        \caption{One-dimensional auto-spectra $P(k_\nu)$ for the \chess\ division subsets, $s_{\rm A}$ to $s_{\rm D}$ from left to right, computed for different levels of cleaning (see colour bar). For simplicity, we fix $\Nfgl=\Nfgs$.}
        \label{fig:radial_auto}
\end{figure*}

We show the results in \cref{fig:mPCA_variance}: they match expectations. It is interesting to see how the frequency behaviour of noise differs from subset to subset: data (and contaminant separation processes) are indeed distinct for each of them.

Another estimate comes from the comparison of the measured auto-spectrum of each subset, $P_{ii}$, and the cross-spectra in which the considered subset is involved, $P_{ij}$, since they should only differ by the additive noise term in the autos $P_{ii}$. The $\Delta^P_{ij}=P_{ii}-P_{ij}$ subtraction also informs us whether the `extra' noise term in the autos comes from thermal noise only (flat spectrum) or deviates from that. To compute $\smash{\Delta^P_{ij}}$, we measure the auto- and cross-spectra on a broader $k$-range, using $16$ bins spanning from $k_{\rm min} = 0.07\,\kum$ to $k_{\rm max} = 0.35\,\kum$, without any cut in $k_\parallel$ or $k_\perp$. Firstly, we point out that the condition of Gaussian noise is not met at large scales, as illustrated in the plots of \cref{fig:autodiff}, where the trend for different levels of cleaning is shown. Indeed, we experience important excess of signal if $\Nfgl<12$--$15$ modes are removed: the lack of a modelling for such a noise term---most likely due to systematics and coupled sky components---was one of the reasons why we excluded the auto-spectra from the analysis of this work. Instead, when the beam suppression is dominant (at $k>0.2\,\kum$), we are indeed left with thermal noise, as the match with the ABBA estimate indicates (dashed magenta lines in \cref{fig:autodiff}). Additionally, also these plots show how thermal noise is lost after increasingly aggressive cleanings.

Finally, the radial power spectra introduced in \cref{sec:pknu_residual} offer an additional comparison with the noise level. For a strong cleaning, the auto-$\smash{P(k_\nu)}$ would be dominated by the noise and we show in \cref{fig:radial_auto} that there is a indeed a good agreement between the ABBA estimate and the trend of the $\smash{P(k_\nu)}$ of each subset, except for the very first bins that, however, we exclude from the analysis due to the potential contamination left (see \cref{sec:pknu_residual}). 

%%%%%%%%%%%%%%%%%%%%%%%%%%%%%%%%%%%%%%%%%%%%%%%%%%
%================== APPENDIX B ==================%
%%%%%%%%%%%%%%%%%%%%%%%%%%%%%%%%%%%%%%%%%%%%%%%%%%

\section{The radial power spectrum as diagnostic for residual foregrounds} \label{app:radial_valid}

In this appendix, we show and validate the radial power spectrum $\smash{P(k_\nu)}$ as a diagnostic for the presence of residual astrophysical foregrounds in the cleaned data.

\begin{figure}
    \centering
    \includegraphics[width=\columnwidth]{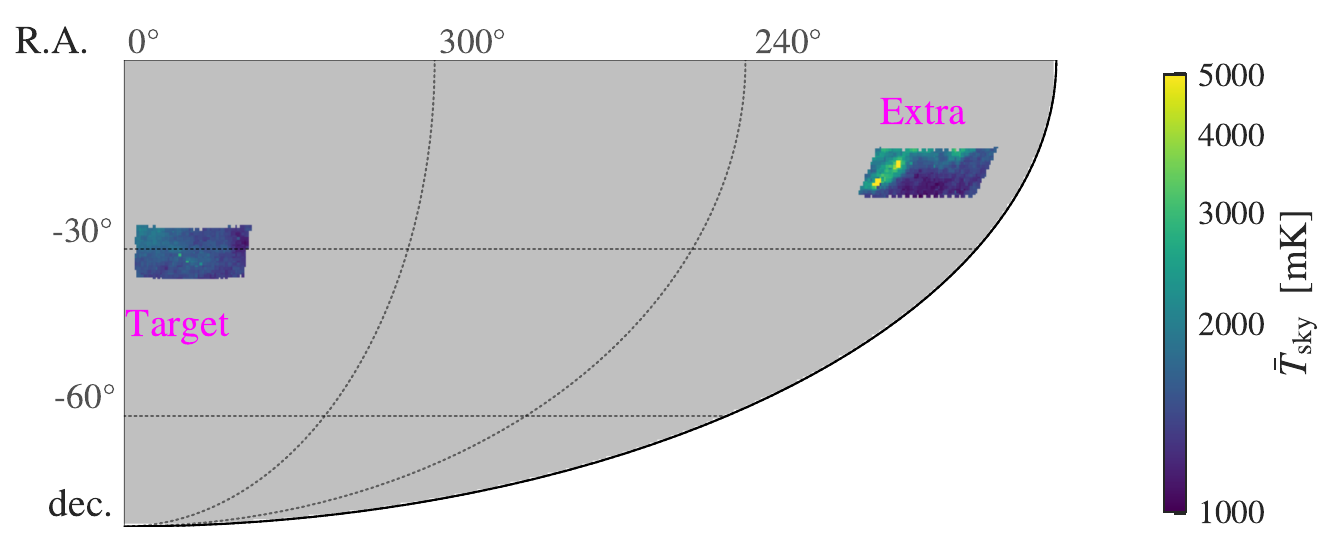}
    \caption{Location and size of the \textit{target} and \textit{extra} region, plotted here with all the contributions in the simulation: cosmological signal, contaminants and thermal noise.}
    \label{fig:TFpatches}
\end{figure}

For this purpose, we rely on the publicly available foreground simulations of \citet{carucci_dataset}, which provide 400-channel full-sky HEALPix maps of foregrounds of Galactic (synchrotron and free-free emission) and extragalactic (point sources) origin, as well as polarization leakage in the frequency range $900$--$1300\,\fum$; i.e., the channel width is roughly 5 times larger than the observation one. For the \hi\ signal, we generate a log-normal realisation within the same redshift range, which we then project onto a HEALPix map for each frequency channel using Monte-Carlo sampling methods. The maps are also smoothed with a frequency-dependent Gaussian filter to simulate the impact of the telescope beam and we add white noise realisations using \cref{eq:radiometer}. We extract the cosmological signal, foregrounds, polarization leakage and noise from the region corresponding to the actual observed field (the \textit{target} region shown in \cref{fig:TFpatches}). Furthermore, we select a batch of 200 frequency channels centred at the median frequency of the observations, hence comparable to the number of channels of the true, observed data set. Indeed, since simulations cannot achieve the same narrow channel width as the observations, the 200-channel simulated cube allows us to apply the estimator under comparable conditions and, most importantly, to probe a range of scales shifted to lower values of $k_\nu$, albeit losing resolution at smaller scales. This is precisely what we are interested in, as we use the $\smash{P(k_\nu)}$ estimator as a diagnostic for contamination and systematics affecting the largest radial scales.

\begin{figure}
    \centering
    \includegraphics[width=\columnwidth]{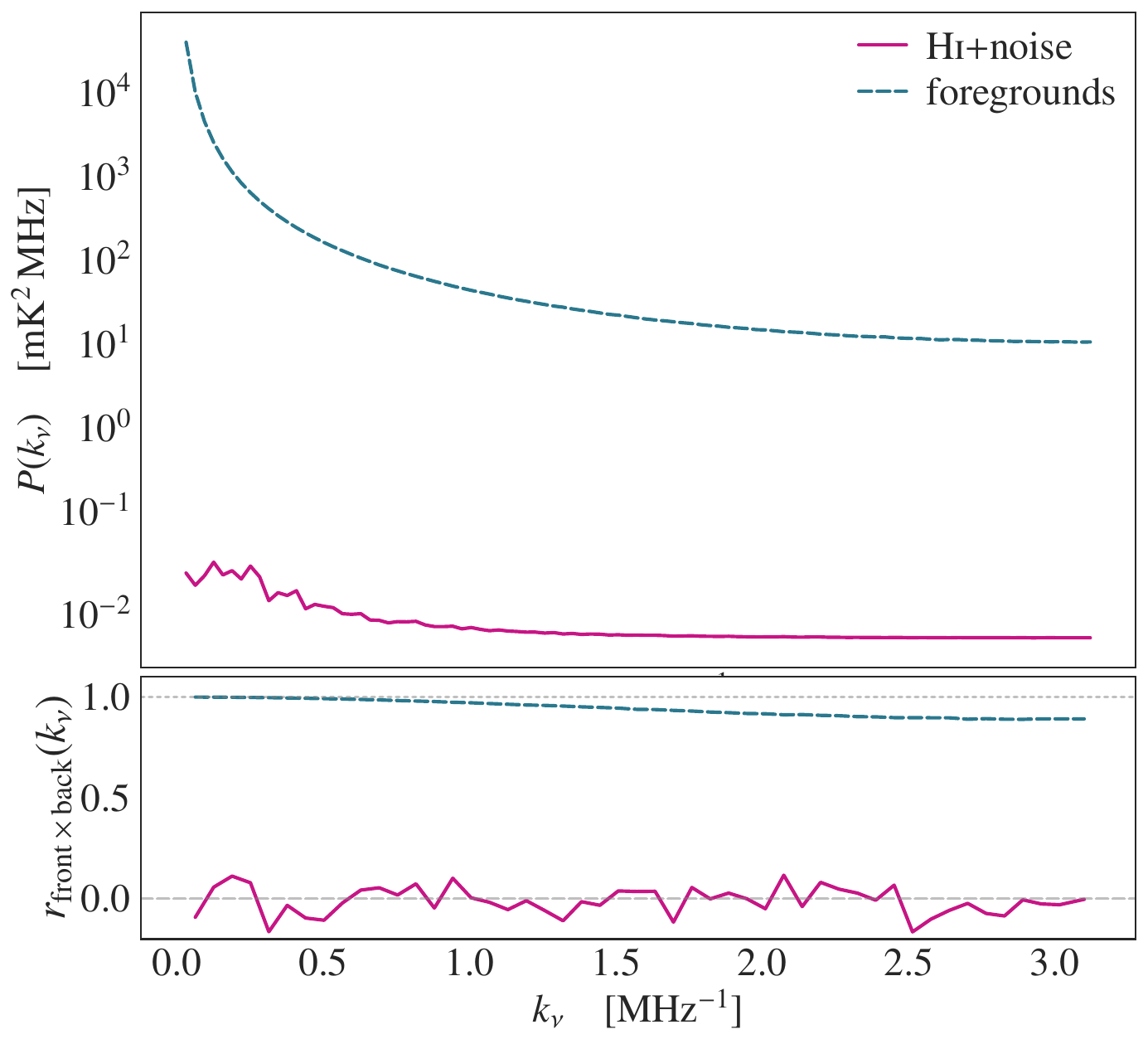}
    \caption{\textit{Top panel: } The radial power spectra computed for the 200-channel simulation. We use a symmetric logarithmic scale (with a threshold for switching to a linear scale at $\left|P(k_\nu)\right|<0.04\,{\rm mK^2\,MHz}$ to better visualise the scaling of the signal arising from the auto-correlation of foregrounds or of \hi\ and noise . \textit{Bottom panel}: cross-correlation coefficient between the front and the back batches in frequency (as in \cref{eq:r_frontback}), for the \hi+noise signal and foregrounds. Highlighted with a dashed and a dotted grey line respectively the cases of absence of correlation and total correlation respectively.}
    \label{fig:Pknu_sims}
\end{figure}

For a better understanding of the behaviour of the different components, we keep separate the contaminants from the cosmological signal and noise. Following the same steps described in \cref{sec:pknu_residual}, we compute the \hi+noise and the foreground power spectra both for the full frequency range or for the front and back halves of the simulations. In \cref{fig:Pknu_sims} we show the trend of these $\smash{P(k_\nu)}$. Foregrounds are orders of magnitude stronger than the cosmological signal and exhibit a much sharper scale dependence compared to \hi\ (and noise). In the right panel, we show the ratios between the cross-$\smash{P(k_\nu)}$ between the disjoint frequency ranges and the geometrical average of the auto-$\smash{P(k_\nu)}$ of the two halves. For a pure cosmological \hi\ signal, the cross-correlation vanishes, and this holds also in the presence of noise (as long as the noise is Gaussian). Oppositely, this is not the case when considering foregrounds, for which the cross-correlation signal is of the same order of the auto-correlations, as a consequence of their frequency coherence. 

%%%%%%%%%%%%%%%%%%%%%%%%%%%%%%%%%%%%%%%%%%%%%%%%%%
%================== APPENDIX C ==================%
%%%%%%%%%%%%%%%%%%%%%%%%%%%%%%%%%%%%%%%%%%%%%%%%%%

\section{Signal loss and its transfer function} \label{app:tf_valid}

\begin{figure*}
    \centering
    \includegraphics[width=\textwidth]{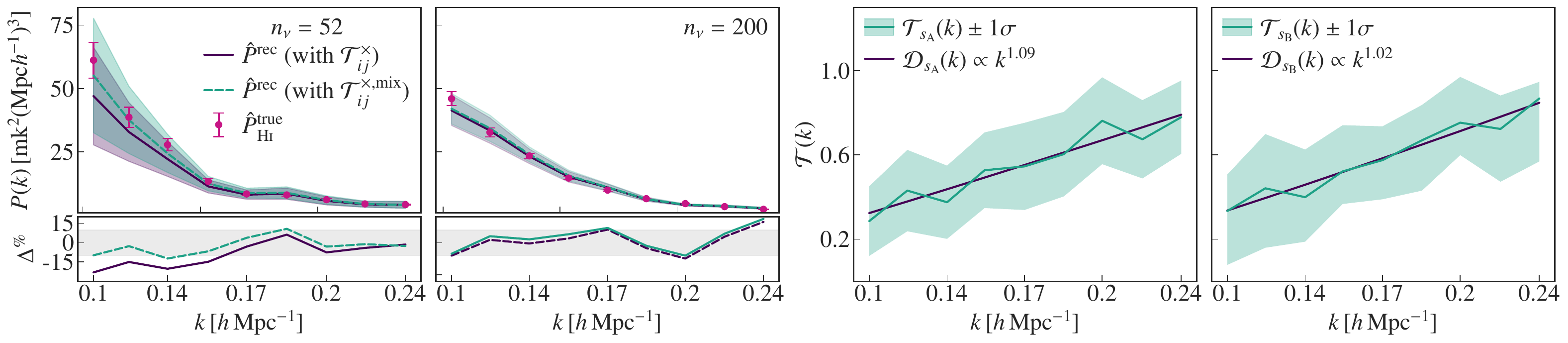}
    \caption{\textit{Left panels:} Performance of the transfer function $\mathcal{T}^{\times}_{ij}$ defined in \cref{eq:TFX} (dark lines) or by its generalisation $\mathcal{T}^{\times, {\rm mix}}_{ij}$ defined in \cref{eq:TFdefmix} (light lines) for the 52- and 200-channel configurations, focusing on the cross-correlation between the realisation made with the foregrounds and polarization leakage from the \textit{target} region (cleaned with $\Nfgl=6$, $\Nfgs=3$) and the one with foregrounds and no polarization leakage from the \textit{extra} region (cleaned with $\Nfgl=3$, $\Nfgs=6$). The number of principal components removed of each subset was selected in order to show a case where the involved subsets underwent a different cleaning, but all the other cases were tested and lead to similar results. In the upper panels, we show the reconstructed power spectra with their corresponding uncertainties (shaded regions) and we compare them to the true \hi\ power spectrum, depicted with analytical error bars in magenta. In the bottom panels we show the percentage difference between the reconstructed power spectra and the pure \hi\ power spectra, highlighting the $10\%$ region with a the grey shaded area. \textit{Right panels:} The agreement between the transfer function corresponding to subsets $s_{\rm A}$ and $s_{\rm B}$, cleaned as in the main analysis, and a best-fit power law function in the same form of the signal-loss damping kernels presented in \cref{sec:model}, whose slope parameter $\beta$, reported in the legend, is obtained via a least-square fit to the computed transfer function.}
    \label{fig:TF_sims}
\end{figure*}

A scale-dependent transfer function has been proposed and adopted in previous measurements of line-intensity mapping power spectra to quantify the signal loss due to foreground cleaning and correct for it at the power spectrum level \citep[\citetalias{MKxGAMA};][]{Masui2013,Switzer2013,Switzer2015,Anderson2018,Wolz2022,Cunnington2023,MKxWZ}. However, a transfer function was never applied to a power spectrum estimated from two independent \hi\ cubes with independent cleanings---hence different loss (see Fig. 12 in \citet{Cunnington2023} and its related discussion). For these reasons, and because work is ongoing in the community to test and improve the transfer function approach in general \citep{Cunnington2023,Carucci2024,Zhaoting2025QE}, we opted to forward-model the signal loss at the power spectrum level (\cref{sec:signal_loss}) and marginalise over it during the inference step (\cref{sec:MCMCfit}). In this appendix, we validate the functional form chosen for the damping kernel of \cref{eq:Dsl} and corroborate the framework of the signal injection tests performed in \cref{sec:injection}.

We adapt the recipe from \citet{Cunnington2023} to our context. We start from the \hi\ power spectrum model described in \cref{sec:model} to produce \hi\ lognormal mocks (that we dub $m$ as opposed to the real observations $s$) on a Cartesian grid with the same resolution as the comoving boxes on which the data are re-gridded to fully embed the MeerKLASS field. The pure \hi\ mocks $m_\hi$ are then projected to celestial coordinates (and to frequency space along the radial directions) by means of a Monte-Carlo sampling technique analogous to the one employed for the gridding (\cref{sec:regrid}), but reversing the process (i.e., the sampling particles are assigned to each Cartesian cell and then converted to celestial coordinates) and the mock maps are smoothed with a Gaussian beam with $\theta_{\rm FWHM}=1.27\,\deg$, corresponding to the median frequency of the observed band. The \hi\ mocks are then injected into the observed maps of each subsets and, for consistency, the $n$th mock relative to each subset is built injecting the same \hi\ realisation into each subset's real data, since, also in observations, the underlying cosmological signal is common to all subsets. 
We apply the same contaminant separation procedure to those $m_{\rm obs}=m_{\hi}+s$ cubes and estimate the signal lost from the mocks. We refer to the maps with the injected signal and actual data after applying the same foreground cleaning procedure with $m_{{\rm clean},i}$. 

We can compute a transfer function for each subset $i$:
\begin{equation}\label{eq:TFdef}
    \pazocal{T}_i(k)=\left\langle\frac{\pazocal{P}\left(m_{{\rm clean},i},m_\hi\right)}{\pazocal{P}\left(m_\hi,m_\hi\right)}\right\rangle_{N_{\rm mock}} \;,
\end{equation}
where $\left\langle\cdot\right\rangle_{N_{\rm mock}}$ stands for the average over the $N_{\rm mock}=100$ mocks and the operator $\pazocal{P}$ is the power spectrum estimator (the same used for the data, \cref{sec:pestim}, including the spherical averaging). However, only one transfer function enters the formula for estimating the loss in the measured power spectrum (\cref{eq:TF_std}): which one should be used when the spectrum involves two different cubes $s_i$ and $s_j$? We test different configurations with a fully under-control simulated setup. 

We use the same set of simulations presented in \cref{app:radial_valid}. In order to reproduce the same conditions as for the data, we not only need to properly define the observed cosmic volume, but it is also essential to replicate the possibility of cross-correlations of independent \hi\ maps that might be influenced by foregrounds with different intensities or by distinct systematics. We adopt the following strategy: on top of the cosmological signal from the \textit{target} region, we add three possible foreground realisations that differ in the components included or in the sky area covered; in particular, two realisations are selected from the \textit{target} region but including or not the contamination by polarization leakage, and a third one comes instead from a different patch of the sky (the \textit{extra} patch, with the same size as the \textit{target}) and with the polarization leakage taken into account (see \cref{fig:TFpatches}). Independent white noise realisations are added to each of the final patches.
In this way, we effectively mimic the division of our observations into multiple subsets that share the same cosmological information, receive independent thermal noise contributions, and are contaminated to potentially different extents. Moreover, since it has been demonstrated that the severity of the signal loss and the cleaning performance are determined by the number of frequency channels available \citep{Carucci2020} and, more specifically, their ratio to the number of pixels \citep{Switzer2013}, we decide to use the 200-channel simulation as well as a smaller $52$-channel only. Both scenarios are close to the real data set, albeit for different reasons: in $52$-channel case, we focus on the same frequency range as the one observed, whereas when using 200 channels we are mostly interested in preserving the same ratio between the number of frequency channels and number of pixels. To ensure consistency with the analysis of the observed subsets, we use the same pipeline as for the data. Finally, we evaluate the signal loss with the recipe described at the beginning of this appendix and 100 lognormal mocks of the \hi\ signal.

Eventually, we find that the most robust approach to estimate the loss of power in the cross-correlation is to use the transfer function corresponding to the subset that underwent the strongest large-scale cleaning (\cref{fig:TF_sims}). In practice, in the case of the cross-correlation of two subsets $s_i$ (cleaned with $N_{{\rm fg,L,s}_i}$) and $s_j$ (cleaned with $N_{{\rm fg,L,s}_j}$), the signal loss can be estimated by computing
\begin{equation} \label{eq:TFX}
    \pazocal{T}^\times_{ij}(k)=\begin{cases}
            \pazocal{T}_i(k) \quad \text{if} \; N_{{\rm fg,L,s}_i} \geq N_{{\rm fg,L,s}_j} \\
            \pazocal{T}_j(k) \quad \text{if} \; N_{{\rm fg,L,s}_i} < N_{{\rm fg,L,s}_j}
        \end{cases} \;. 
\end{equation}
We interpret the result above as follows. The cross-spectrum measures the clustering signal common to the involved subsets, regardless of whether it originates from cosmological, noise, or contaminant sources. Consequently, if one of the subsets has less cosmological signal, the signal retained in the other subset cannot contribute to the power spectrum, as it lacks a counterpart to correlate with. The key question is then what determines the amount of cosmological signal that is left and, most importantly, the amount of left cosmological signal in common between the subsets. Since the number of principal components removed during the contaminant separation is a reliable indicator of the amount of signal lost \citep{Carucci2020,cunnington2021,Cunnington2023}, we can use it to pin down the mostly signal-deprived subset. 

Having defined the transfer function, it is therefore possible to compute the reconstructed power spectrum by applying the transfer function $\pazocal{T}^\times_{ij}(k)$ as in \cref{eq:TF_std}. We do so for the mock-injection tests of \cref{sec:injection}. The remaining issue concerns which uncertainties to associate with the reconstructed power spectrum samples. In this work, we do not use them, as the reconstructed spectra are compared with the inferred spectra at the inferred confidence level. However, for completeness, here we go through our findings in this respect.

To account for the variances related to both involved subsets, we define another transfer function, thereby generalising \cref{eq:TFdef}:
\begin{equation}\label{eq:TFdefmix}
    \pazocal{T}^{\times,{\rm mix}}_{ij}(k)=\left\langle\frac{\pazocal{P}\left(m_{{\rm clean},i},m_{{\rm clean},j}\right)}{\pazocal{P}\left(m_\hi,m_\hi\right)}\right\rangle_{N_{\rm mock}} \;.
\end{equation}
It is important to note that in the transfer function defined above it is crucial to subtract the cleaned data from the cleaned mocks, because otherwise there would be an excess of signal in the power spectrum of the cleaned mocks as both the cosmological signal of the lognormal realisation and the one of the real data signal would contribute, resulting in a roughly doubled contribution to the power spectrum. Because of the latter, we find it more challenging for $\mathcal{T}^{\times, {\rm mix}}_{ij}$ to converge when dealing with real observations. However, we believe that the scatter in $\mathcal{T}^{\times, {\rm mix}}_{ij}$ is more representative than the one from $\pazocal{T}^\times_{ij}(k)$ because, as said, it does not underestimate the covariance contribution from either of the two subsets.

In practice, we estimate the covariance of the reconstructed measurement through the scatter of the distribution of the transfer function across the mocks, as shown in \citetalias{MKxGAMA}. By applying error propagation through summation in quadrature, we have
\begin{equation}
    \frac{\delta \widehat P^{\rm rec}_{ij,m}(\kv)}{\widehat P^{\rm rec}_{ij}(\kv)} = \sqrt{\left(\frac{\delta \widehat P_{ij,m}(\kv)}{\widehat P_{ij}(\kv)}\right)^2 + \left(\frac{\delta \pazocal{T}^{\times,{\rm mix}}_{ij,m}(\kv)}{\pazocal{T}^\times_{ij}(\kv)}\right)^2} \;,
\end{equation}
where the suffix $m$ runs over the $N_{\rm mock}$ mocks. For each mock $m$ the scatter of its distribution can be estimated as
\begin{equation}
    \delta\pazocal{T}^{\times,{\rm mix}}_{ij,m}(\kv) = \pazocal{T}^{\times,{\rm mix}}_{ij,m}(\kv) - \pazocal{T}^{\times,{\rm mix}}_{ij}(\kv) \;.
\end{equation}
To get the final estimate of the covariance, we can assume that the dominant source of scatter in the reconstructed power is the transfer function distribution across the mocks while the contribution contribution from $\delta \widehat P_{ij,m}$ is negligible. Having set these conditions, the estimate of the error on the reconstructed power spectrum is 
\begin{multline}\label{eq:TFcov}
    \varSigma_{\rm TF}[\widehat P^{\rm rec}_{ij}(k),\widehat P^{\rm rec}_{ij}(k')]
    = \frac{N_{\rm mock}-1}{N_{\rm mock}-n_k-2} \\ 
    \times\widehat P^{\rm rec}_{ij}(k)\,\widehat P^{\rm rec}_{ij}(k')
    \left\langle
    \frac{\delta\mathcal{T}^{\times,{\rm mix}}_{ij}(k)}{\mathcal{T}^\times_{ij}(k)}
    \;
    \frac{\delta\mathcal{T}^{\times,{\rm mix}}_{ij}(k')}{\mathcal{T}^\times_{ij}(k')}
    \right\rangle_{N_{\rm mock}} \; .
\end{multline}
In summary, we associate the covariance in \cref{eq:TFcov} to the reconstructed spectra with the transfer function described in \cref{eq:TFX}. We point out that using one transfer function for the correction and another for the covariance was already in place for the standard case \citep{Cunnington2023}, which here we have generalised. 

We assess the performance of $\mathcal{T}^{\times}_{ij}$ and $\mathcal{T}^{\times, {\rm mix}}_{ij}$ considering the uncertainties defined in \cref{eq:TFcov}, both on the 200- and 52-channel configurations in the two left panels of \cref{fig:TF_sims}. Overall, what impacts the accuracy of the correction the most is the number of frequency channels, as expected. 

Finally, we use the transfer functions computed here and in the main text as a testing ground for the functional form of the damping kernel $\mathcal{D}(k)$ described in \cref{sec:signal_loss}. A couple of examples are in the last two panels of \cref{fig:TF_sims}; the agreement with the power-law form is solid across all cases explored.

%%%%%%%%%%%%%%%%%%%%%%%%%%%%%%%%%%%%%%%%%%%%%%%%%%

% Don't change these lines
\bsp	% typesetting comment
\label{lastpage}
\end{document}